\pdfoutput=1

\documentclass[11pt,twoside,a4paper,cmspaper,final,collab]{cms-tdr}
\def\svnVersion{0368a96-D}\def\svnDate{2019/10/29}\def\cmsCernNoTag{CERN-EP-2019-007}\def\cmsCernDate{\today}\def\cmsMessage{"Published in the European Physical Journal C as \href{http://dx.doi.org/10.1140/epjc/s10052-019-7499-4}{\doi{10.1140/epjc/s10052-019-7499-4}.}"}

\begin{document}\cmsNoteHeader{GEN-17-001}

\hyphenation{had-ron-i-za-tion}
\hyphenation{cal-or-i-me-ter}
\hyphenation{de-vices}
\RCS$HeadURL$
\RCS$Id$
\newlength\cmsFigWidth
\ifthenelse{\boolean{cms@external}}{\setlength\cmsFigWidth{0.85\columnwidth}}{\setlength\cmsFigWidth{0.4\textwidth}}
\ifthenelse{\boolean{cms@external}}{\providecommand{\cmsLeft}{top\xspace}}{\providecommand{\cmsLeft}{left\xspace}}
\ifthenelse{\boolean{cms@external}}{\providecommand{\cmsRight}{bottom\xspace}}{\providecommand{\cmsRight}{right\xspace}}
\ifthenelse{\boolean{cms@external}}{\providecommand{\cmsTable}[1]{#1}}{\providecommand{\cmsTable}[1]{\resizebox{\textwidth}{!}{#1}}}
\newcommand{\MG} {\textsc{mg5}\_a\textsc{mc}}
\newcommand{\eff} {\ensuremath{\sigma_{\text{eff}}}}
\newcommand*{\tmax}{transMAX}
\newcommand*{\tmin}{transMIN}
\newcommand*{\ptmax}{\ensuremath{\pt^\text{max}}}
\newcommand*{\ptsum}{\ensuremath{\pt^\text{sum}}}
\newcommand*{\etaphi}{\ensuremath{\eta\text{-}\phi}}
\newcommand*{\delphi}{\Delta\phi}
\newcommand{\FxFx}{FxFx\xspace}
\newcommand{\ktMLM}{\kt--MLM\xspace}
\newcommand{\RIVET}{\textsc{rivet}}
\newcommand{\PROFESSOR}{\textsc{professor}}
\newcommand{\PYTHIAviii}{\textsc{pythia}8\xspace}

\cmsNoteHeader{GEN-17-001}
\title{Extraction and validation of a new set of CMS \PYTHIAviii tunes from underlying-event measurements}

\date{\today}

\abstract{
New sets of CMS underlying-event parameters (``tunes") are presented for the \PYTHIAviii event generator. These tunes use the NNPDF3.1 parton distribution functions (PDFs) at leading (LO), next-to-leading (NLO), or next-to-next-to-leading (NNLO) orders in perturbative quantum chromodynamics, and the strong coupling evolution at LO or NLO. Measurements of charged-particle multiplicity and transverse momentum densities at various hadron collision energies are fit simultaneously to determine the parameters of the tunes. Comparisons of the predictions of the new tunes are provided for observables sensitive to the event shapes at LEP, global underlying event, soft multiparton interactions, and double-parton scattering contributions. In addition, comparisons are made for observables measured in various specific processes, such as multijet, Drell--Yan, and top quark-antiquark pair production including jet substructure observables. The simulation of the underlying event provided by the new tunes is interfaced to a higher-order matrix-element calculation. For the first time, predictions from \PYTHIAviii obtained with tunes based on NLO or NNLO PDFs are shown to reliably describe minimum-bias and underlying-event data with a similar level of agreement to predictions from tunes using LO PDF sets.}

\hypersetup{
pdfauthor={CMS Collaboration},
pdftitle={Extraction and validation of a new set of CMS PYTHIA8 tunes from underlying-event measurements},
pdfsubject={CMS},
pdfkeywords={CMS, underlying event tune, Monte Carlo event generators, parton shower codes}}

\maketitle
\section{Introduction}
\label{sec:introduction}
{\tolerance=800 Monte Carlo (MC) simulation codes describe hadron-hadron collisions with models based on several components. The hard scattering component of the event consists of particles from the hadronization of partons whose kinematics are predicted using perturbative matrix elements (MEs), along with partons from initial-state radiation (ISR) and final-state radiation (FSR) that are simulated using a showering algorithm.
The underlying event (UE)  consists of the beam-beam remnants (BBR) and the particles that arise from multiple-parton interactions (MPI).  The BBR are what remains after a parton is scattered out of each of the two initial beam hadrons. The MPI are additional soft or semi-hard parton-parton scatterings that occur within the same hadron-hadron collision. Generally, observables sensitive to the UE also receive contributions from the hard-scattering components. Accurately describing observables that are sensitive to the UE not only requires a good description of BBR and MPI, but also a good modeling of hadronization, ISR, and FSR. Standard MC event generators, such as \PYTHIAviii~\cite{pythia8}, \HERWIG~\cite{herwigpp,Bellm:2015jjp}, and \SHERPA~\cite{Gleisberg:2008ta} have adjustable parameters to control the behavior of their event modeling.  A set of these parameters, which has been adjusted to better fit some aspects of the data, is referred to as a tune. \par}

{\tolerance=1200 In a previous study~\cite{Khachatryan:2015pea}, we presented several \PYTHIAviii and \HERWIGpp UE tunes constructed for a center-of-mass energy $\sqrt{s}$ lower than 13\TeV.
The CMS \PYTHIAviii tune CUETP8M1 is based on the Monash tune~\cite{Skands:2014pea}, using both the NNPDF2.3LO parton distribution function (PDF) set~\cite{Carrazza:2013axa}. The CMS \PYTHIAviii tune CUETP8S1-CTEQ6L1 is based instead on the tune 4C~\cite{Corke:2010yf}.
Both tunes CUETP8M1 and CUETP8S1-CTEQ6L1 were constructed by fitting the CDF UE data at $\sqrt{s}=900\GeV$ and 1.96\TeV~\cite{Aaltonen:2015aoa} together with CMS UE data at $\sqrt{s}=7\TeV$~\cite{CMS:2012zxa}. A similar procedure was used for the determination of the \HERWIGpp tune (CUETHppS1) with the CTEQ6L1 PDF set \cite{cteq6l1}. A collection of previously published tunes is documented in~\cite{Skands:2014pea,ATL-PHYS-PUB-2014-021,Corke:2010yf,ATLAS:2012uec,Aad:2014xaa,ATL-PHYS-PUB-2014-021,Fischer:2016zzs}.\par}

In this paper, a new set of tunes for the UE simulation in the \PYTHIAviii (version 8.226) event generator is obtained by fitting various measurements sensitive to soft and semi-hard MPI at different hadron collision energies~\cite{Aaltonen:2015aoa,CMS:2012zxa}, including data from $\sqrt{s}=13\TeV$~\cite{CMS:2015zev}. These tunes are constructed with the leading order (LO), next-to-leading order (NLO), and next-to-next-to-leading order (NNLO) versions of the NNPDF3.1 PDF set \cite{Ball:2017nwa} for the simulation of all UE components. Typically, the values of strong coupling used for the simulation of the hard scattering are chosen consistent with the order of the PDF set used.

{\tolerance=800 The new tunes are obtained by fitting CDF UE data at $\sqrt{s}=1.96\TeV$~\cite{Aaltonen:2015aoa}, together with CMS UE data at $\sqrt{s}=7\TeV$~\cite{CMS:2012zxa} and at 13\TeV~\cite{CMS:2015zev,Khachatryan:2015jna}. For the first time, we show that predictions obtained with tunes based on higher-order PDF sets are able to give a reliable description of minimum-bias (MB) and UE measurements with a similar level of agreement to predictions from tunes using LO PDF sets. We also compare the predictions for multijet, Drell--Yan, and top-antiquark (\ttbar) processes from \PYTHIAviii with new tunes in ME-parton shower (PS) merged configurations. \par}

In Section 2 we describe observables that are sensitive to MB and UE: diffractive processes~\cite{Collins:1977jy}, where one or both protons remain intact after the collision; and double-parton scattering (DPS), where two hard scatterings occur within the same collision. In Section 3, we compare the tunes that were constructed before the data at $\sqrt{s}=13\TeV$ were available (``Pre-13\TeV{}" tunes) with UE data measured at 13\TeV. Section 4 is dedicated to a general discussion of the choice of PDF sets and strong coupling values for the UE simulation. In Section 5 we describe the new tunes. Section 6 shows the validation of the new CMS \PYTHIAviii tunes for multijet, Drell--Yan, \ttbar, and DPS processes. Section 7 is the summary and conclusions.

\section{Observables for characterizing minimum bias, underlying event, and double-parton scattering}

Minimum bias is a generic term that refers to inelastic events that are collected with a loose event selection that has the smallest bias possible. The MB observables are constructed from data with little or no additional selection requirements. The majority of MB collisions are soft, with a typical transverse momentum scale $\pt\lesssim2\GeV$.
The UE is defined as the activity that is not associated with the particles originating from the hard scattering of two partons and is generally studied in events that contain a hard scattering with $\pt\gtrsim2\GeV$.
The main contribution to the UE comes from color exchanges between the beam partons and is modeled in terms of MPI, BBR, and color reconnection (CR). The MB and UE observables have quite different kinematic properties because they are affected by different mixtures of hard and soft scattering processes.

\begin{figure}[ht!]
\centering
\includegraphics[width=0.4\textwidth]{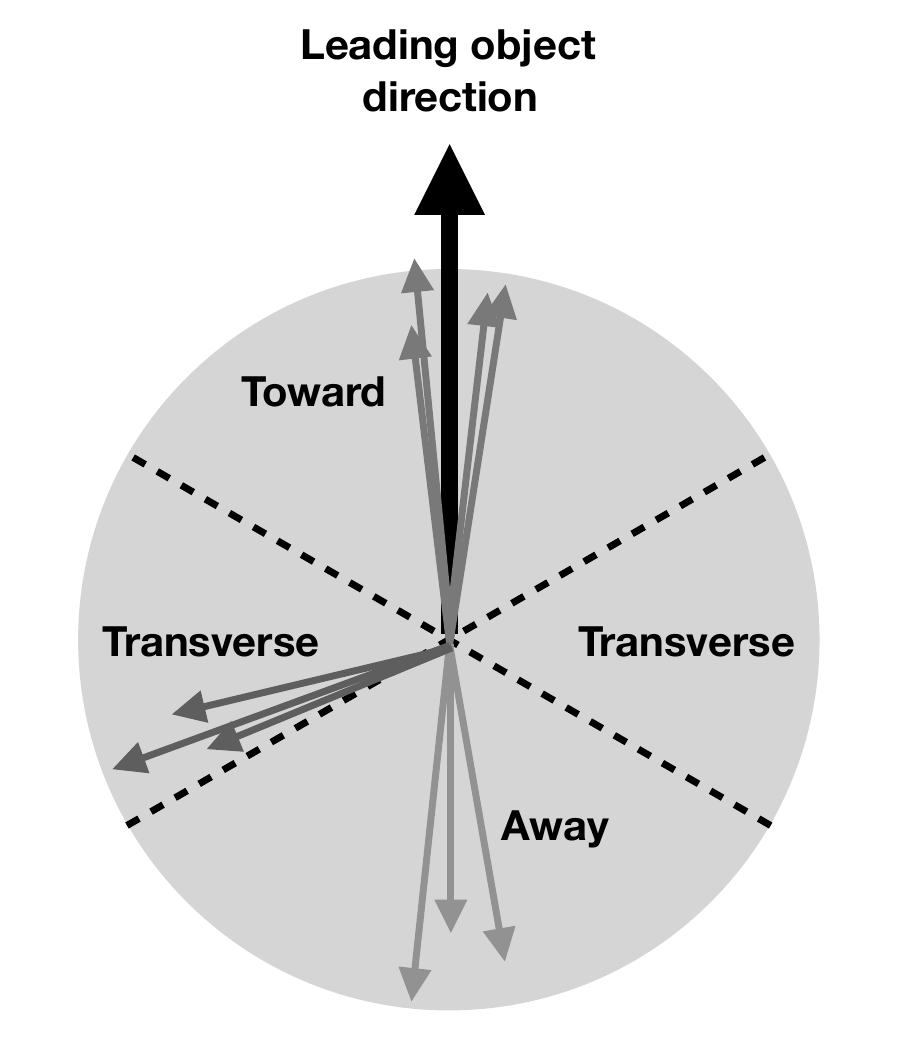}
\caption{Illustration of several $\phi$ regions relative to the leading object
that are sensitive to the
underlying event. See the text for the details on the definitions of the
regions.
}
\label{fig:regions}
\end{figure}

As illustrated in Fig.~\ref{fig:regions}, one can use the topological structure of a typical hard hadron-hadron collision to study the UE experimentally. On an event-by-event basis, a   leading object    is used to define regions of \etaphi~space that are sensitive to the modeling of the UE, where $\eta$ is the pseudorapidity and $\phi$ is the azimuthal scattering angle defined in the $xy$ plane.
The azimuthal separation between charged particles and the leading object, $\delphi\  = \phi - \phi_{\text{max}}$, is used to define the UE-sensitive regions.
Here $\phi_{\text{max}}$ is the azimuth of the leading object and $\phi$ is the azimuth angle of an outgoing charged particle. The regions are labelled as `toward' ($\abs{\delphi}\leq 60^{\circ}$), `away' ($\abs{\delphi}<120^{\circ}$), and `transverse' ($60^{\circ}<\abs{\delphi}\leq 120^{\circ}$).
The transverse region can further be separated into   \tmax\    and  \tmin. On an event-by-event basis, \tmax\ (\tmin) is defined as the transverse region having the maximum (minimum) of either the number of charged particles, or scalar \pt\ sum of charged particles (\ptsum), depending on the quantity under study.

Published UE studies used the charged-particle jet with the largest \pt~\cite{CMS:2015zev}, the dilepton system in DY~\cite{Chatrchyan:2012tb,Sirunyan:2017vio}, or \ttbar~\cite{Sirunyan:2018avv} events as the leading (\ie, the highest \pt) objects. The tunes from CDF and CMS data~\cite{Aaltonen:2015aoa,CMS:2012zxa} made use of the charged particle with the largest \pt ($\ptmax$) as the ``leading object", and use only charged particles with $\pt > 0.5 \GeV$ and $\abs{\eta} < 0.8$ to characterize the UE. The toward region contains the leading object, and the away region is expected to include the object recoiling against the leading one. Most of the UE contributions, \ie, PS and MPI, are contained in the two transverse regions. For events with multiple ISR or FSR emissions, transMAX often contains a third hard jet, while both transMAX and transMIN receive contributions from the MPI and BBR components.  Typically, the transMIN observables are more sensitive to the MPI and BBR components of the UE.

{\tolerance=1800 Observables sensitive to UE contributions are the charged-particle multiplicity and the charged-particle scalar-\pt sum densities in the \etaphi\ space, measured in \tmin\ and \tmax.  The tunes that are constructed by fitting such UE-sensitive observables are referred to as ``UE tunes". \par}

{\tolerance=800 The \PYTHIAviii MC event generator also simulates single-diffractive (SD) dissociation, double-diffractive (DD) dissociation, central-diffractive (CD), and nondiffractive (ND) processes~\cite{Navin:2010kk}, which contribute to the inelastic cross section in hadron-hadron collisions. In SD, CD, and DD  events, one or both of the beam particles are excited into color singlet states, which then decay.
The SD and DD processes correspond to color singlet exchanges between the beam hadrons, while CD corresponds to double color singlet exchange with a diffractive system produced centrally. For ND processes, color exchanges occur, the outgoing remnants are no longer color singlets, and a multitude of particles is produced. All processes except SD are defined as nonSD (NSD) processes.
An NSD-enhanced sample is required to have an energy deposit in both the backward ($-5 < \eta < -3$) and the forward ($3 < \eta < 5$) regions of the detector. The details of the selection for different types of diffractive events can be found in Ref.~\cite{Sirunyan:2018zdc}. \par}

Generally, MC models such as \PYTHIAviii regularize the contributions of the primary hard-scattering processes and MPI to the differential cross section by using a threshold parameter $\pt^0$.
The primary hard-scattering processes and the MPI are regularized in the same way with this parameter. This threshold is expected to have a dependence on the center-of-mass energy of the hadron-hadron collision, $\sqrt{s}$. The threshold at a reference center-of-mass energy $\sqrt{s}=7\TeV$ is called \texttt{pT0Ref}. In \PYTHIAviii
the energy dependence is parameterized
using a power law function with a reference energy parameter $s_0$ and an exponent $\epsilon$.
At a given center-of-mass energy, the amount of MPI depends on the threshold $\pt^0$, the PDF, and the overlap of the matter distributions of the two colliding hadrons. Smaller values of $\pt^0$ result in larger MPI contributions because of a higher MPI cross section.
Each MPI adds colored partons to the final state, creating a dense net of color lines that spatially overlap with the fields produced by the partons of the hard scattering and with each other. All the generated color lines may connect to each other according to the CR model.

Since \PYTHIAviii regularizes both the cross section for MPI and the cross section of collisions with low-\pt exchange using the $\pt^0$ parameter, one can model the overall  ND cross section by letting the \pt ~of the primary hard scattering become small.  In this simple approach, the UE in a hard-scattering process is related to MB collisions.  At the same center-of-mass energy, the activity in the UE of a hard-scattering process is greater than that of an average MB collision. In \PYTHIAviii, this is caused by the higher MPI activity in hard-scattering processes compared to a typical MB collision. By demanding a hard scattering, one forces the collision to be more central, \ie, with a small impact parameter between the protons, and this increases the probability of MPI. For MB collisions, peripheral collisions, where the impact parameter between the two colliding protons is large, are most common.

Typically MPI interactions contain particles with substantially lower \pt (``softer"). However, occasionally two hard 2-to-2 parton scatterings can occur within the same hadron-hadron collision.  This is referred to as DPS.
Tunes that are constructed by fitting DPS-sensitive observables are referred to as ``DPS tunes". Ultimately, one universal tune that simultaneously accurately describes observables in hard scattering events, as well as MB collisions, is desirable.

{\tolerance=1200
The goals of this paper are to produce improved 13\TeV\  \PYTHIAviii tunes with well-motivated parameters, and to provide an investigation of the possible choices that can be made in \PYTHIAviii which simultaneously describe a wide range of UE and MB measurements and are suitable for merged configurations, where a ME calculation is interfaced to the simulation of UE contributions.
\par}

\section{Comparisons of predictions for UE observables from previous tunes to measurements at \texorpdfstring{13\TeV}{13 TeV}}
\label{sec:pre13TeV}
\begin{figure*}[ht!]
\centering
\includegraphics[width=0.49\textwidth]{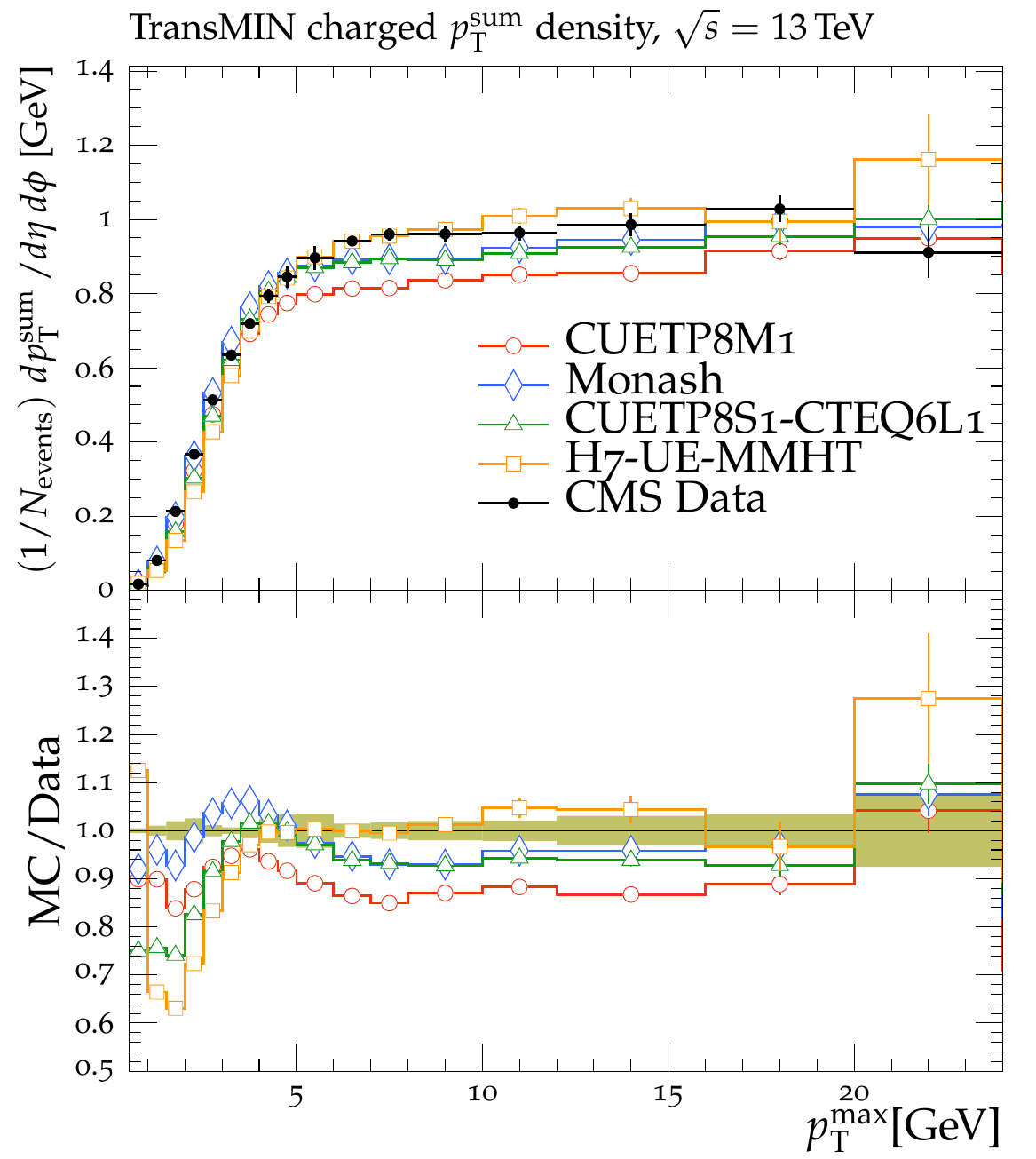}
\includegraphics[width=0.49\textwidth]{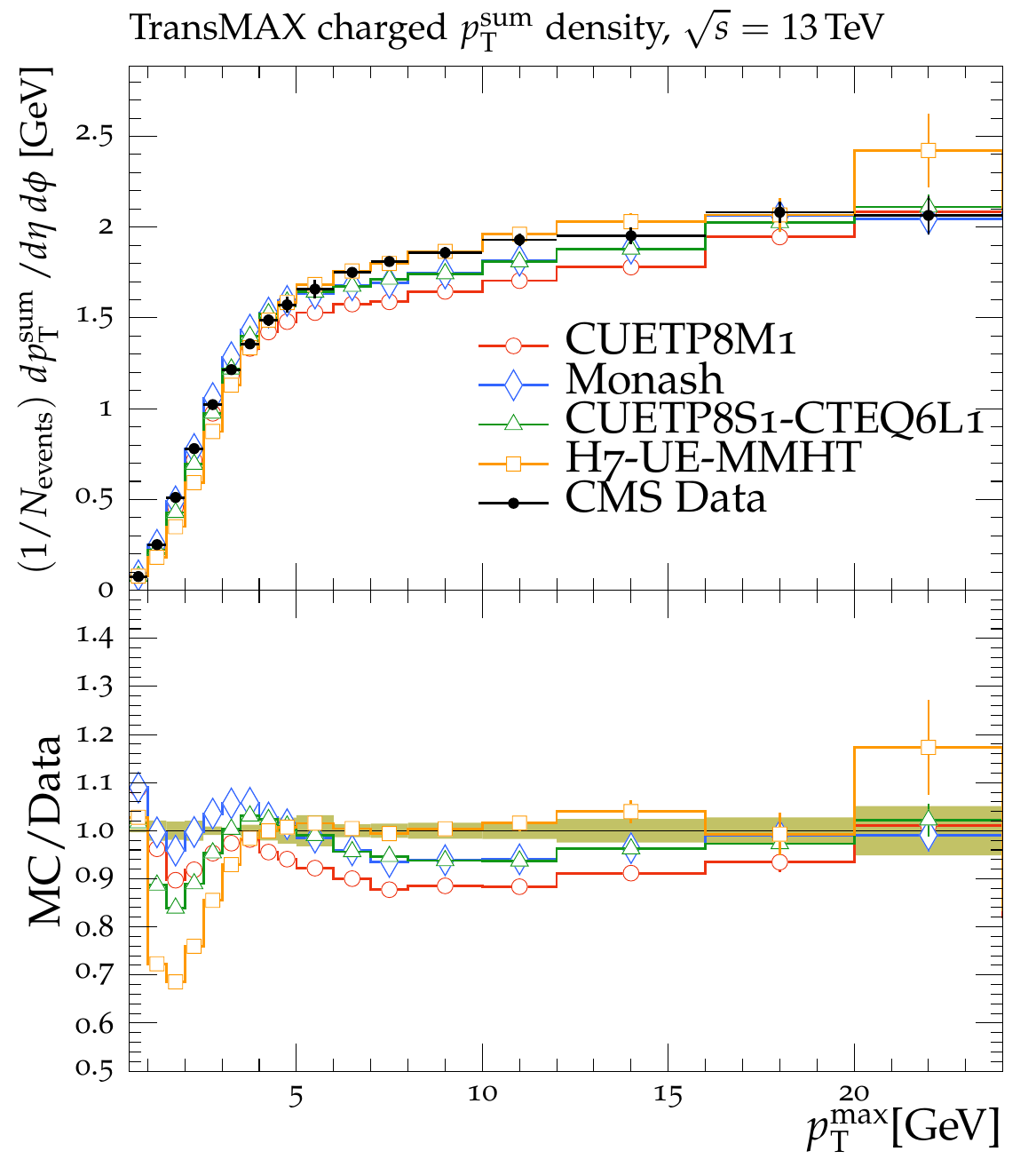}
\includegraphics[width=0.49\textwidth]{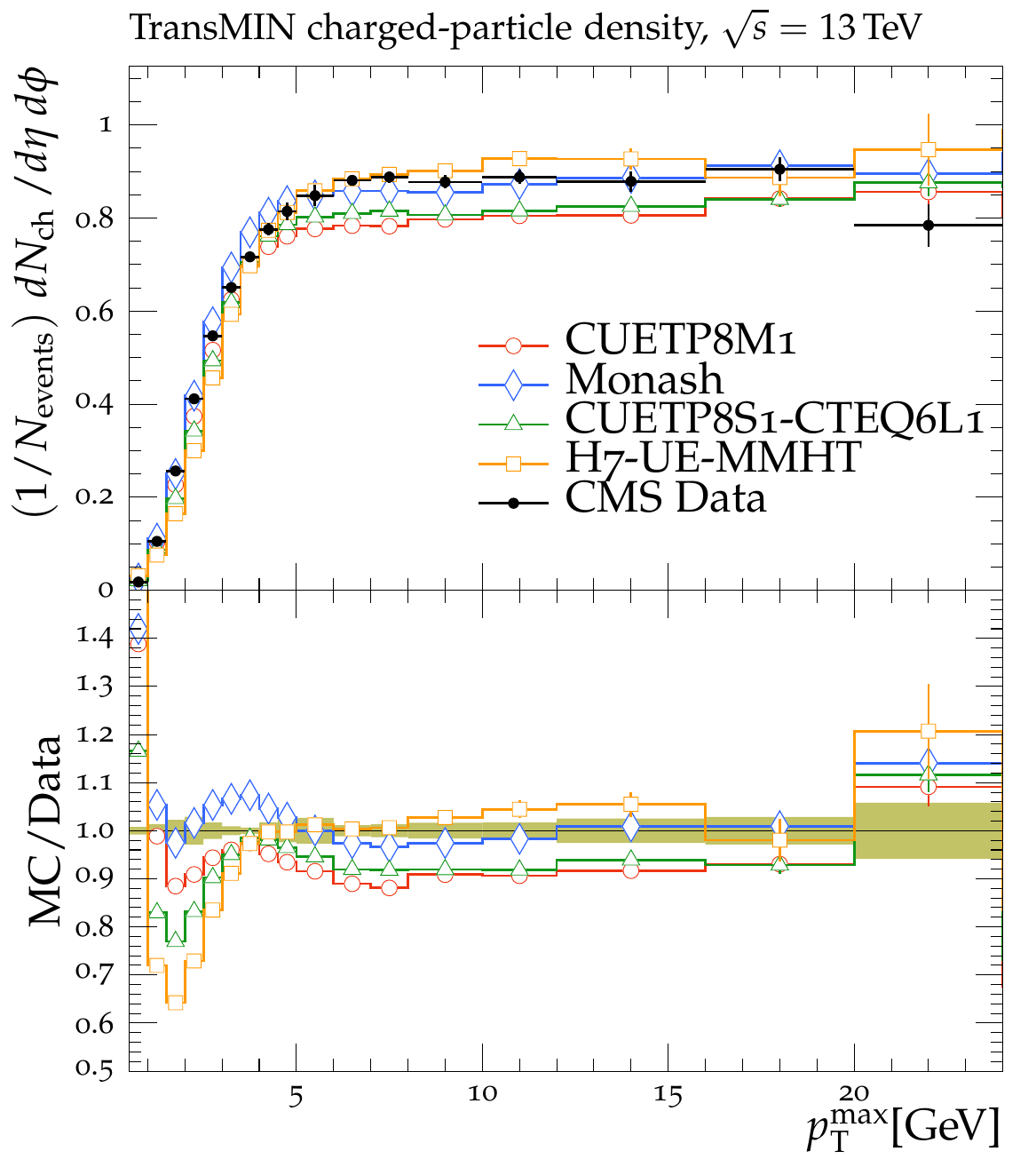}
\includegraphics[width=0.49\textwidth]{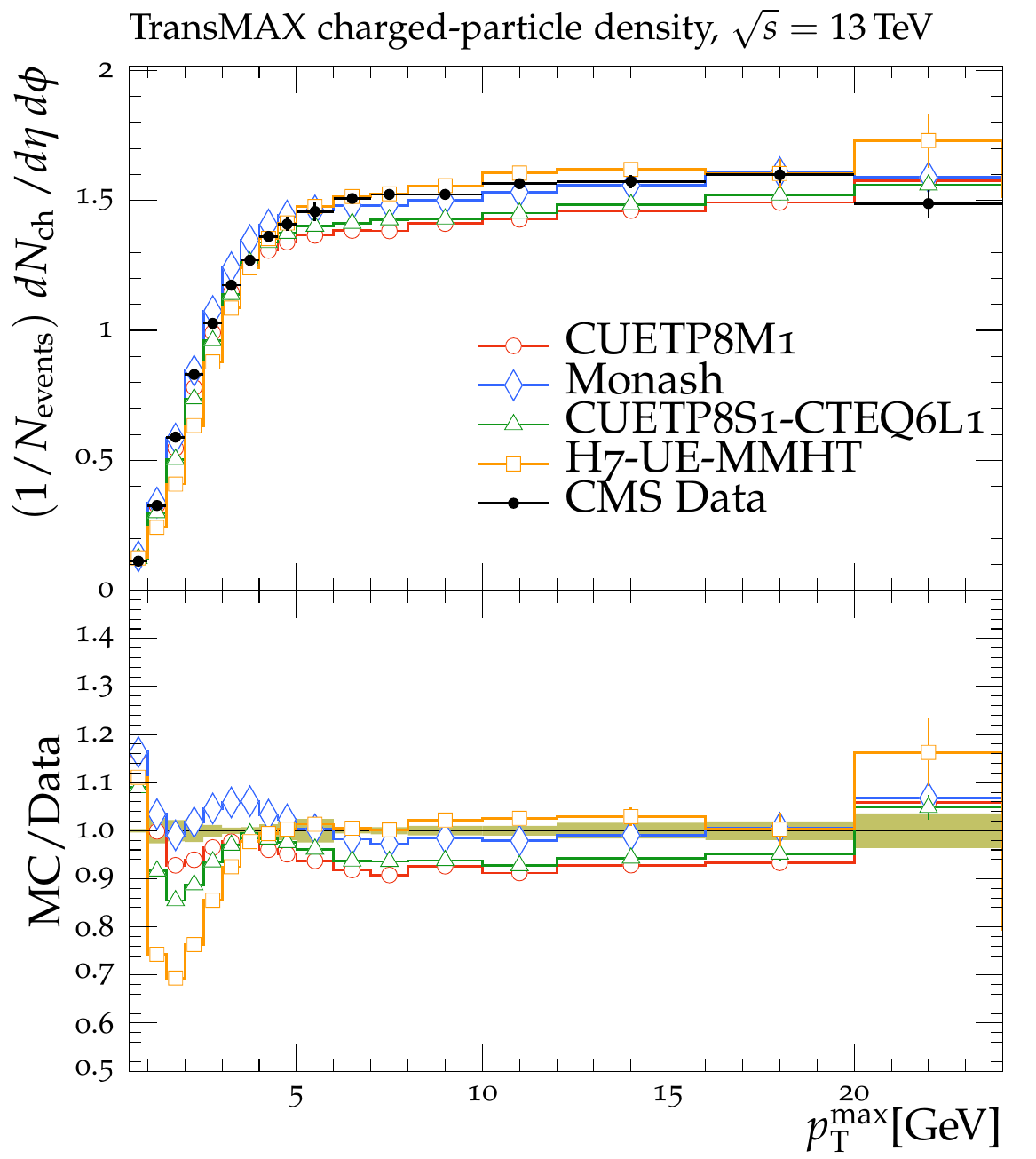}
\caption{The (left column) \tmin\ and (right column) \tmax\, charged-particle \ptsum\ (upper row), and multiplicity (lower row) densities for particles with $\pt>0.5\GeV$ in $\abs{\eta} < 2.0$, as a function of the transverse momentum of the leading charged particle ($\ptmax$), from the CMS $\sqrt{s}=13\TeV$ analysis~\cite{CMS:2015zev}.
The data are compared with the \PYTHIAviii tune Monash, the CMS \PYTHIAviii tunes CUETP8S1-CTEQ6L1 and CUETP8M1, and the {\HERWIG}7 (labelled as ``H7") tune UE-MMHT.
The ratios of the simulations to the data (MC/Data) are also shown, where the shaded band indicates the total experimental uncertainty in the data. Vertical lines drawn on the data points refer to the total uncertainty in the data. Vertical lines drawn on the MC points refer to the statistical uncertainty in the predictions. Horizontal bars indicate the associated bin width.}
\label{fig:rick1}
\end{figure*}

{\tolerance=800 In this section,   comparisons  are presented between data collected at $\sqrt{s}=13\TeV$ and predictions from tunes obtained using fits to measurements performed at lower center-of-mass energies. Figure~\ref{fig:rick1} displays comparisons of CMS data at 13\TeV~\cite{CMS:2015zev} for the \tmin\ and \tmax\  charged-particle \ptsum\ densities, as functions of the leading charged-particle $\ptmax$. The data are compared with predictions from the \PYTHIAviii tunes CUETP8S1-CTEQ6L1~\cite{Khachatryan:2015pea}, CUETP8M1~\cite{Khachatryan:2015pea}, and Monash~\cite{Skands:2014pea}. \par}

{\tolerance=1200 The CMS Monash-based tune CUETP8M1 does not describe the central values of the data at $\sqrt{s}=13\TeV$ well, nor does the original Monash tune.  For example, CUETP8M1 and Monash tunes do not predict enough UE activity in the region with $\ptmax > 5\GeV$ (the ``plateau" region) of transMIN at 13\TeV, with a disagreement of $\approx$10\% and $\approx$5\%, respectively.  The   transMIN    observables are very sensitive to MPI, which suggests that tune CUETP8M1 does not produce enough charged particles at 13\TeV.  In addition, CUETP8M1 does not provide a good fit to the jet multiplicity in \ttbar production either at 8\TeV or at 13\TeV~\cite{TOP-16-007, CMS-PAS-TOP-16-021}. High jet multiplicity \ttbar events are sensitive to the modeling of the ISR.  Hence, CUETP8M1 may not have the proper mixture of MPI and ISR. The ATLAS collaboration has also observed some discrepancies between the predictions of the A14 tune~\cite{ATL-PHYS-PUB-2014-021}, used as standard tune for analyses of 7 and 8\TeV data, and the data at 13\TeV~\cite{Aaboud:2017fwp}.  The CMS UE tunes were constructed by fitting CDF UE data at $\sqrt{s}=900\GeV$ and 1.96\TeV, together with CMS UE data at $\sqrt{s}=7\TeV$. \par}

In Fig.~\ref{fig:rick1} the CMS UE tunes provide a fairly good description of the 13\TeV UE data. Because the CMS UE tunes were obtained by fitting UE observables at various collision energies ($\sqrt{s}=900$, 1960, and 7000\GeV), they underestimate the data at $\sqrt{s}=13\TeV$. This might be an indication of the need to improve the energy extrapolation function implemented in \PYTHIAviii~\cite{Gunnellini:2018kug}. Predictions obtained with the Monash tune, which is the default \PYTHIAviii tune, slightly better reproduce the 13\TeV UE data, but is somewhat worse at describing the UE observables at $\sqrt{s}=900$ and 1960\GeV than the CMS UE tunes.

Predictions from the {\HERWIG}7.1 tune UE-MMHT~\cite{Bellm:2015jjp} are also shown. The H7-UE-MMHT tune was obtained by fitting UE data at $\sqrt{s}=0.9$ and 7\TeV.
This tune is based on the MMHT2014 PDF set~\cite{Harland-Lang:2014zoa} and is able to describe the plateau region of the UE observables at $\sqrt{s}=13\TeV$.
The part of the spectrum at $\ptmax > 5\GeV$ is not well reproduced in the range of the leading charged-particle $\ptmax$ between 2 and 7\GeV, with differences of up to 30\% with respect to the data. The predictions from {\HERWIG}7 achieve an overall good agreement with measurements at $\sqrt{s}=7\TeV$~\cite{Gieseke:2016fpz}, while the disagreement observed for measurements at $\sqrt{s}=13\TeV$ might indicate the need for further tuning of the new soft MPI model~\cite{Gieseke:2016fpz}. Since many parameters related to PS changed between \HERWIGpp and {\HERWIG}7, the CMS tunes extracted for \HERWIGpp with the CTEQ6L1 PDF set and documented in Ref.~\cite{Khachatryan:2015pea} are not updated and should not be used with {\HERWIG}7.

Since no currently available tune is able to optimally reproduce the UE data at $\sqrt{s}=13\TeV$, we aim to produce improved \PYTHIAviii UE tunes.

\section{PDF and strong coupling values for the tunes}
Two of the basic input parameters to the predictions are the choice of the order of the PDF sets and values of the strong coupling $\alpS$.   These appear in the  hard partonic MEs, the PS model, and the MPI model. The $\alpS$ values used in simulations at LO or NLO
are typically different. Traditionally, the perturbative order of the PDF is matched to the order of the ME calculation. Merged calculations capture some higher-order corrections with respect to the formal order of the ME calculation.
Merging schemes, such as the \ktMLM~\cite{mlm} or CKKW~\cite{Catani:2001cc,Krauss:2002up}, allow the combination of predictions of jet production using ME calculations with those from PS emissions for soft and collinear parton radiation at leading-log accuracy without double counting or dead regions. Merging can be applied also for processes generated at NLO. Using the same PDF set and $\alpS$ value in the ME calculations and in the simulation of the various components of the PS is advocated in Ref.~\cite{Cooper:2011gk}, and by the {\HERWIG}7 and \SHERPA Collaborations, especially when the PS simulation is merged with calculations of higher-order MEs.
The PDF used for the hard process is constrained by the accuracy of the ME calculation.
If we require the PDF to match between the ME and PS, simulations with a (N)NLO ME will also require a (N)NLO PDF in the PS. Depending on the process, this may not have a significant effect. For PS MC event generators, different strategies are adopted; CMS~\cite{Khachatryan:2015pea} and ATLAS~\cite{ATL-PHYS-PUB-2014-021} tunes are traditionally based on LO PDFs, \PYTHIAviii~\cite{Skands:2014pea} tunes are mostly based on LO PDFs, new \SHERPA~\cite{Gleisberg:2008ta} tunes are based on NNLO PDFs, and {\HERWIG}7~\cite{Gieseke:2016fpz} provide tunes based on NLO PDFs.
The usage of a LO PDF set in the UE simulation is motivated by the fact that MPI processes occur at very low energy scales, where a physical (positive) gluon distribution is required by the parton shower. However, there is no consensus on the choice of the order of the PDF. For example, in the NNPDF3.1 set at NNLO, the gluon distribution remains physical even at very low scales.

In the \PYTHIAviii tunes produced prior to this paper, the values used for $\alpS$ were often not the same as those used in the PDFs. For example, in the Monash tune, the FSR $\alpS(m_\cPZ)$, set to 0.1365, is obtained by fitting \PYTHIAviii predictions to LEP event-shape measurements~\cite{Skands:2014pea}, the ISR $\alpS(m_\cPZ)$ is assumed to be equal to FSR $\alpS(m_\cPZ)$, and the hard scattering and MPI $\alpS(m_\cPZ)$ is set to 0.13 according to the value used in the LO PDF set. Even though the $\alpS$ values are free parameters in event generators and various possibilities are viable, the usual course is to choose them consistent with the value used by the PDF set.

In this paper, a collection of new tunes is presented for PDF sets that are evaluated at different accuracies and tested against observables of MB, UE, and hard processes. The NNPDF3.1 PDF sets at the LO, NLO, and NNLO accuracy are used~\cite{Ball:2017nwa}. The LO PDF set uses an $\alpS(m_\cPZ)$ value of 0.13, while 0.118 \cite{PhysRevD.98.030001} is the $\alpS(m_\cPZ)$ value used for the NLO and NNLO PDF sets. None of the central values of the PDF sets have negative values for any parton flavor in the phase space relevant for comparisons.  Special care is required when applying these tunes at high-x regions, where the parton distributions in NNPDF3.1 NLO and NNLO PDF may become negative, which implies an unphysical (negative) value of the calculated cross sections.

{\tolerance=800 The UE simulation is performed by \PYTHIAviii, together with PS merged with a calculation of a higher-order or a multileg ME provided by external programs, such as \POWHEG~\cite{powhegv23} or \textsc{madgraph5}\_a\textsc{mc@NLO} (\MG)~\cite{Alwall:2014hca}. The issue of combining external ME calculations with PS contributions is addressed by the merging procedure.
The procedures considered in this paper are the ``FxFx"~\cite{fxfx} or the ``POWHEG"~\cite{powhegv22} methods for merging higher-order (NLO) MEs to PS and the ``MLM" method~\cite{mlm}. \par}

During this study, we also investigated the effect of imposing an additional rapidity ($y$) ordering to ISR in these merging calculations.
The \PYTHIAviii Monash tune includes a rapidity ordering for both ISR and MPI. The rapidity ordering acts as an extra constraint on the \pt-ordered emissions, thus reducing the phase space for parton emission.

\section{\texorpdfstring{New CMS \PYTHIAviii tunes at 13\TeV}{New CMS PYTHIA8 tunes as 13 TeV}}

In the following, a set of new 13\TeV \PYTHIAviii.226 tunes is presented with different choices of values of the strong coupling used in the modeling of the ISR, FSR, hard scattering, and MPI, as well as the order of its evolution as a function of the four-momentum squared $Q^2$. We distinguish the new tunes according to the order of the PDF set used: LO-PDF, NLO-PDF, or NNLO-PDF. The tunes are labeled as CPi, where CP stands for ``CMS \PYTHIAviii{}" and i is a progressive number from 1 to 5. Only five parameters related to the simulation of MPI, to the overlap matter distribution function~\cite{Sjostrand:1987su}, and to the amount of CR are constrained for the new CMS tunes. In all tunes, we use the MPI-based CR model~\cite{Sjostrand:2004pf}.
The CP tunes are multipurpose tunes, aiming for a consistent description of UE and MB observables at several collision energies and a reliable prediction of the UE simulation in various processes when merged with higher-order ME calculations.

The settings, used in the determination of the new CMS \PYTHIAviii UE tunes, are as follows:
\begin{itemize}
\item Tune CP1 uses the NNPDF3.1 PDF set at LO, with $\alpS$ values used for the simulation of MPI, hard scattering, FSR, and ISR equal to, respectively, 0.13, 0.13, 0.1365, and 0.1365, and running according to an LO evolution.
\item Tune CP2 is a slight variation with respect to CP1, uses the NNPDF3.1 PDF set at LO, with $\alpS$ values used for the simulation of MPI, hard scattering, FSR, and ISR contributions equal to 0.13, and running according to an LO evolution.
\item Tune CP3 uses the NNPDF3.1 PDF set at NLO, with $\alpS$ values used for the simulation of MPI, hard scattering, FSR, and ISR contributions equal to 0.118, and running according to an NLO evolution.
\item Tune CP4 uses the NNPDF3.1 PDF set at NNLO, with $\alpS$ values used for the simulation of MPI, hard scattering, FSR, and ISR contributions equal to 0.118, and running according to an NLO evolution.
\item Tune CP5 has the same settings as CP4, but with the ISR emissions ordered according to rapidity.
\end{itemize}

The parameters related to the simulation of the hadronization and beam remnants are not varied in the fits and are kept fixed to the values of the Monash tune.
The overlap distribution between the two colliding protons is modeled according to a double-Gaussian functional form with the parameters \texttt{coreRadius} and \texttt{coreFraction}. This parametrization of the transverse partonic overlap of two protons identifies an inner, denser part, the so-called core, and an outer less dense part.
The \texttt{coreRadius} parameter represents the width of the core and the \texttt{coreFraction}, the fraction of quark and gluon content enclosed in the core.
A double-Gaussian function is preferred for modeling the proton overlap over the negative exponential  used in some previous tunes. Tunes using a double-Gaussian function tend to better reproduce the cross sections measured by the CMS experiment at $\sqrt{s}=7\TeV$~\cite{CMS:2012zxa}, simultaneously as a function of charged-particle multiplicity and transverse momenta.

The parameter that determines the amount of simulated CR in the MPI-based model is varied in the fits. A small (large) value of the final-state CR parameter tends to increase (reduce) the final particle multiplicities.

The new CMS \PYTHIAviii tunes are extracted by varying the parameters listed in Table~\ref{table2} and by fitting UE observables at various collision energies.
In the fitting procedure, we use the charged-particle and \ptsum~densities, measured in \tmin\ and \tmax\ regions as a function of $\ptmax$, as well as the charged-particle multiplicity as a function of pseudorapidity $\eta$, measured by CMS at $\sqrt{s}=13\TeV$~\cite{CMS:2015zev,Khachatryan:2015jna}.
In addition, we also use the charged-particle and \ptsum\ densities as a function of the leading charged-particle \pt, measured in \tmin\ and \tmax\ by CMS at $\sqrt{s}=7\TeV$~\cite{CMS:2012zxa} and by CDF at $\sqrt{s}=1.96\TeV$~\cite{Aaltonen:2015aoa}.

\begin{table*}[htbp]
\renewcommand{\arraystretch}{1.2}
\centering
\topcaption{Parameters in the \PYTHIAviii MC event generator together with the PDFs determine the energy dependence of MPI, the overlap matter distribution function, and the amount of simulated color reconnection. The parameter ranges used for the fits are also listed.}
\label{table2}
\cmsTable{
\begin{tabular}{l c c c}
 Parameter description				& Name in \PYTHIAviii &  Range considered  \\
 \hline
  MPI threshold [\GeVns{}], \texttt{pT0Ref}, at $\sqrt{s}=\sqrt{s_0}$  & \texttt{MultipartonInteractions:pT0Ref} &  1.0--3.0\\
  Exponent of $\sqrt{s}$ dependence, $\epsilon$	& \texttt{MultipartonInteractions:ecmPow} &  0.0--0.3\\
  Matter fraction contained in the core & \texttt{MultipartonInteractions:coreFraction} &  0.1--0.95\\
  Radius of the core & \texttt{MultipartonInteractions:coreRadius} &  0.1--0.8\\
  Range of color reconnection probability & \texttt{ColorReconnection:range} &  1.0--9.0\\
\end{tabular}
}
\end{table*}

Tunes are determined by generating sets of predictions using the \RIVET~\cite{Buckley:2013} (version 2.5.2) and the \PROFESSOR~\cite{Buckley:2009bj} (version 1.4.0) frameworks with around 150 different choices of the five parameter values used in the event simulation.
The predictions form a grid in the five-dimensional parameter space which is fitted using a third-order polynomial function.
The uncertainty introduced in the fitted parameters due to the interpolation procedure is negligible compared with the quoted tune uncertainty.
Results are found to be stable if one decreases this number to 100 or increases to 200, or uses a fourth-order polynomial function for the grid interpolation.
The generated inelastic events include ND and diffractive (DD$+$SD$+$CD) contributions. The UE observables used to determine the tunes are sensitive to diffractive contributions only at very small $\ptmax$ values ($<$3\GeV). The ND component is dominant for $\pt^\text{max}$ values greater than $\approx$3.0\GeV, since the cross section of the diffractive components rapidly decreases as a function of the exchanged \pt. Minimum-bias observables, such as the inclusive charged-particle multiplicity as a function of $\eta$, are sensitive to all contributions over the whole spectrum.

The fit is performed by minimizing the $\chi^2$ function
\begin{equation}\label{chi}
\chi^2(p)=\sum_{\textrm{O}_j} \sum_{i}\frac{(f_{i,\textrm{O}_j}(p)-R_{i,\textrm{O}_j})^2}{\Delta_{i,\textrm{O}_j}^2}
\end{equation}
where the sum runs over each bin $i$ of every observable O$_\mathrm{j}$. The $f_\mathrm{i}(p)$ functions represent a parametrization of the dependence of the predictions in bin $i$ on the tuning parameters, $R_\mathrm{i}$ is the value of the measured observable in bin $i$, and $\Delta_\mathrm{i}$ is the total experimental uncertainty of $R_\mathrm{i}$. The best fit values of the tuned parameters are shown in Table~\ref{tab:lo_tunes} for CP1 and CP2, \ie, the tunes using LO PDF sets, and in Table~\ref{tab:nlo_tunes} for CP3, CP4, CP5, \ie, the tunes using NLO or NNLO PDF sets. Uncertainties in the parameters of these tunes are discussed in Appendix A. No correlation across bins is included in the minimized $\chi^2$ function.

\begin{figure*}[ht!]
\centering
\includegraphics[width=0.49\textwidth]{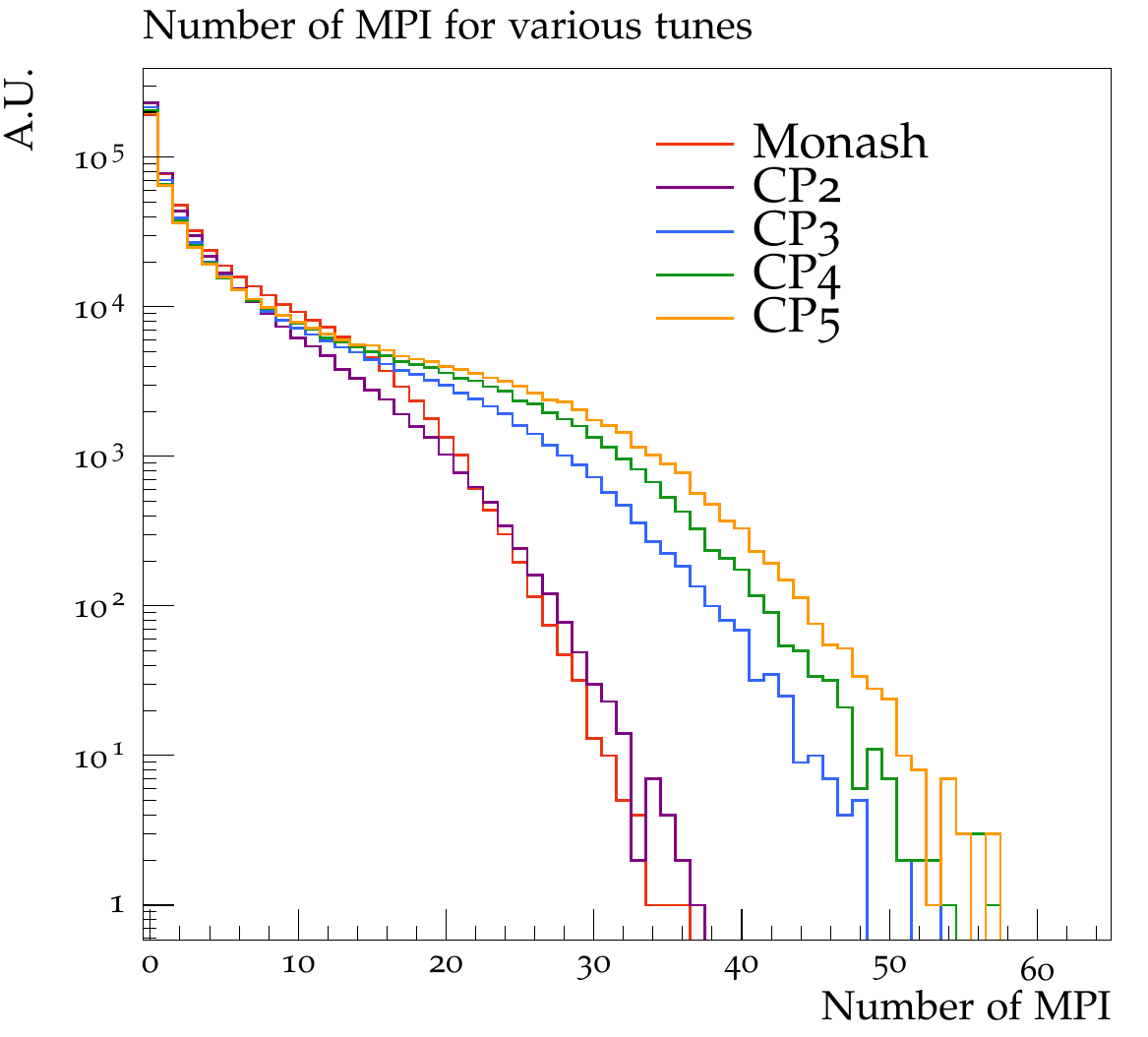}
\includegraphics[width=0.49\textwidth]{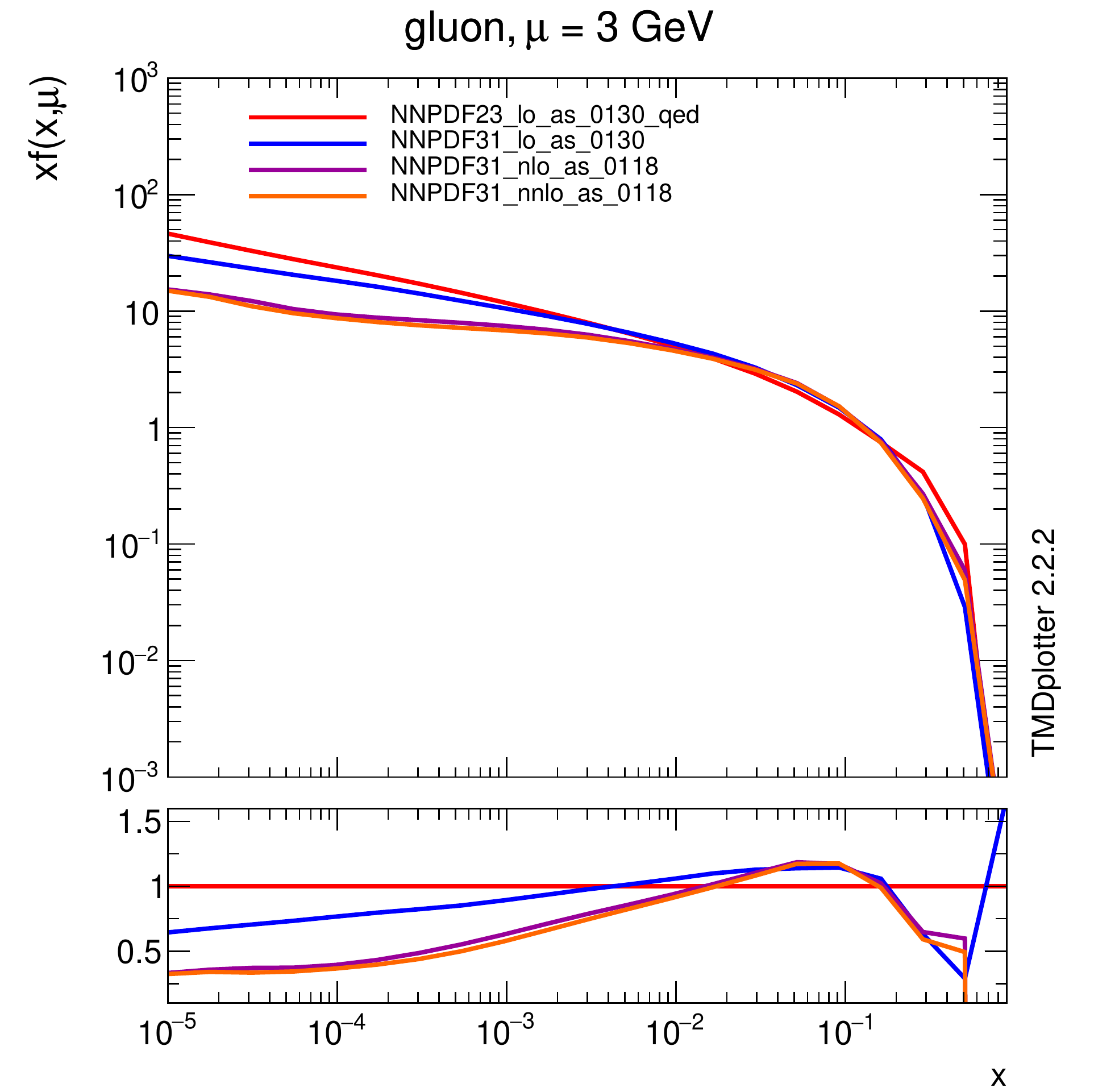}
\caption{Distribution of number of MPI simulated by the tunes Monash, CP2, CP3, CP4, and CP5 (left). Gluon distribution function at a reference scale of $\mu = 3\GeV$ (right) for the NNPDF2.3LO PDF set and the different versions of the NNPDF3.1 PDF set: LO, NLO, and NNLO.
The ratio of NNPDF3.1 gluon distribution functions to the NNPDF2.3LO gluon distribution function are also shown. }
\label{fig:MPI}
\end{figure*}

The value of \texttt{pT0Ref} and its energy dependence is very different between tunes based on LO PDF sets and tunes based on NLO or NNLO PDFs. While \texttt{pT0Ref} is around 2.3--2.4\GeV for CP1 and CP2 tunes with $\epsilon\approx0.14$--0.15, CP3, CP4, and CP5 tunes prefer much lower values for both \texttt{pT0Ref} ($\approx$1.4--1.5) and $\epsilon$ ($\approx$0.03--0.04).
A value of $\epsilon$ of $\approx$0.03--0.04 corresponds to a very weak energy dependence of the threshold of the MPI cross section.
These results can be understood by considering the shapes of the gluon densities at small $x$ for the different PDF sets. In order to describe the UE observables, the rapidly increasing  gluon densities at small $x$ in LO PDF sets favor large values of \texttt{pT0Ref}. Meanwhile NLO and NNLO PDF sets, whose gluon densities are more flat at low $x$, need higher contributions of MPI, \ie, a small value of \texttt{pT0Ref}. Figure~\ref{fig:MPI} shows the number of MPI observed for the various tunes and the gluon distribution at a reference scale of $\mu = 3\GeV$ for various NNPDF versions. The larger number of simulated MPI for NLO and NNLO tunes with respect to LO tunes is apparent.

We have found that the values of \texttt{pT0Ref} and $\epsilon$ also depend on the order of the running used for $\alpS$.
In particular, fits based on NLO or NNLO PDF sets, \ie, CP3, CP4, or CP5, with an LO $\alpS$ running prefer even smaller values for both \texttt{pT0Ref} and $\epsilon$ than the ones in the tunes obtained with an NLO $\alpS$ running.
This is because $\alpS$ runs faster at NLO than at LO.
When $\alpS$ is run from the same value at the same scale ($m_\cPZ$), the effective coupling at low scales is larger for NLO running than for LO running. Therefore, a lower \texttt{pT0Ref} is needed for NLO $\alpS$ running than for LO $\alpS$ running to obtain a similar number of MPI.

For tunes based on NLO and NNLO PDF sets, the value of \texttt{pT0Ref} is as low as the initial scale of the PDF $Q_\text{min}^2$. For interactions occurring at $Q^2$ which are lower than $Q_\text{min}^2$, the value of the PDF is left frozen to the value assumed at the initial scale.

\begin{table*}[pht!]
\centering
\topcaption{CMS \PYTHIAviii LO-PDF tunes CP1 and CP2. Both the values at $Q = m_\cPZ$ and the order of running with $Q^2$ of the strong coupling $\alpS(m_\cPZ)$ are listed. In these tunes, we use the Schuler-Sj\"ostrand diffraction model~\cite{Schuler:1993wr} and also include the simulation of CD processes. The number of degrees of freedom for tunes CP1 and CP2 is 63.}
\label{tab:lo_tunes}
\cmsTable{
\begin{tabular}{lcc}
\PYTHIAviii parameter                   & CP1         & CP2  \\
\hline
PDF Set                             & NNPDF3.1 LO & NNPDF3.1 LO \\
$\alpS(m_\cPZ)$                      & 0.130      & 0.130  \\
\texttt{SpaceShower:rapidityOrder}   &     off & off \\
\texttt{MultipartonInteractions:EcmRef} [\GeVns{}]                    &   7000  & 7000          \\
$\alpS^\mathrm{ISR}(m_\cPZ)$ value/order      &   0.1365/LO   & 0.130/LO  \\
$\alpS^\mathrm{FSR}(m_\cPZ)$ value/order      &   0.1365/LO    & 0.130/LO  \\
$\alpS^\mathrm{MPI}(m_\cPZ)$ value/order       &   0.130/LO     & 0.130/LO  \\
$\alpS^\mathrm{ME}(m_\cPZ)$ value/order       &   0.130/LO     & 0.130/LO  \\
\hline
\texttt{MultipartonInteractions:pT0Ref} [\GeVns{}]                 &     2.4    & 2.3      \\
\texttt{MultipartonInteractions:ecmPow}                       &     0.15   & 0.14       \\
\texttt{MultipartonInteractions:coreRadius}               &    0.54    & 0.38       \\
\texttt{MultipartonInteractions:coreFraction}                &    0.68   & 0.33        \\
\texttt{ColorReconnection:range}                   &     2.63  & 2.32       \\
$\chi^2$/dof                                  & 0.89       & 0.54 \\
\end{tabular}
}
\end{table*}

\begin{table*}[pht!]
\centering
\topcaption{CMS \PYTHIAviii NLO-PDF tune CP3 and NNLO-PDF tunes CP4 and CP5. Both the values at $Q = m_\cPZ$ and the order of running with $Q^2$ of the strong coupling $\alpS$ are listed. In these tunes, we use the Schuler-Sj\"ostrand diffraction model~\cite{Schuler:1993wr} and also include the simulation of CD processes.  The number of degrees of freedom for tunes CP3, CP4, and CP5 is 63.}
\label{tab:nlo_tunes}
\cmsTable{
\begin{tabular}{lccc}
\PYTHIAviii parameter                       & CP3        & CP4    & CP5   \\
\hline
PDF Set                         & NNPDF3.1 NLO           & NNPDF3.1 NNLO &  NNPDF3.1 NNLO     \\
$\alpS(m_\cPZ)$                 & 0.118          &  0.118 &  0.118 \\
\texttt{SpaceShower:rapidityOrder}       &   off & off & on \\
\texttt{MultipartonInteractions:EcmRef} [\GeVns{}]      &    7000        & 7000 & 7000\\
$\alpS^\mathrm{ISR}(m_\cPZ)$ value/order     &   0.118/NLO         & 0.118/NLO  & 0.118/NLO   \\
$\alpS^\mathrm{FSR}(m_\cPZ)$ value/order     &   0.118/NLO            & 0.118/NLO  & 0.118/NLO   \\
$\alpS^\mathrm{MPI}(m_\cPZ)$ value/order     &  0.118/NLO             &  0.118/NLO & 0.118/NLO   \\
$\alpS^\mathrm{ME}(m_\cPZ)$ value/order     &   0.118/NLO            & 0.118/NLO & 0.118/NLO   \\
\hline
\texttt{MultipartonInteractions:pT0Ref} [\GeVns{}]       &    1.52        & 1.48 & 1.41 \\
\texttt{MultipartonInteractions:ecmPow}                            &    0.02        & 0.02 & 0.03\\
\texttt{MultipartonInteractions:coreRadius}                        &    0.54        & 0.60 & 0.76 \\
\texttt{MultipartonInteractions:coreFraction}                        &    0.39       & 0.30 & 0.63\\
\texttt{ColorReconnection:range}               &    4.73       & 5.61 & 5.18\\
$\chi^2$/dof                            &   0.76               & 0.80       & 1.04 \\
\end{tabular}
}
\end{table*}

\begin{figure*}[ht!]
\centering
\includegraphics[width=0.49\textwidth]{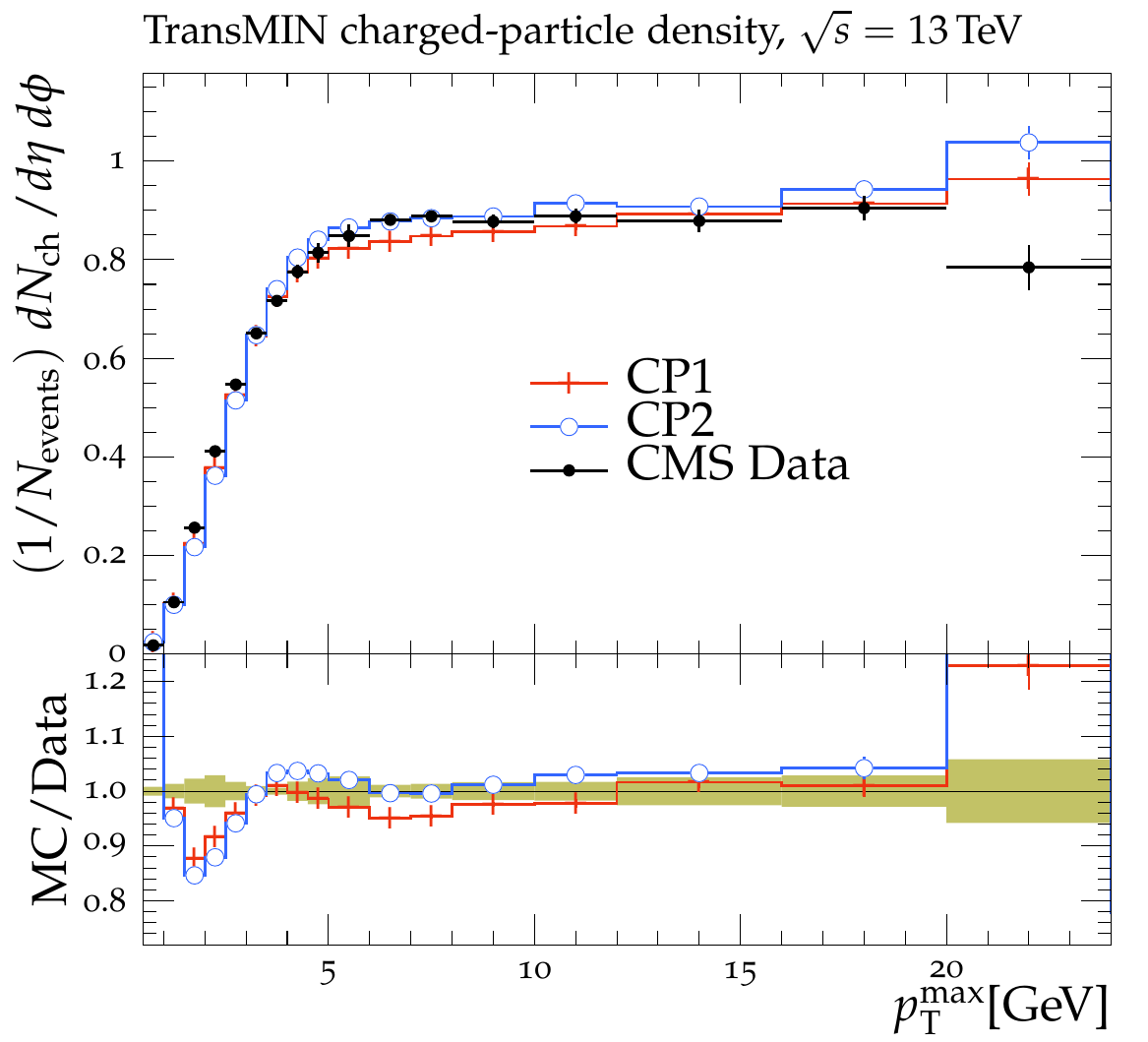}
\includegraphics[width=0.49\textwidth]{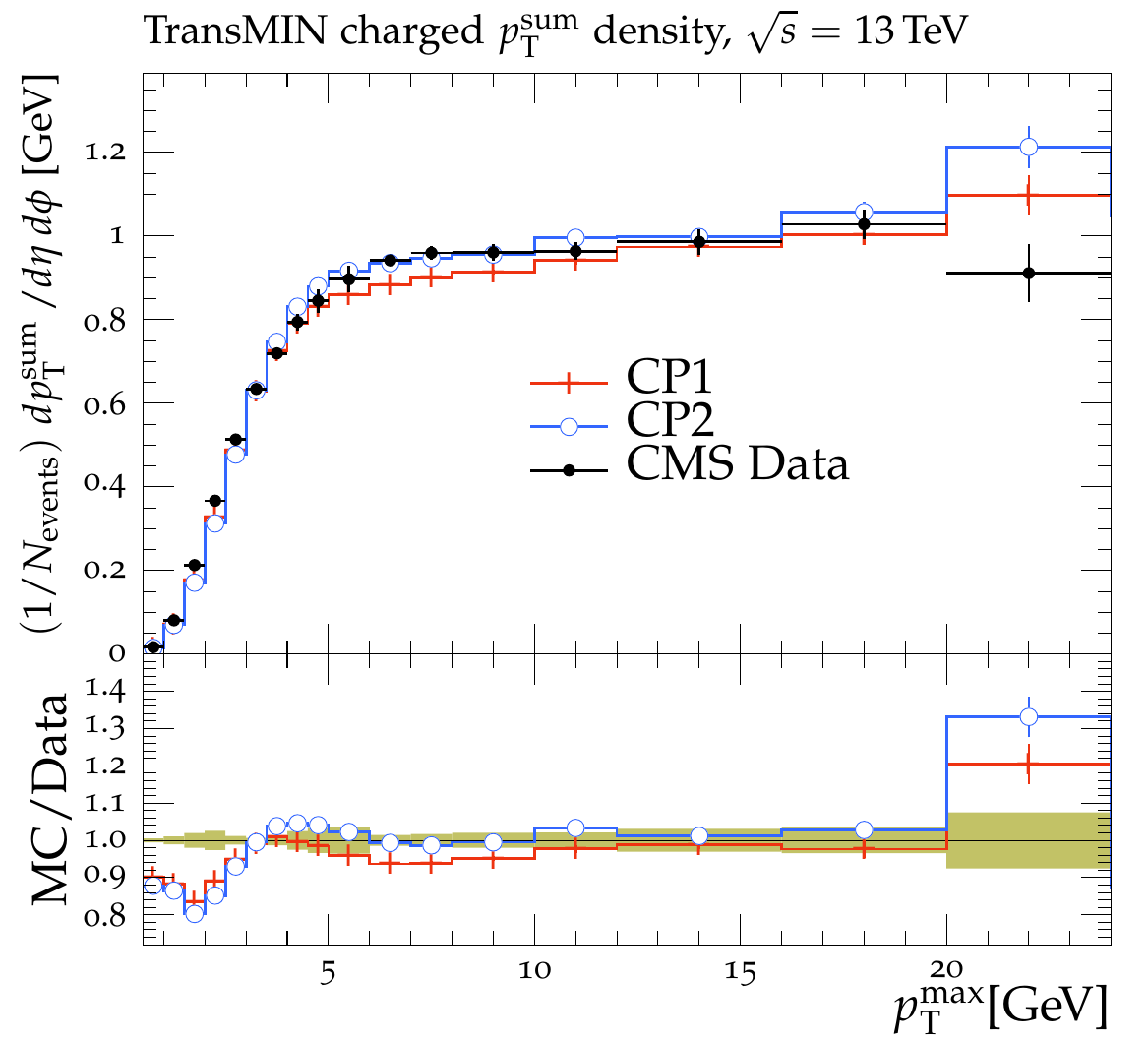}\\
\includegraphics[width=0.49\textwidth]{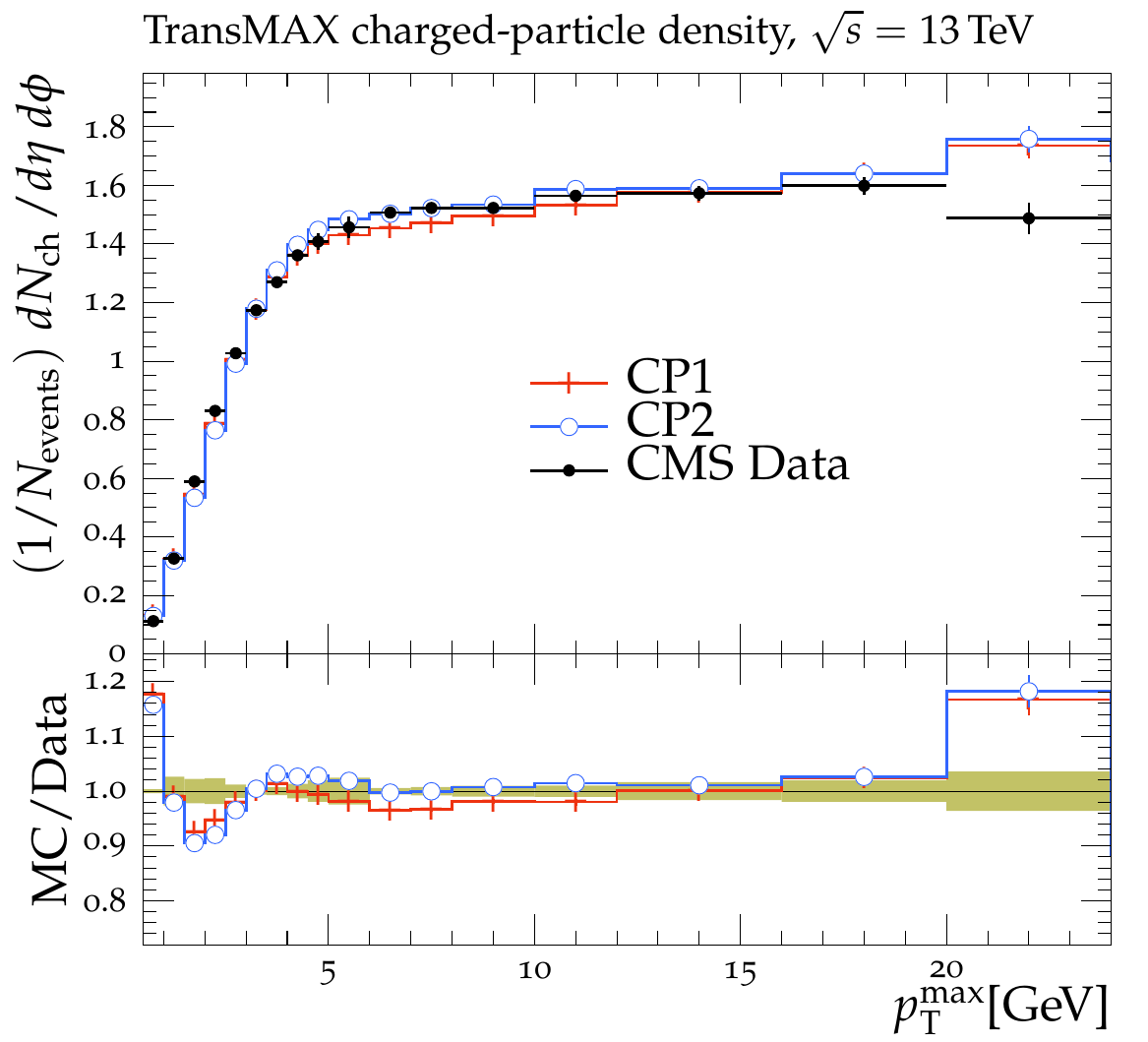}
\includegraphics[width=0.49\textwidth]{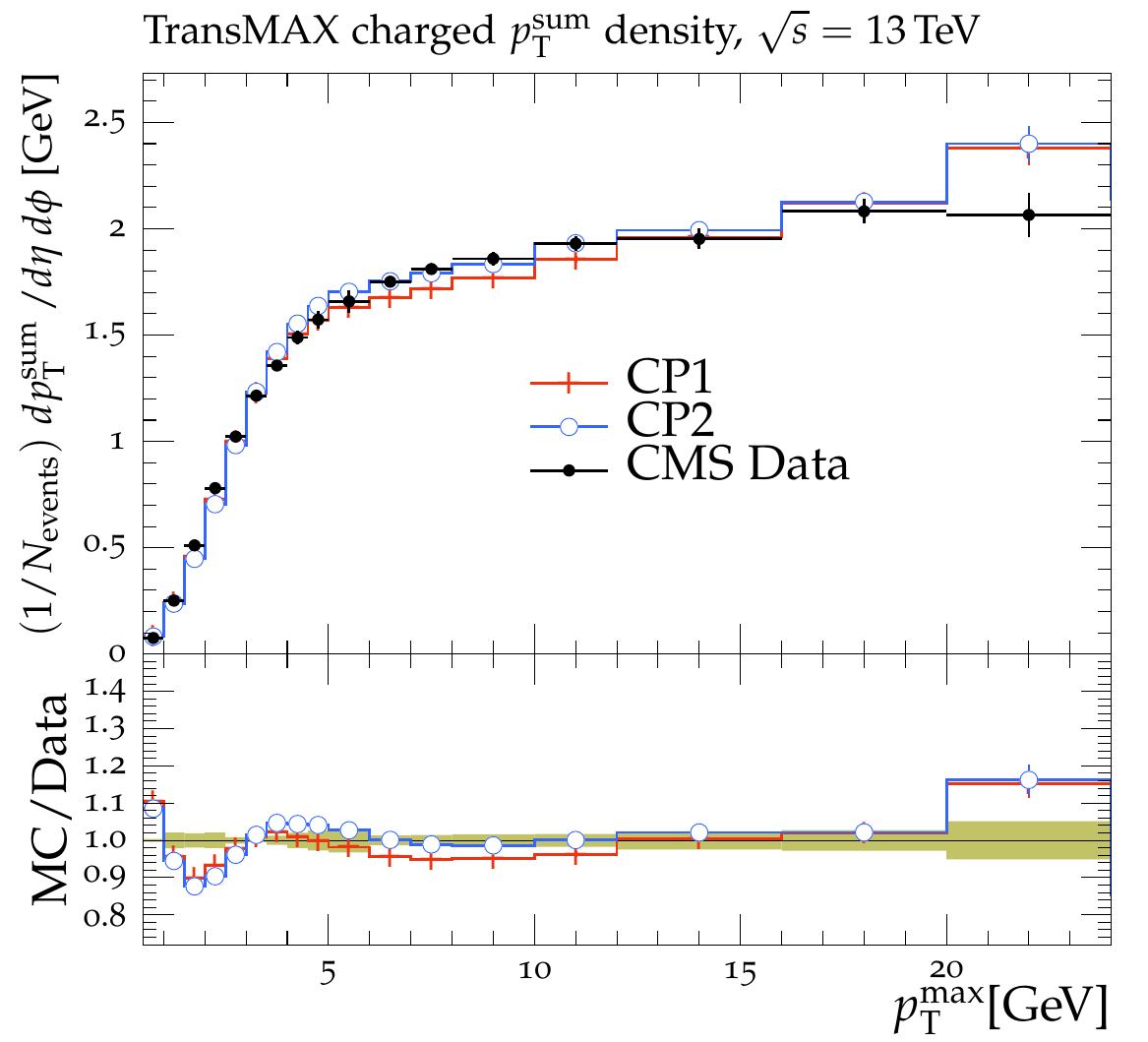}
\caption{
The \tmin\ (upper left) charged-particle and (upper right) charged \ptsum\ densities
and the \tmax\ (lower left) charged-particle and (lower right) charged \ptsum\ densities, as a function of the transverse momentum of the leading charged particle, $\ptmax$, from the CMS $\sqrt{s}=13\TeV$ analysis~\cite{CMS:2015zev}.
Charged hadrons are measured with $\pt>0.5\GeV$ in $\abs{\eta}< 2.0$.
The \tmin\ densities are more sensitive to the MPI, whereas the \tmax\ densities are more sensitive to ISR and FSR.
The data are compared with the CMS \PYTHIAviii LO-PDF tunes CP1 and CP2. The ratios of the simulations to the data (MC/Data) are also shown, where the shaded band indicates the total experimental uncertainty in the data. Vertical lines drawn on the data points refer to the total uncertainty in the data. Vertical lines drawn on the MC points refer to the statistical uncertainty in the predictions. Horizontal bars indicate the associated bin width.}
\label{fig:1-3}
\end{figure*}

\begin{figure*}[ht!]
\centering
\includegraphics[width=0.49\textwidth]{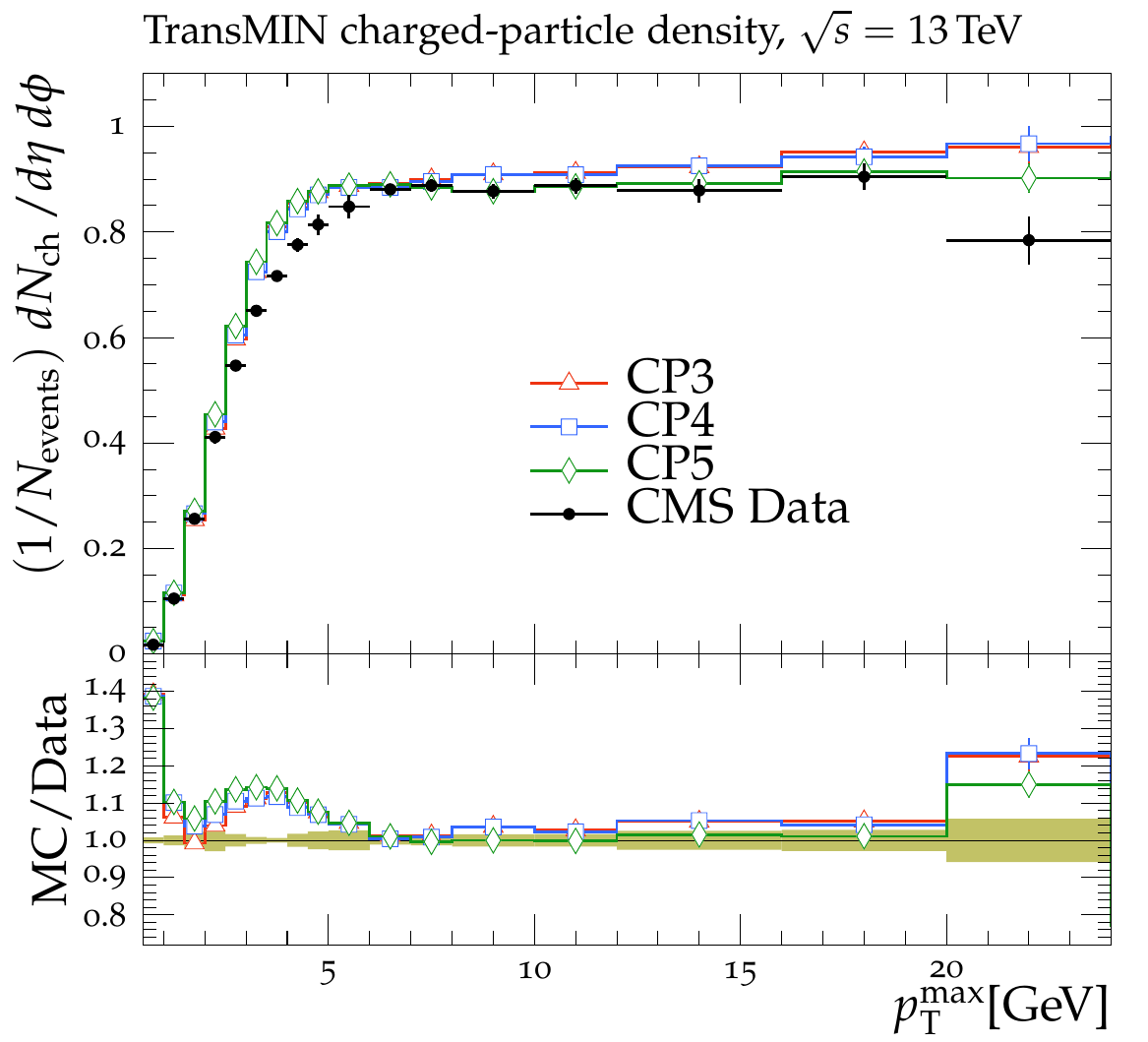}
\includegraphics[width=0.49\textwidth]{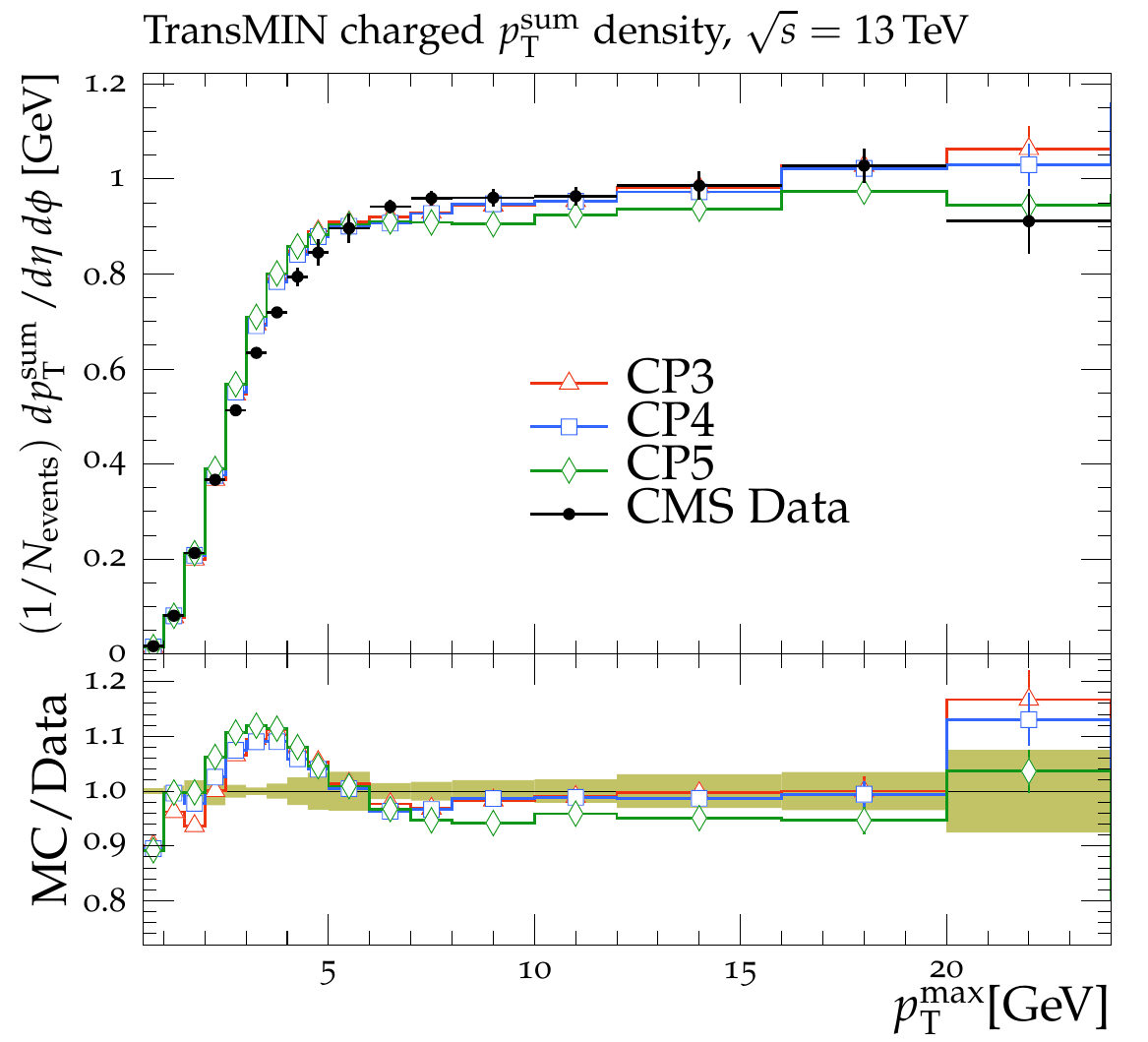}
\includegraphics[width=0.49\textwidth]{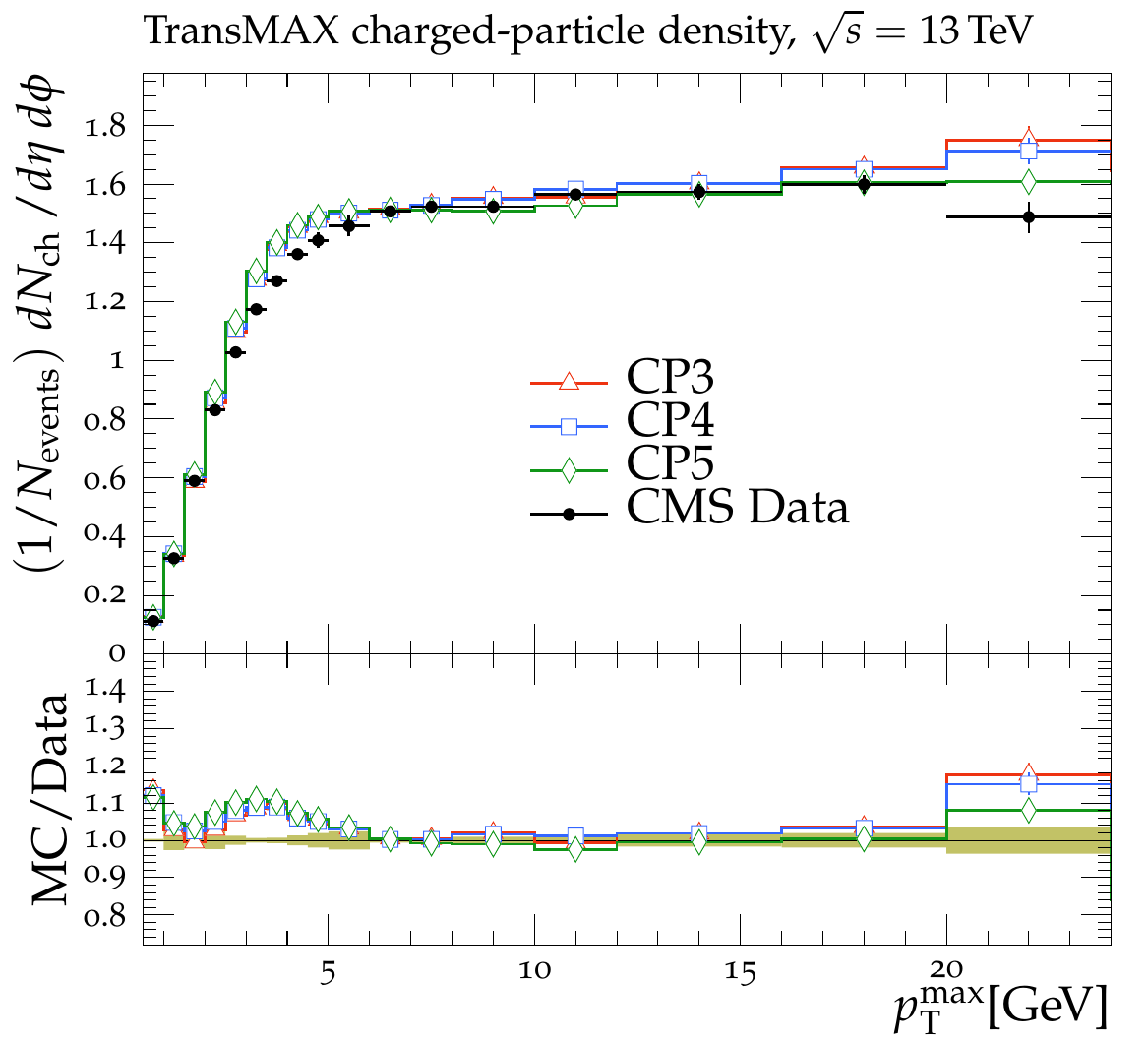}
\includegraphics[width=0.49\textwidth]{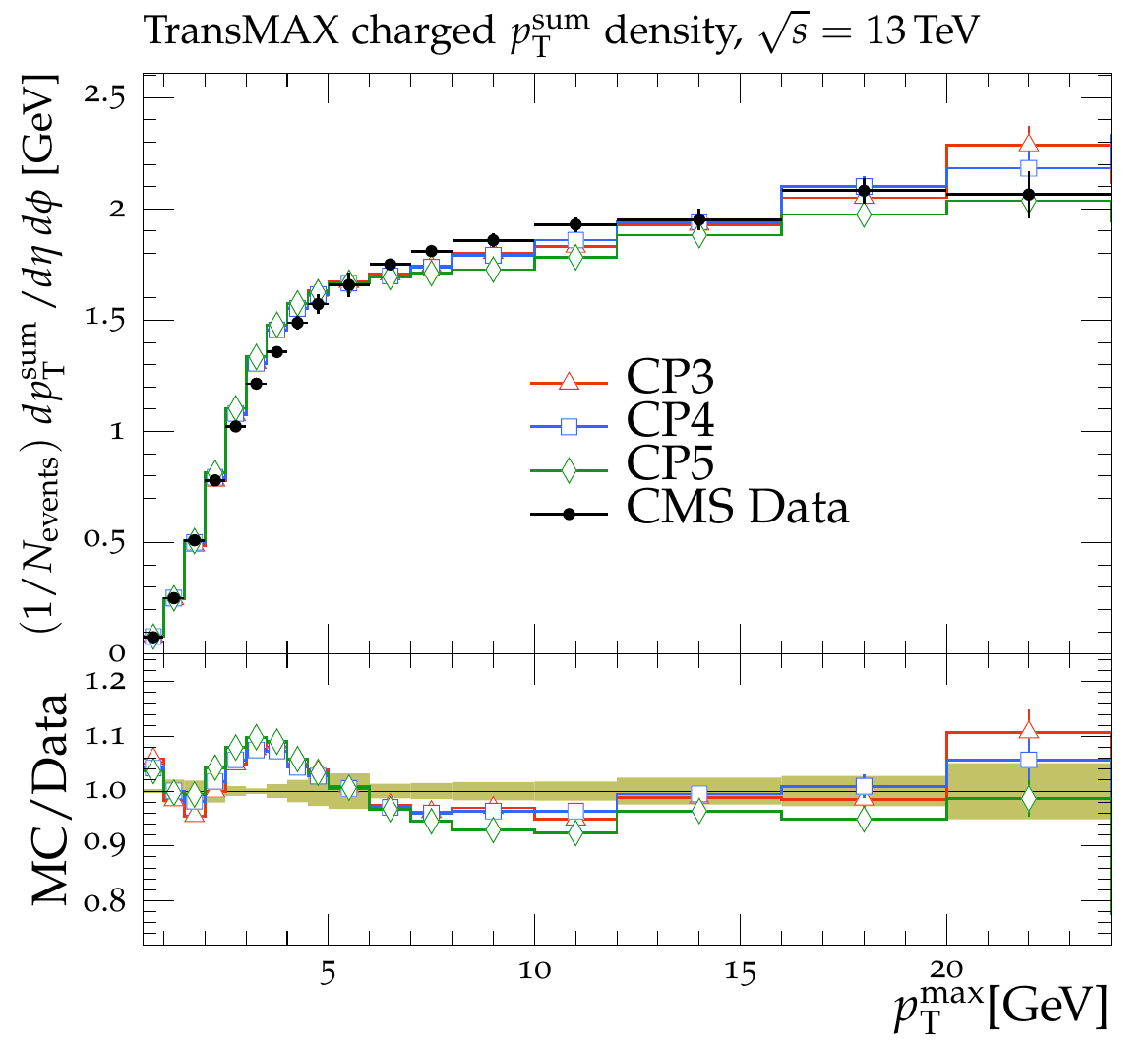}
\caption{The \tmin\ (upper left) charged-particle and (upper right) charged \ptsum\ densities
and the \tmax\ (lower left) charged-particle and (lower right) charged \ptsum\ densities, as a function of the transverse momentum of the leading charged particle, $\ptmax$, from the CMS $\sqrt{s}=13\TeV$ analysis~\cite{CMS:2015zev}. Charged hadrons are measured with $\pt>0.5\GeV$ in $\abs{\eta}< 2.0$. The data are compared with the CMS \PYTHIAviii (N)NLO-PDF tunes CP3, CP4, and CP5. The ratios of simulations to the data (MC/Data) are also shown, where the shaded band indicates the total experimental uncertainty in the data. Vertical lines drawn on the data points refer to the total uncertainty in the data. Vertical lines drawn on the MC points refer to the statistical uncertainty in the predictions. Horizontal bars indicate the associated bin width.}
\label{fig:2-3}
\end{figure*}

\begin{figure*}[ht!]
\centering
\includegraphics[width=0.49\textwidth]{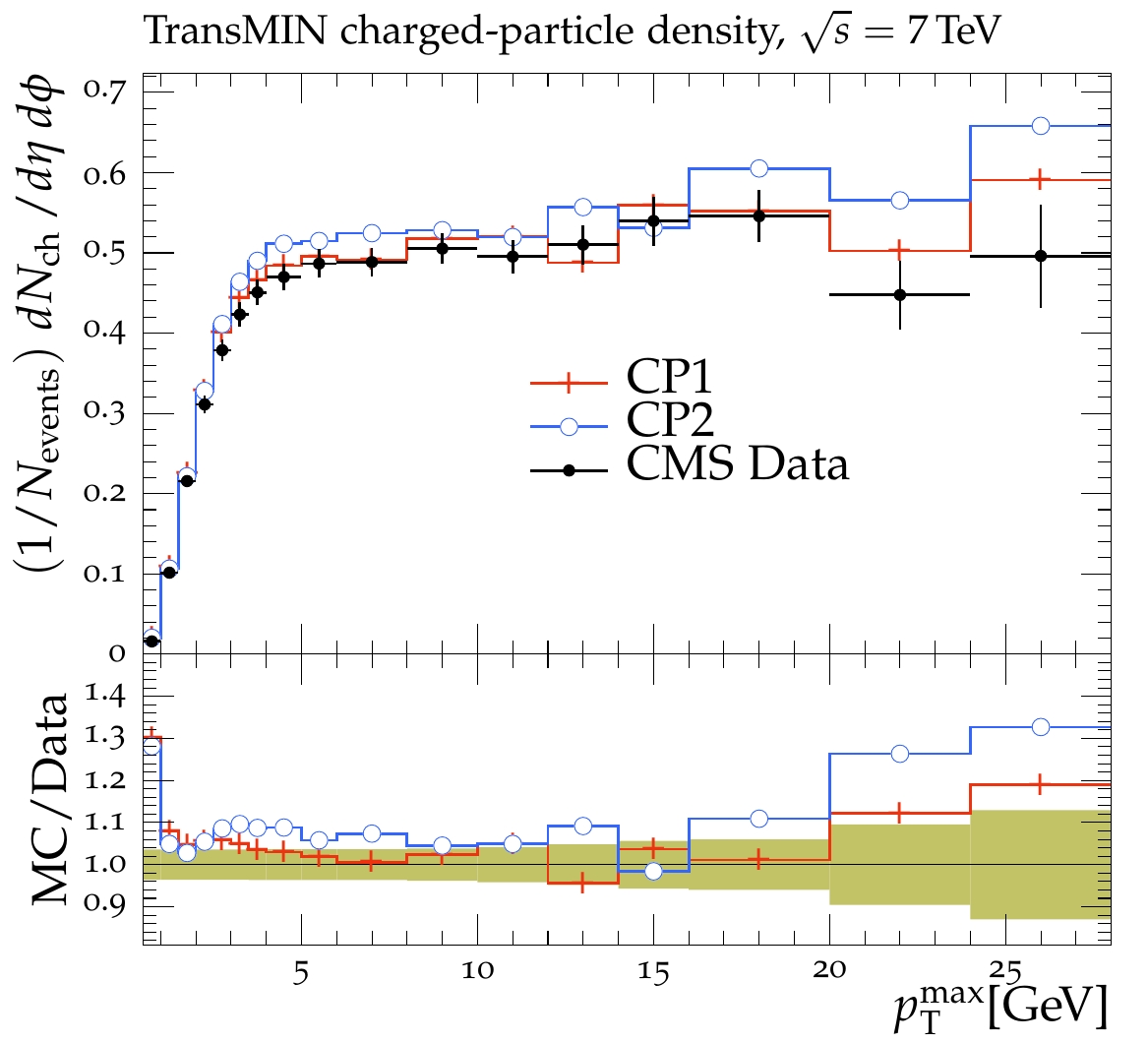}
\includegraphics[width=0.49\textwidth]{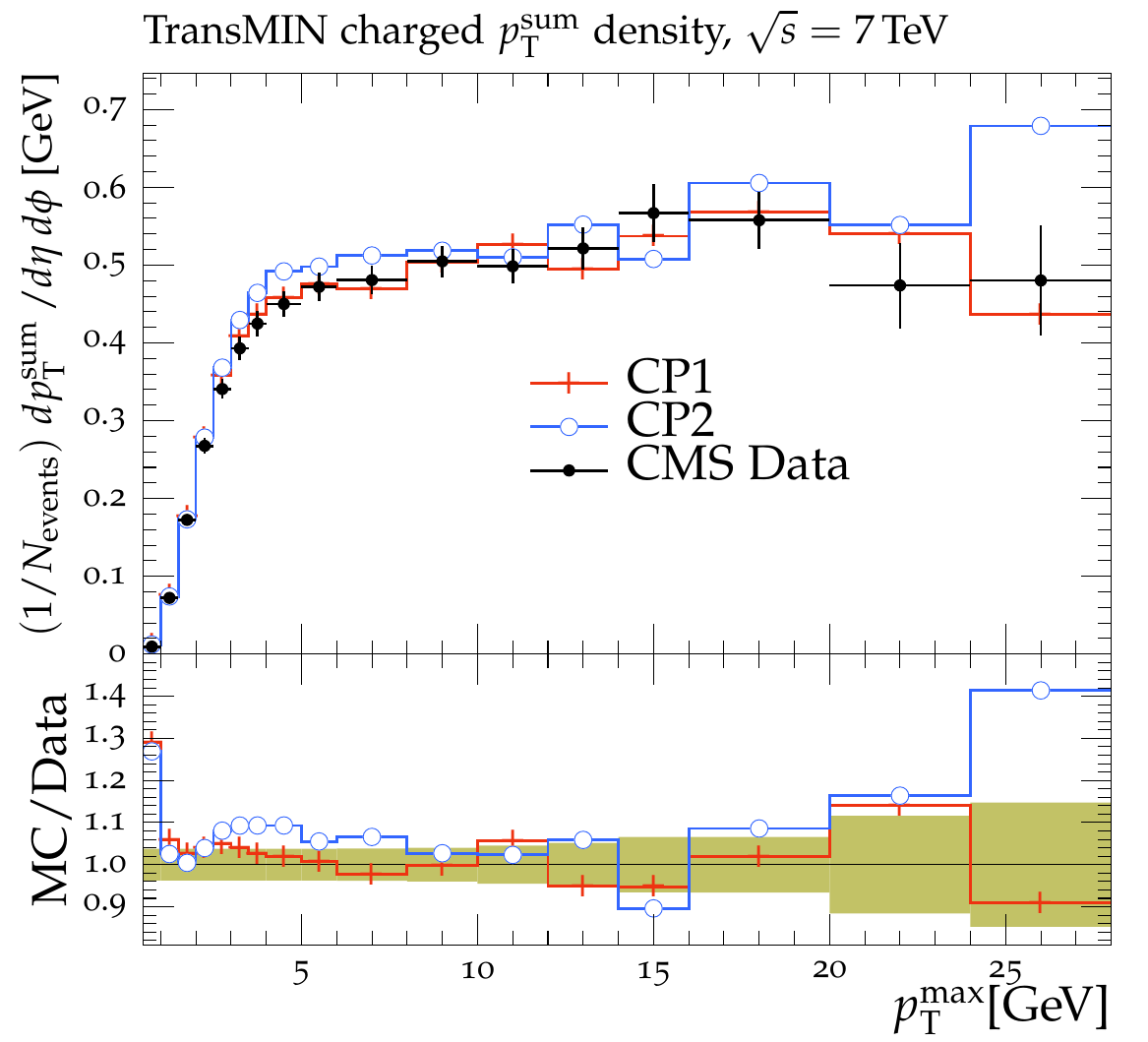}
\includegraphics[width=0.49\textwidth]{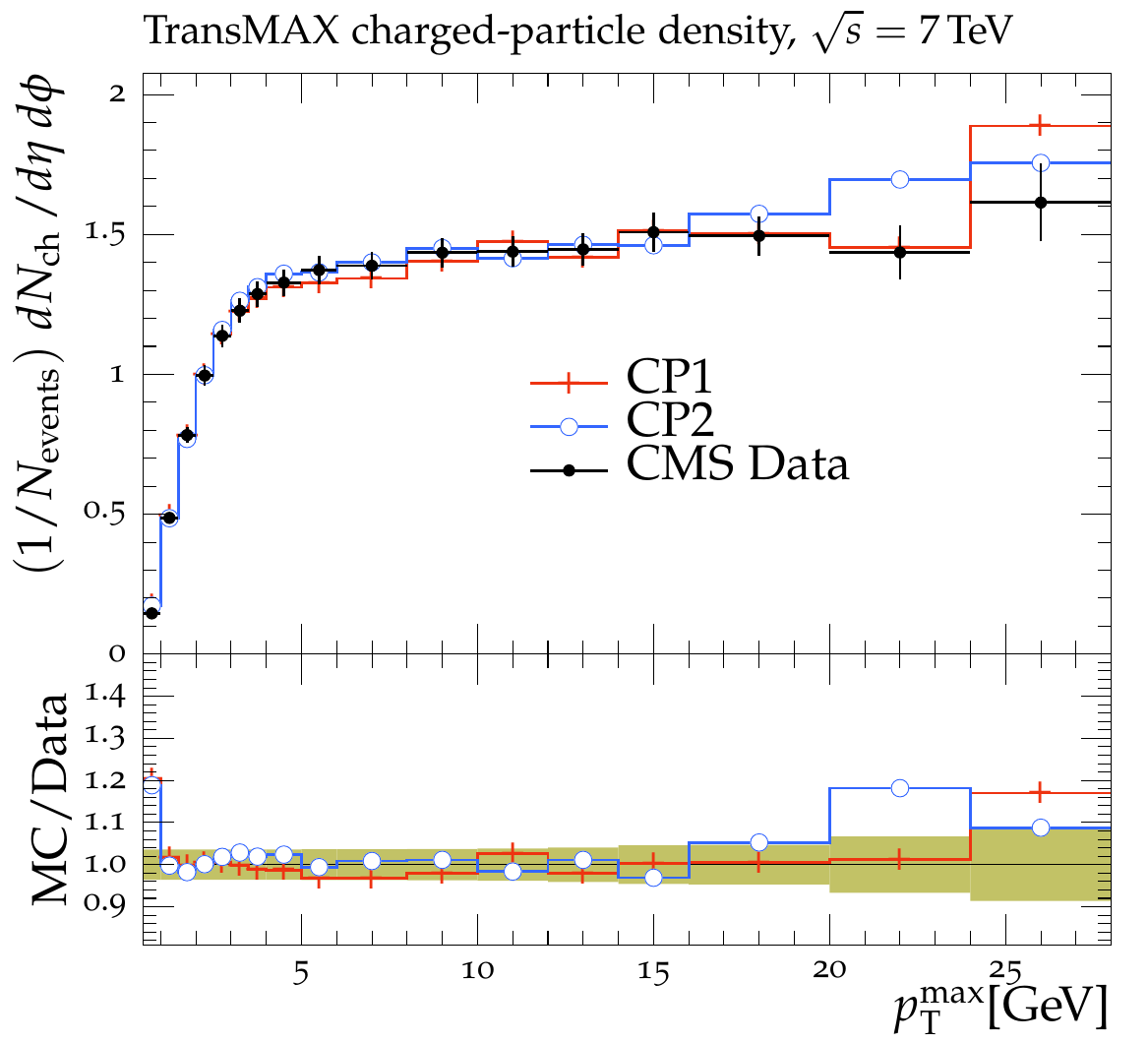}
\includegraphics[width=0.49\textwidth]{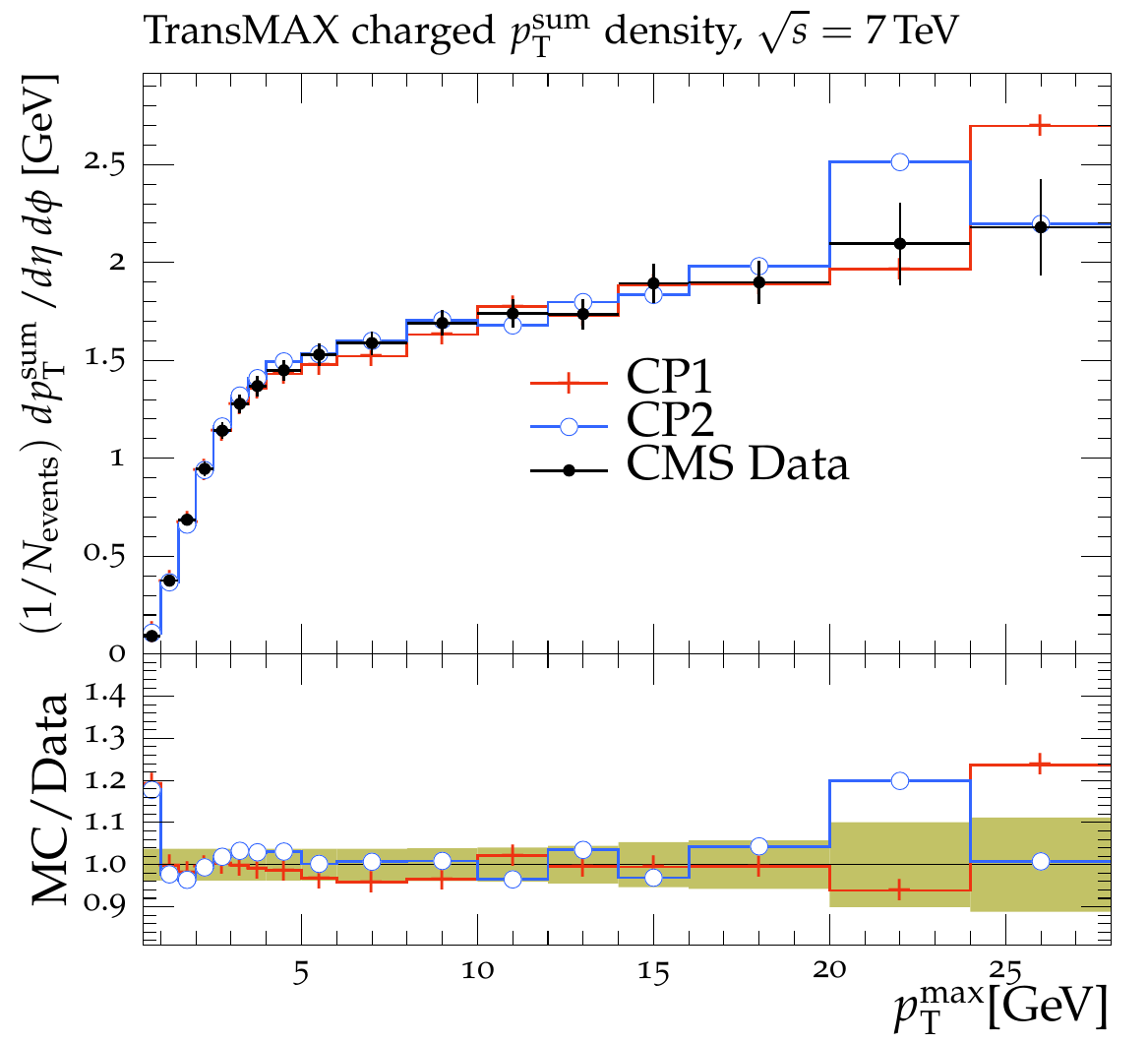}
\caption{The \tmin\ (upper left) charged-particle and (upper right) charged \ptsum\ densities
and the \tmax\ (lower left) charged-particle and (lower right) charged \ptsum\ densities, as a function of the transverse momentum of the leading charged particle, $\ptmax$, from the CMS $\sqrt{s}=7\TeV$ analysis~\cite{CMS:2012zxa}. Charged hadrons are measured with $\pt>0.5\GeV$ in $\abs{\eta}< 0.8$.
The data are compared with the CMS \PYTHIAviii LO-PDF tunes CP1 and CP2. The ratios of simulations to the data (MC/Data) are also shown, where the shaded band indicates the total experimental uncertainty in the data. Vertical lines drawn on the data points refer to the total uncertainty in the data. Vertical lines drawn on the MC points refer to the statistical uncertainty in the predictions. Horizontal bars indicate the associated bin width.}
\label{fig:5-3}
\end{figure*}

\begin{figure*}[ht!]
\centering
\includegraphics[width=0.49\textwidth]{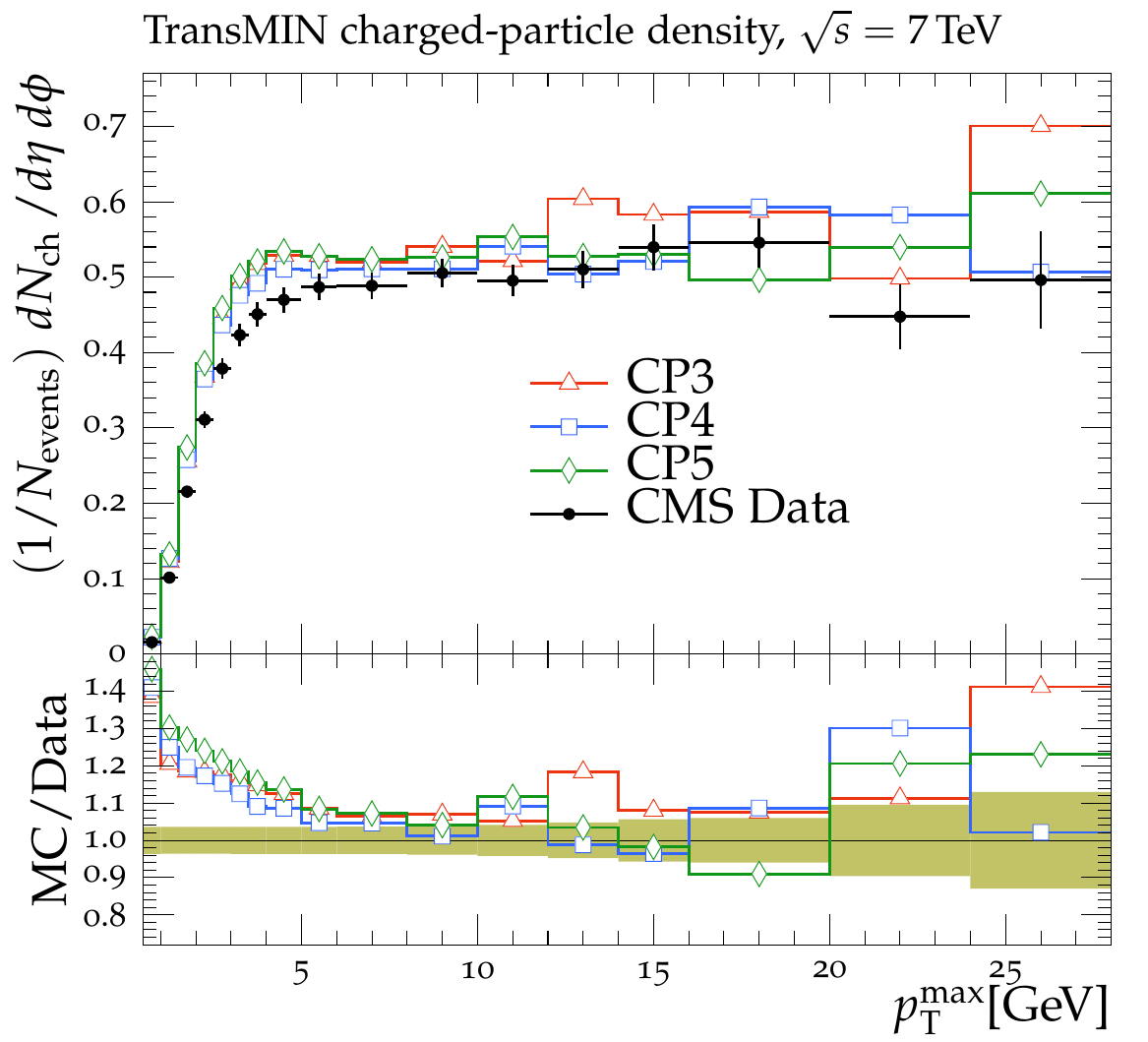}
\includegraphics[width=0.49\textwidth]{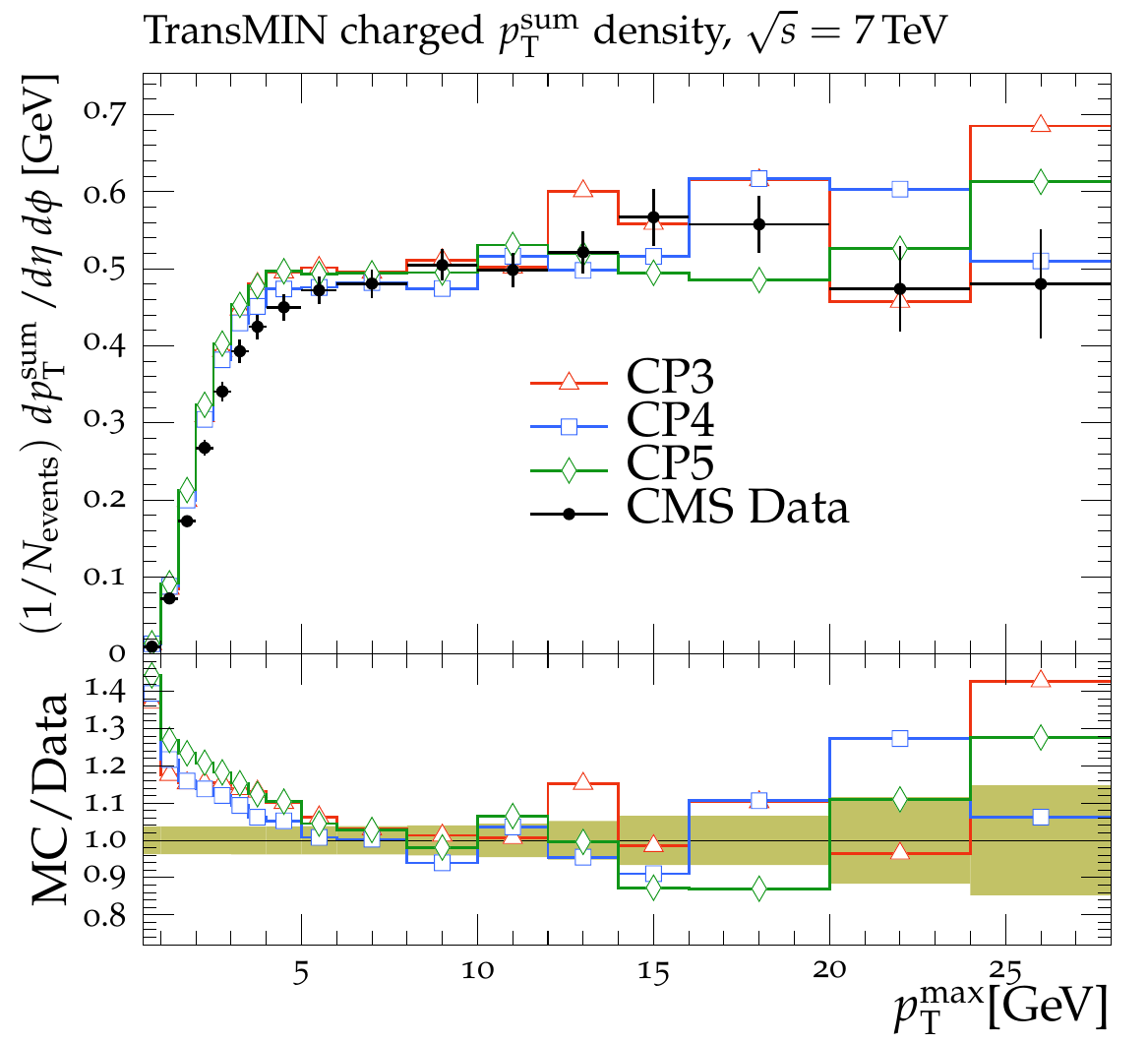}
\includegraphics[width=0.49\textwidth]{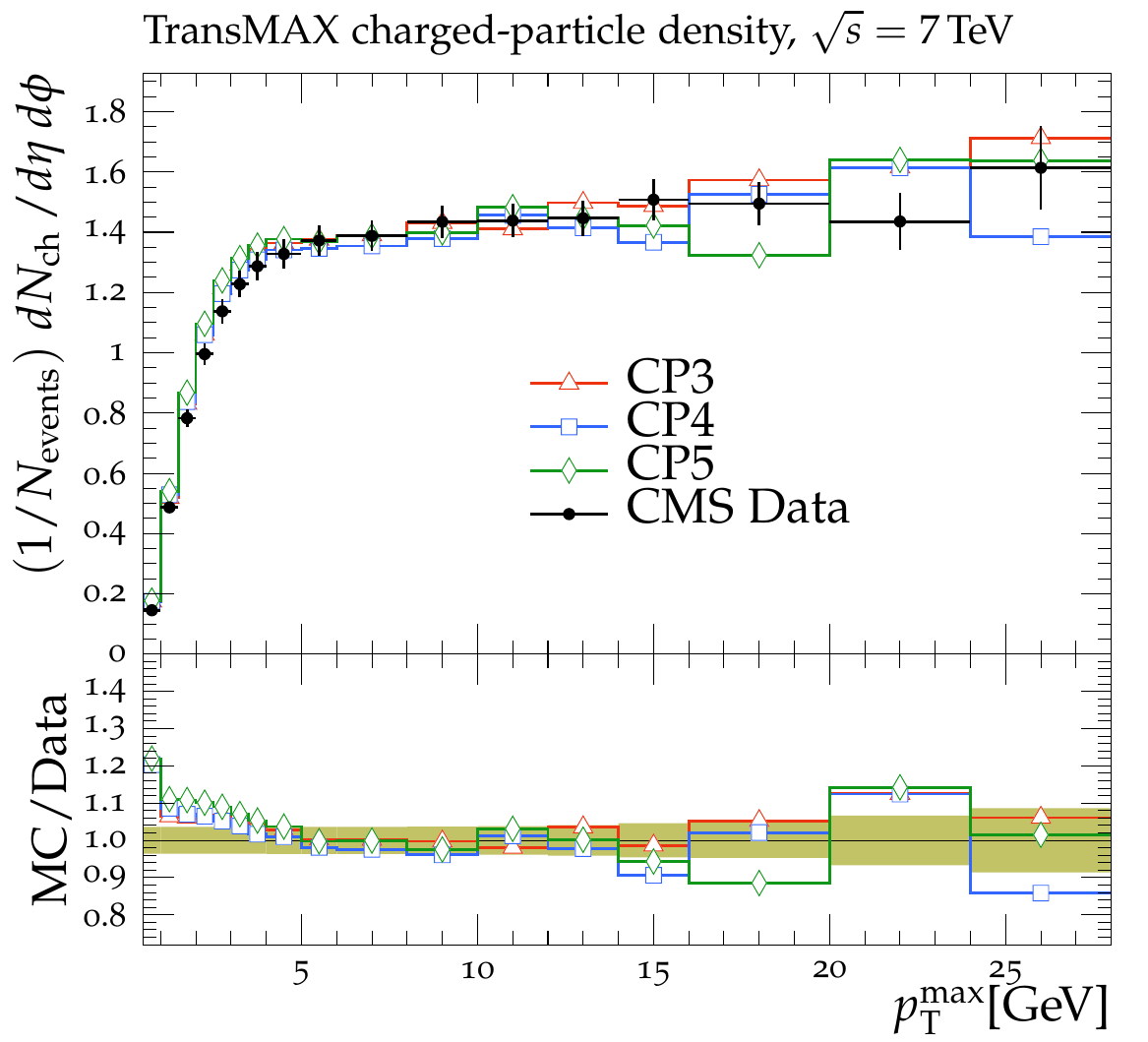}
\includegraphics[width=0.49\textwidth]{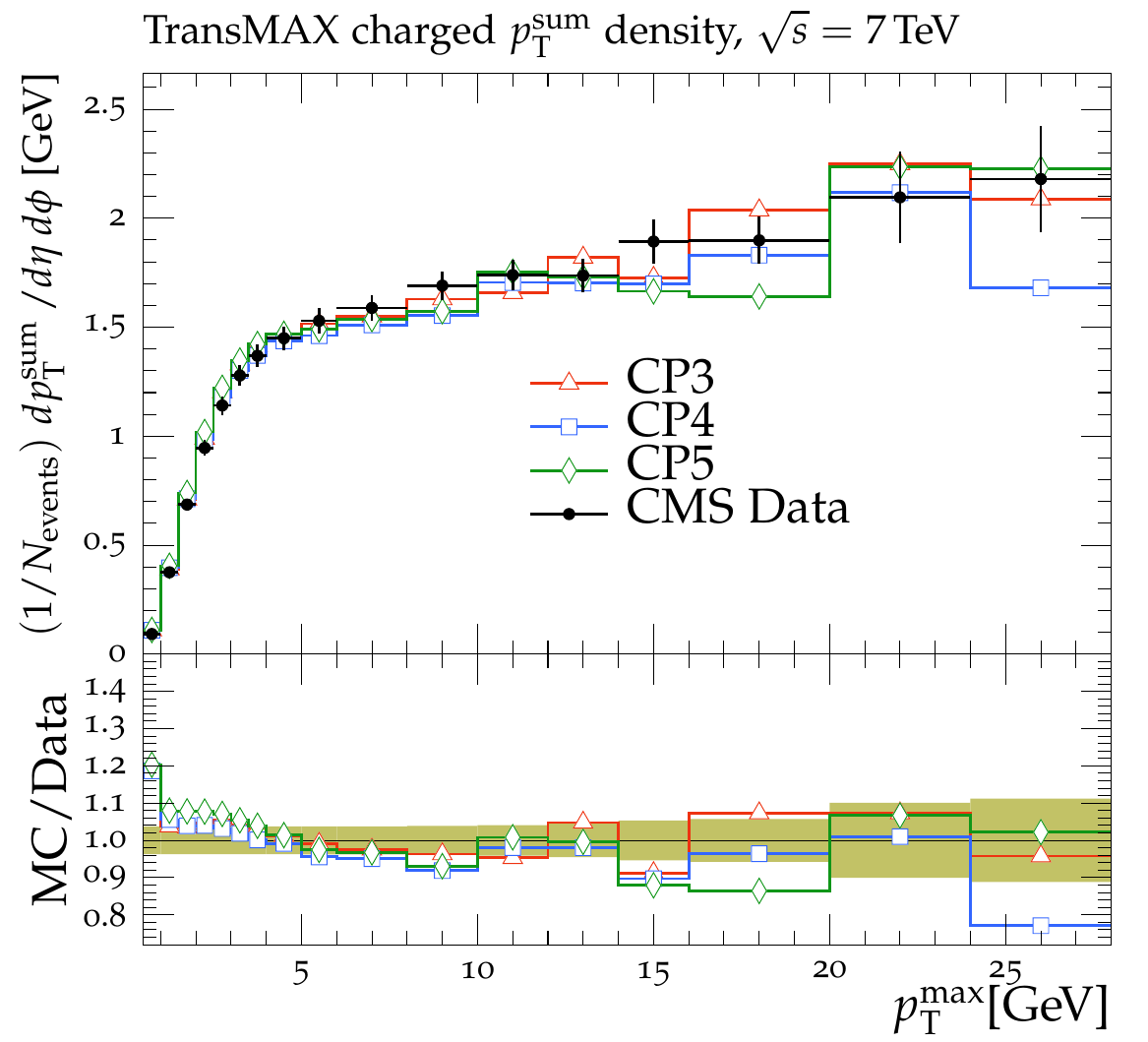}
\caption{The \tmin\ (upper left) charged-particle and (upper right) charged \ptsum\ densities
and the \tmax\ (lower left) charged-particle and (lower right) charged \ptsum\ densities, as a function of the transverse momentum of the leading charged particle, $\ptmax$, from the CMS $\sqrt{s}=7\TeV$ analysis~\cite{CMS:2012zxa}. Charged hadrons are measured with $\pt>0.5\GeV$ in $\abs{\eta}< 0.8$.  The data are compared with the CMS \PYTHIAviii (N)NLO-PDF tunes CP3, CP4, and CP5. The ratios of simulations to the data (MC/Data) are also shown, where the shaded band indicates the total experimental uncertainty in the data. Vertical lines drawn on the data points refer to the total uncertainty in the data. Vertical lines drawn on the MC points refer to the statistical uncertainty in the predictions. Horizontal bars indicate the associated bin width.}
\label{fig:6-3}
\end{figure*}

\begin{figure*}[ht!]
\centering
\includegraphics[width=0.49\textwidth]{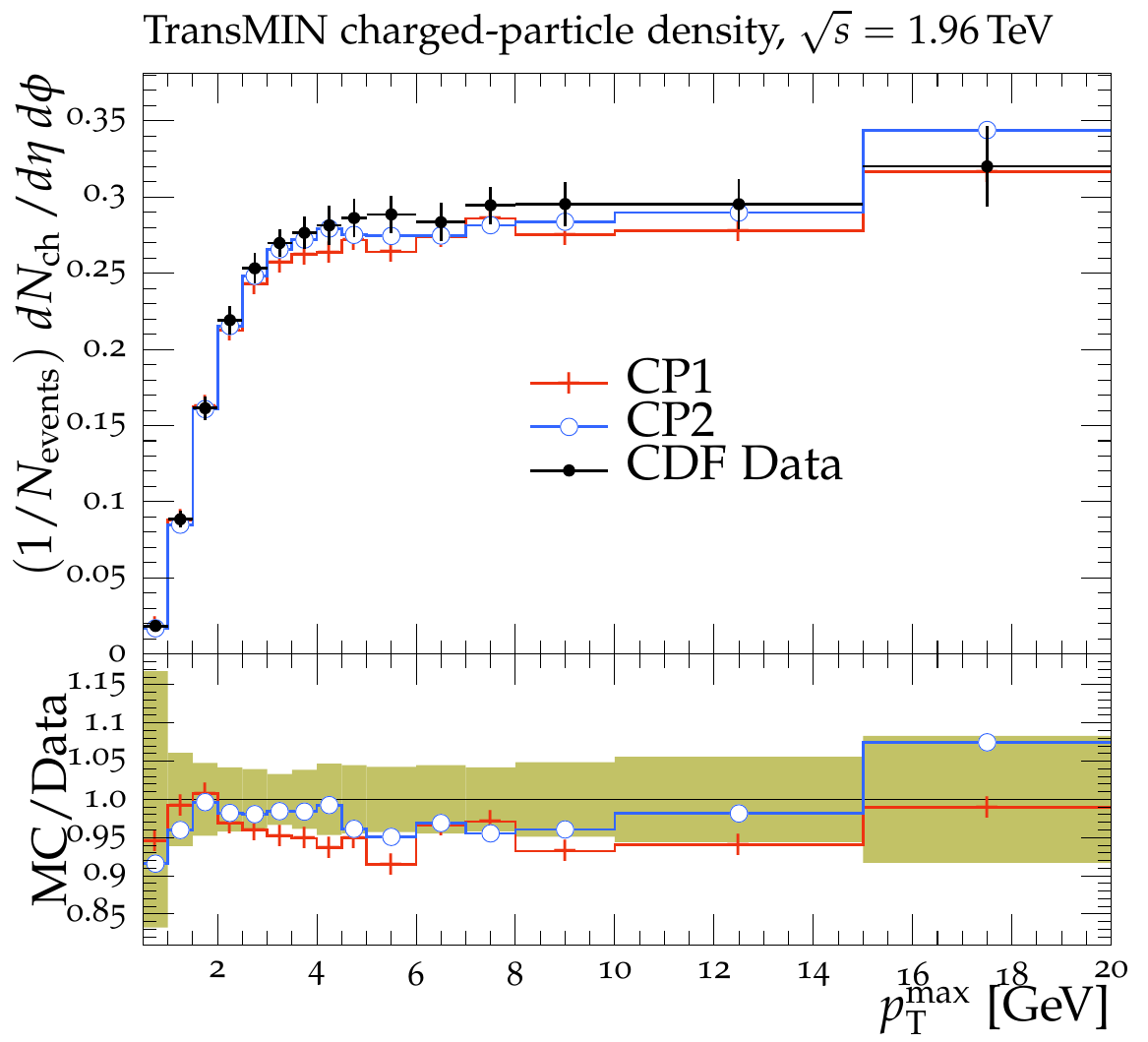}
\includegraphics[width=0.49\textwidth]{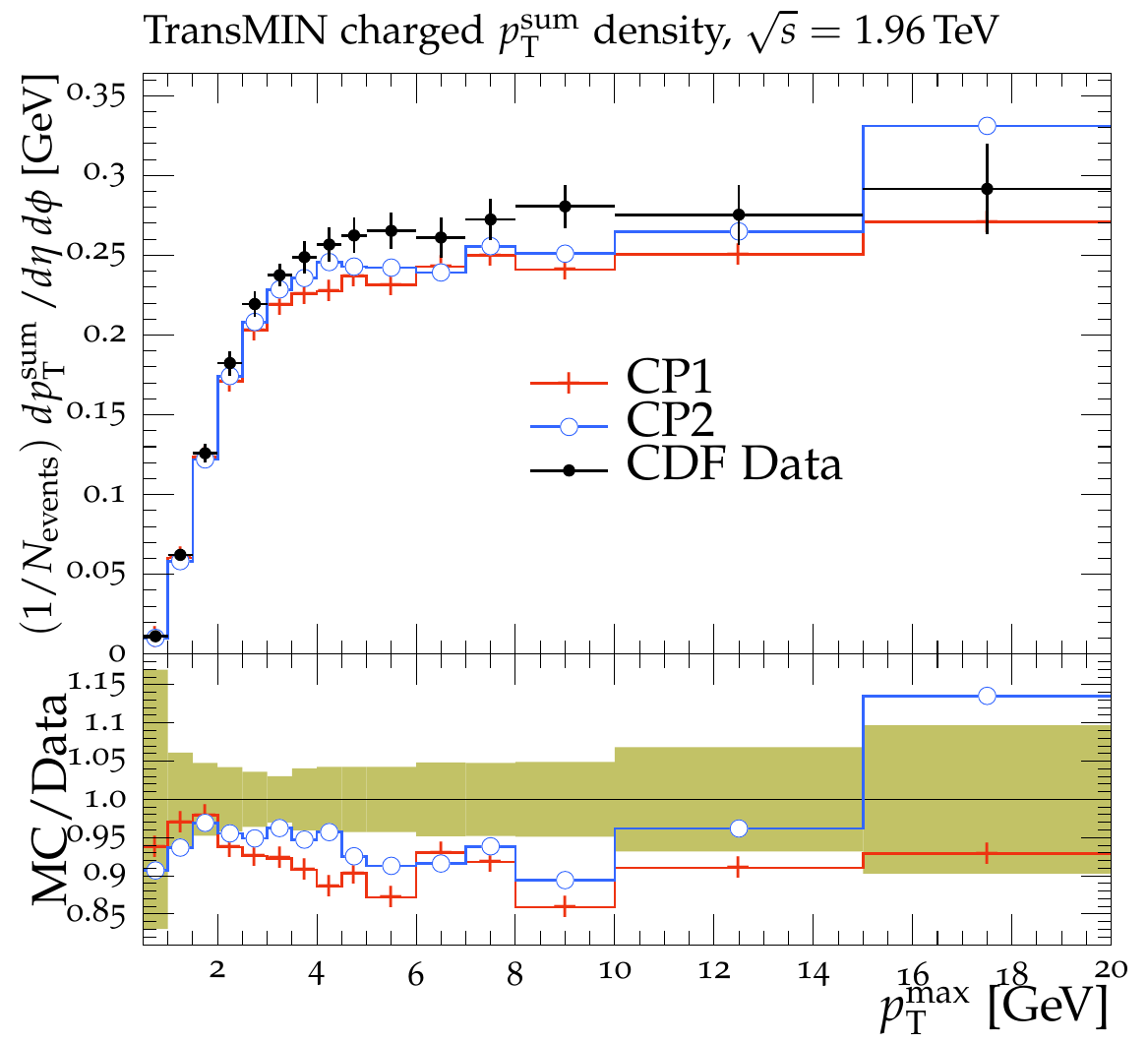}
\includegraphics[width=0.49\textwidth]{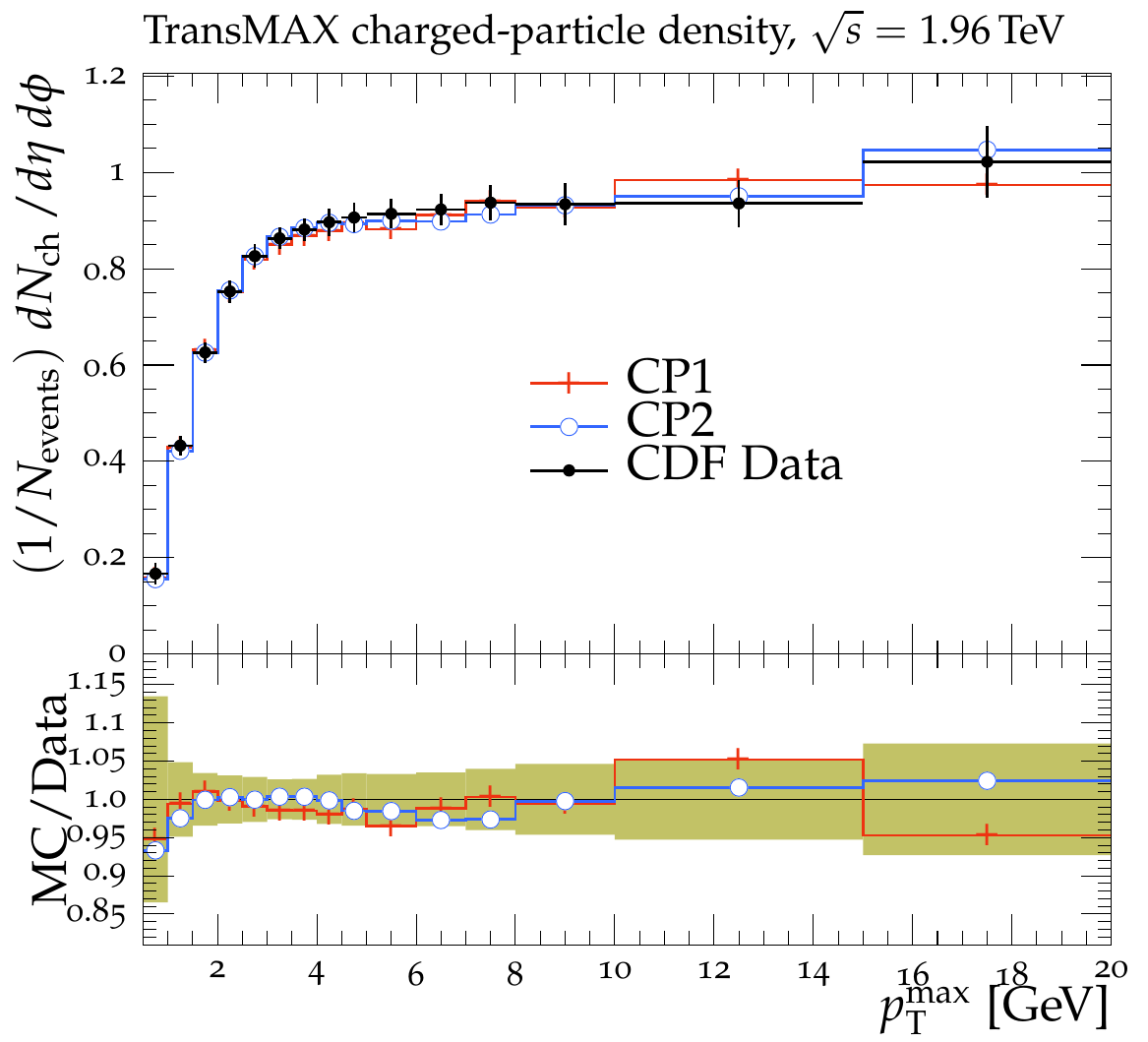}
\includegraphics[width=0.49\textwidth]{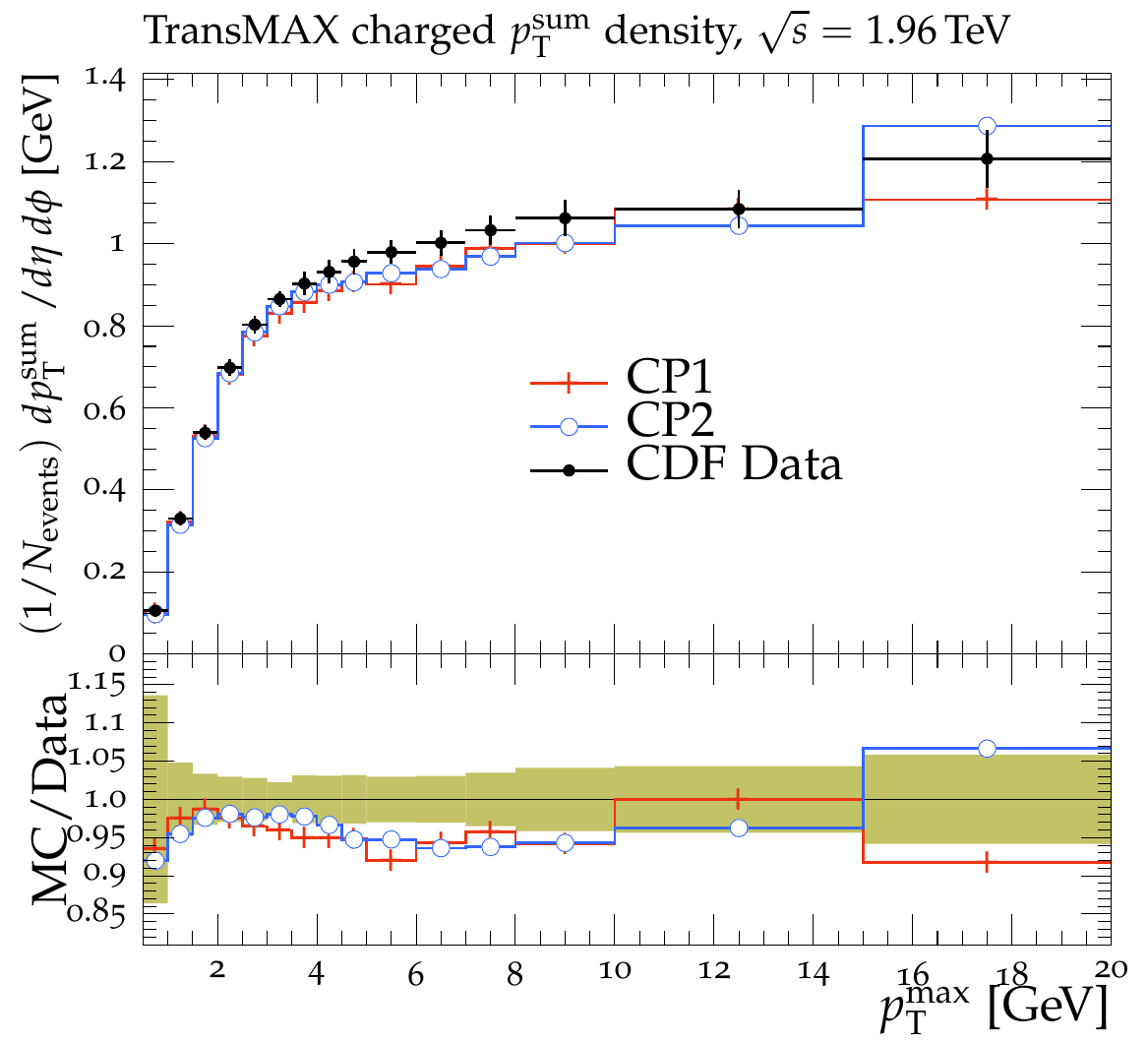}
\caption{The \tmin\ (upper left) charged-particle and (upper right) charged \ptsum\ densities
and the \tmax\ (lower left) charged-particle and (lower right) charged \ptsum\ densities, as a function of the transverse momentum of the leading charged particle, $\ptmax$, from the CDF $\sqrt{s}=1.96\TeV$ analysis~\cite{Aaltonen:2015aoa}. Charged hadrons are measured with $\pt>0.5\GeV$ in $\abs{\eta}< 0.8$. The data are compared with the CMS \PYTHIAviii LO-PDF tunes CP1 and CP2. The ratios of simulations to the data (MC/Data) are also shown, where the shaded band indicates the total experimental uncertainty in the data. Vertical lines drawn on the data points refer to the total uncertainty in the data. Vertical lines drawn on the MC points refer to the statistical uncertainty in the predictions. Horizontal bars indicate the associated bin width.}
\label{fig:3-3}
\end{figure*}

\begin{figure*}[ht!]
\centering
\includegraphics[width=0.49\textwidth]{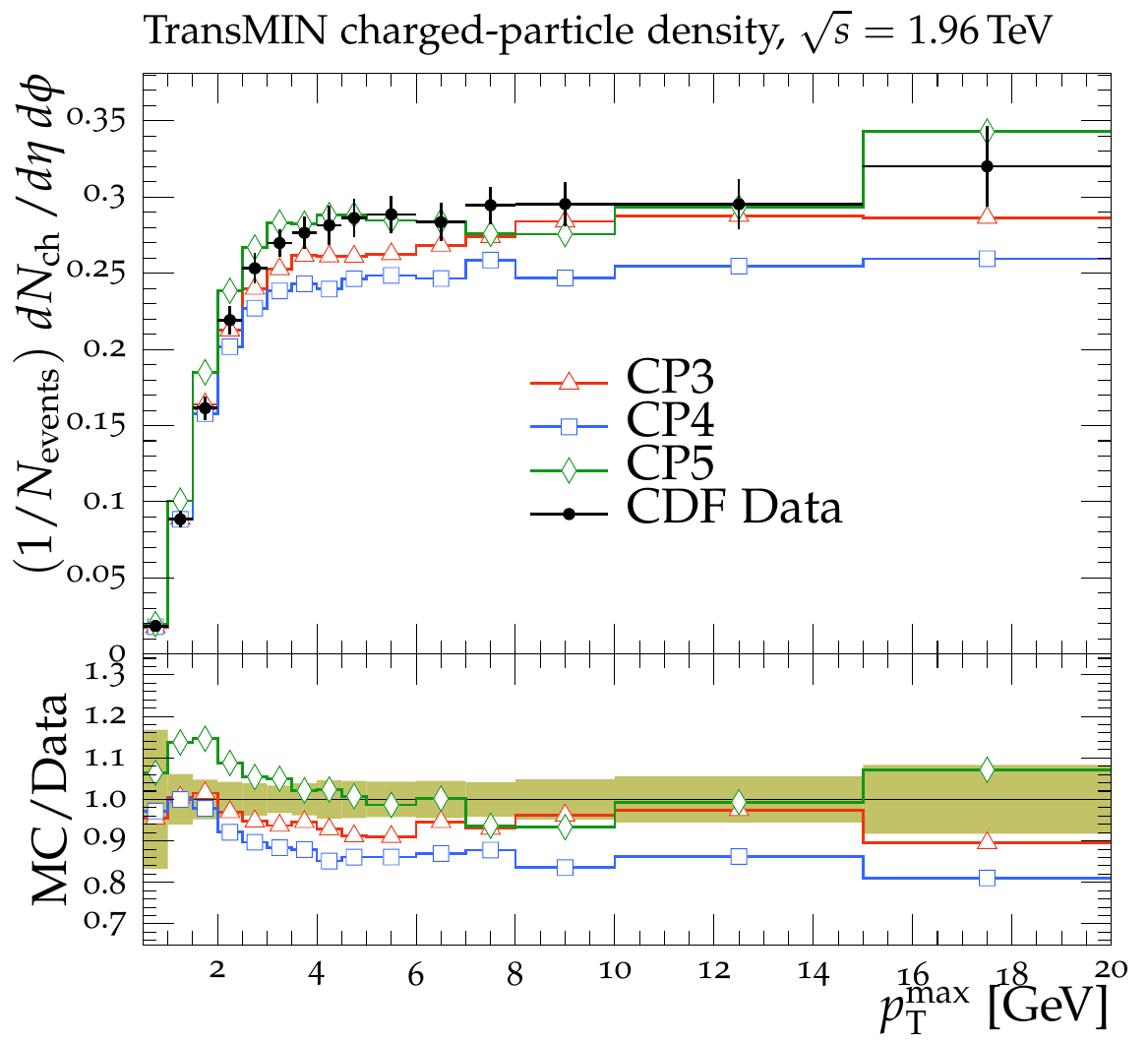}
\includegraphics[width=0.49\textwidth]{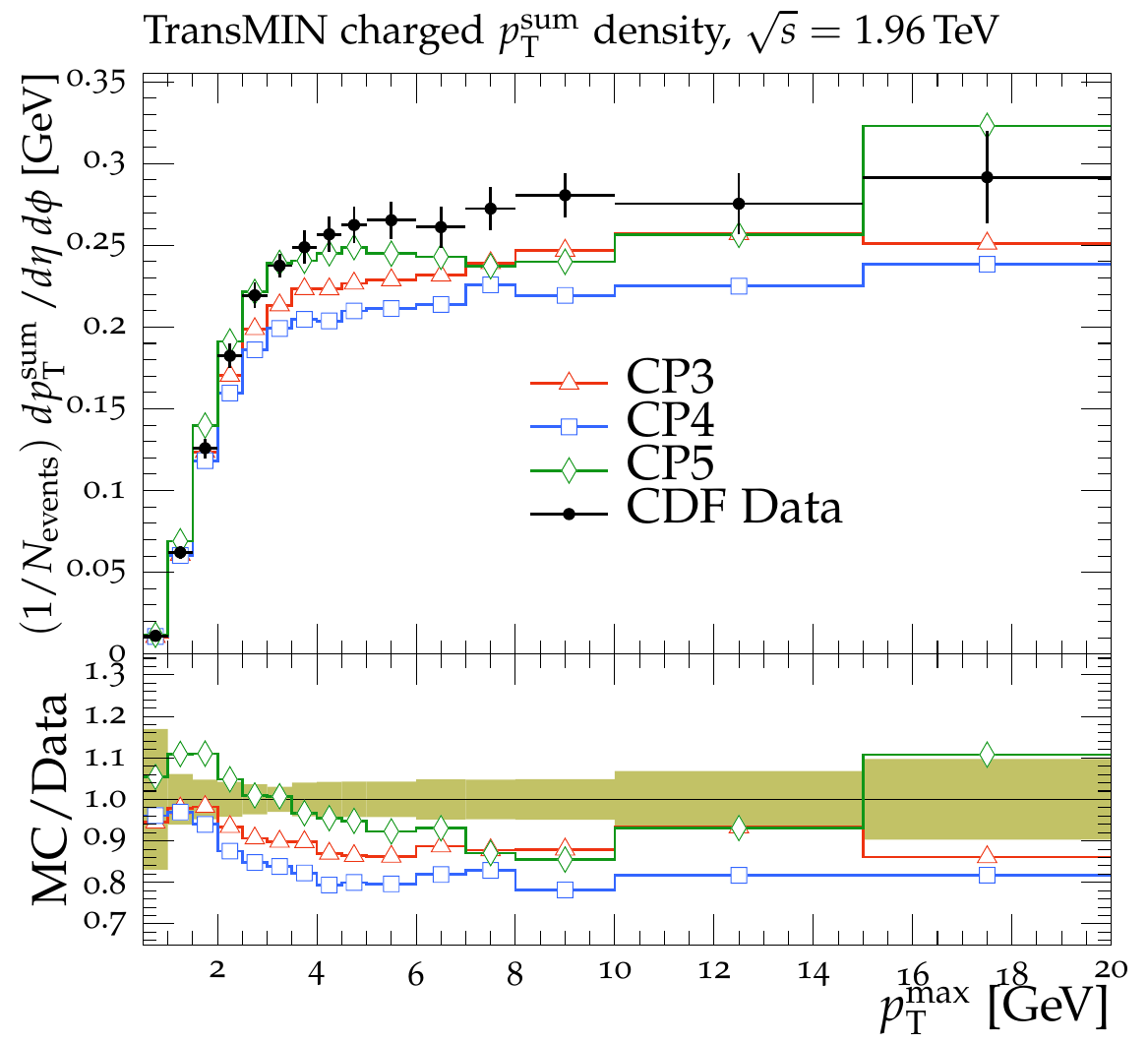}
\includegraphics[width=0.49\textwidth]{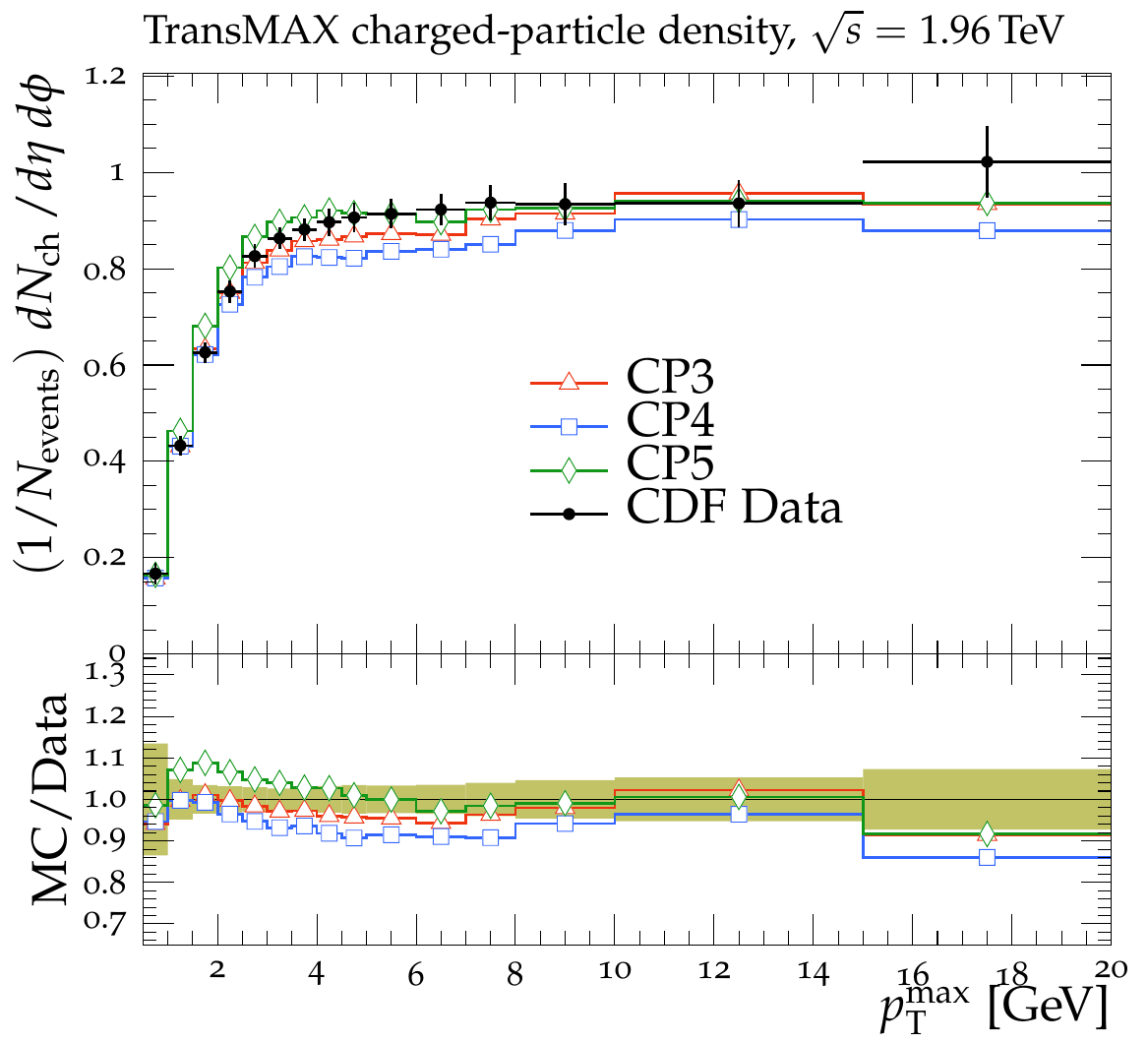}
\includegraphics[width=0.49\textwidth]{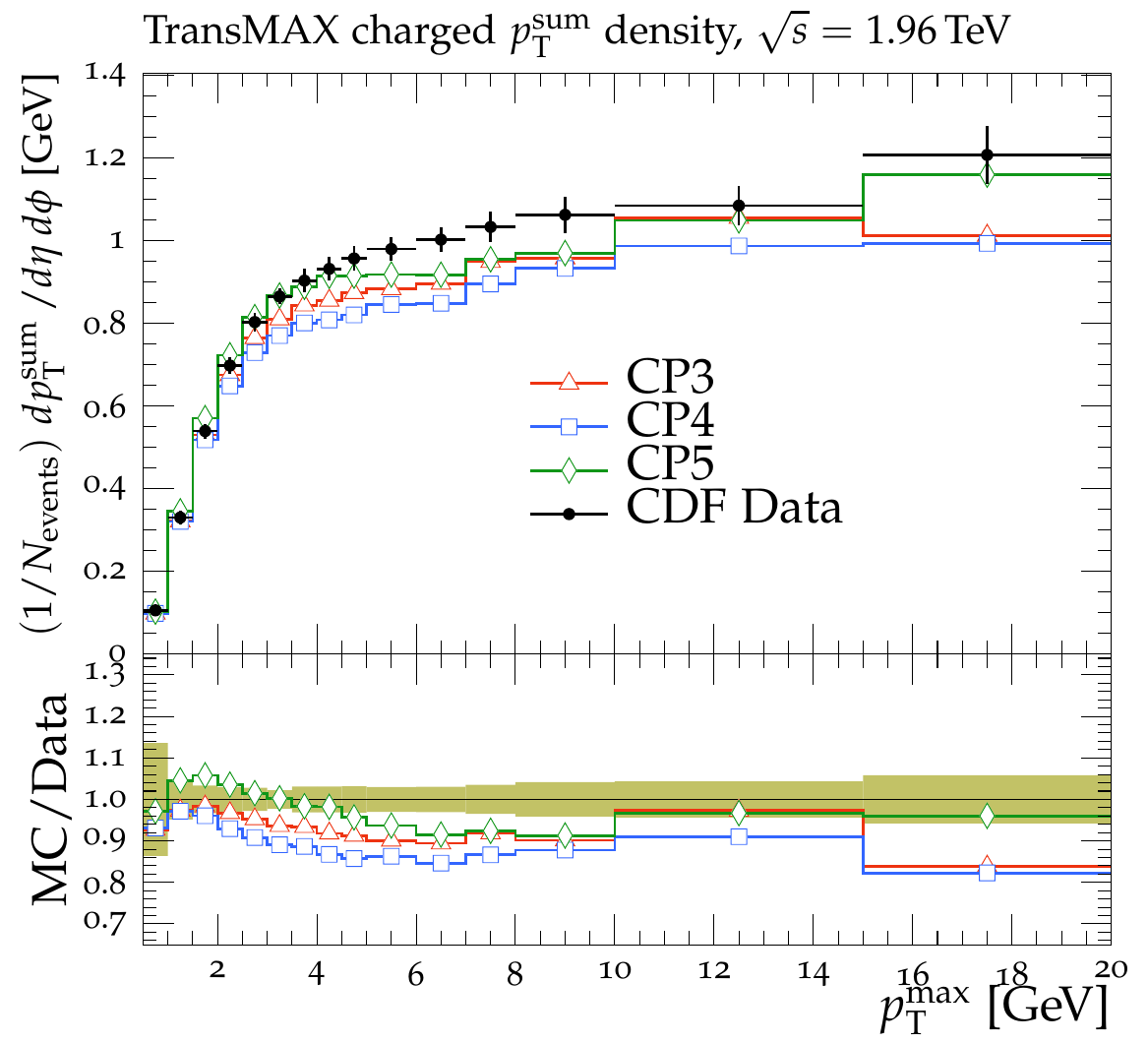}
\caption{The \tmin\ (upper left) charged-particle and (upper right) charged \ptsum\ densities
and the \tmax\ (lower left) charged-particle and (lower right) charged \ptsum\ densities, as a function of the transverse momentum of the leading charged particle, $\ptmax$, from the CDF $\sqrt{s}=1.96\TeV$ analysis~\cite{Aaltonen:2015aoa}. Charged hadrons are measured with $\pt>0.5\GeV$ in $\abs{\eta}< 0.8$. The data are compared with the CMS \PYTHIAviii (N)NLO-PDF tunes CP3, CP4, and CP5. The ratios of simulations to the data (MC/Data) are also shown, where the shaded band indicates the total experimental uncertainty in the data. Vertical lines drawn on the data points refer to the total uncertainty in the data. Vertical lines drawn on the MC points refer to the statistical uncertainty in the predictions. Horizontal bars indicate the associated bin width.}
\label{fig:4-3}
\end{figure*}

\begin{figure*}[ht!]
\centering
\includegraphics[width=0.49\textwidth]{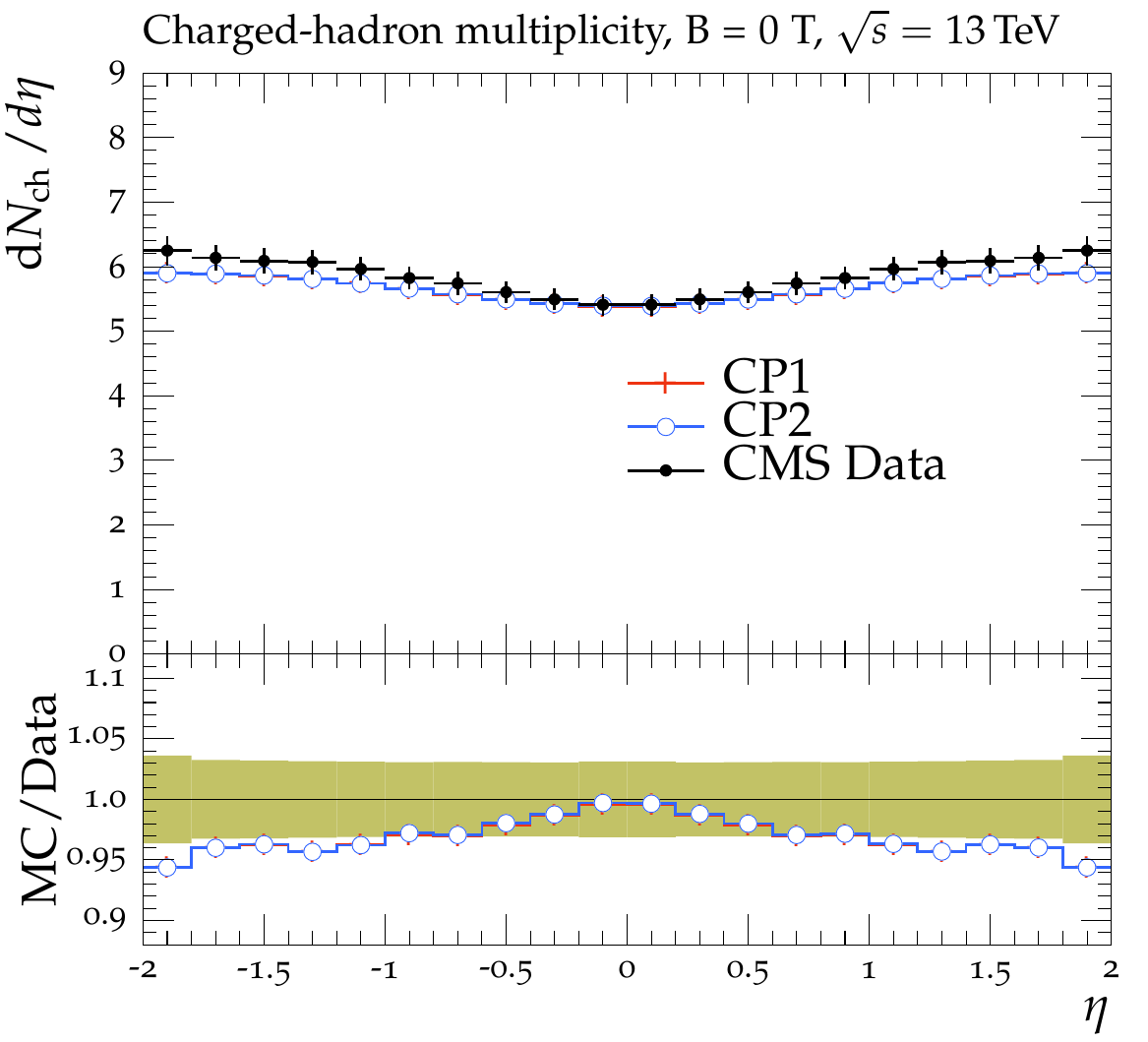}
\includegraphics[width=0.49\textwidth]{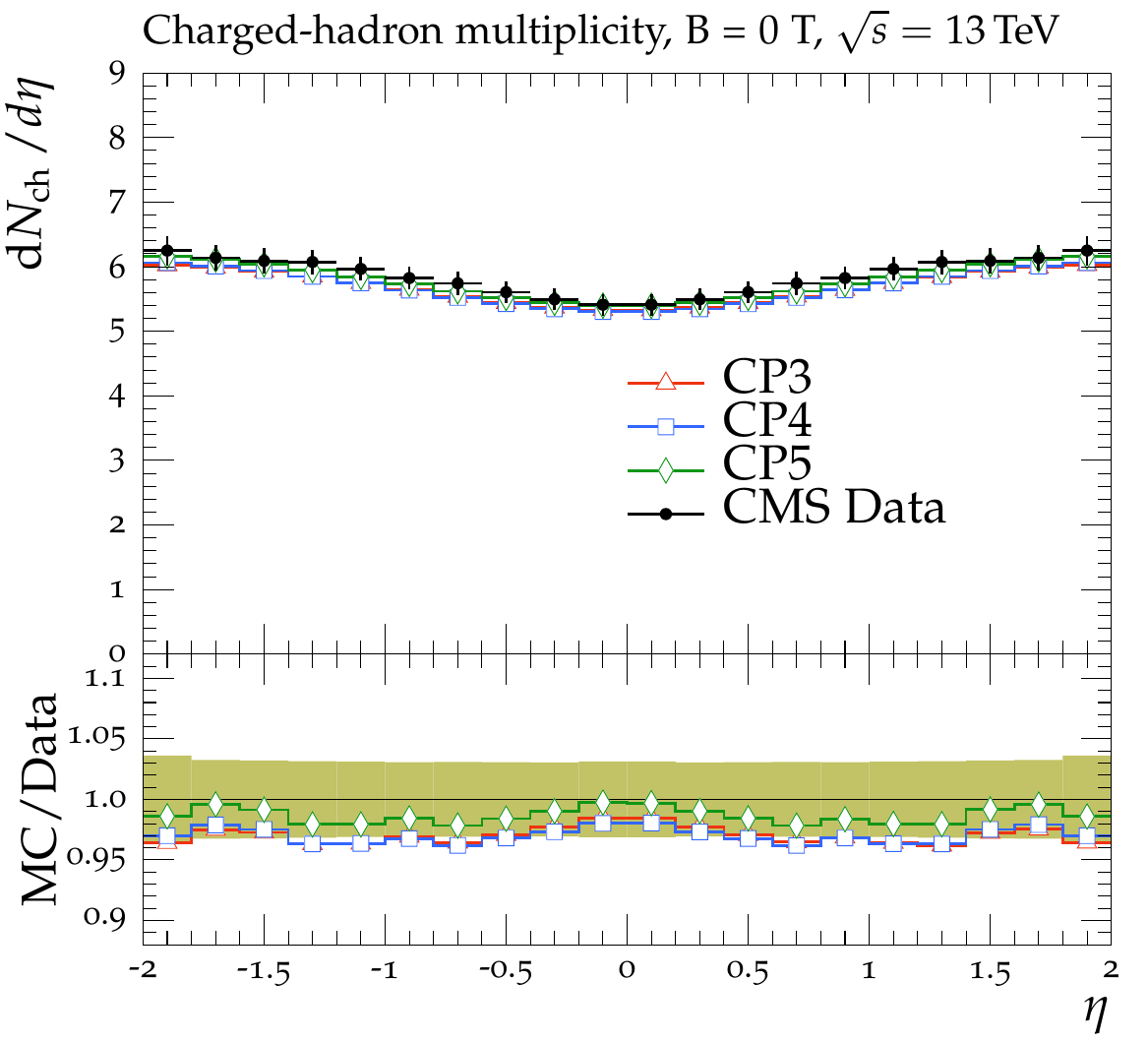}
\caption{The pseudorapidity distribution of charged hadrons measured in $\abs{\eta} < 2$ for an inclusive selection in inelastic proton-proton collisions, with zero magnetic field strength (B = 0 T), from the CMS $\sqrt{s}=13\TeV$ analysis~\cite{Khachatryan:2015jna}. The data are compared with (left) the CMS \PYTHIAviii LO-PDF tunes CP1 and CP2, and with (right) the CMS \PYTHIAviii NLO-PDF tune CP3 and the CMS \PYTHIAviii NNLO-PDF tunes CP4 and CP5. The ratios of simulations to the data (MC/Data) are also shown, where the shaded band indicates the total experimental uncertainty in the data. Vertical lines drawn on the data points refer to the total uncertainty in the data. Vertical lines drawn on the MC points refer to the statistical uncertainty in the predictions. Horizontal bars indicate the associated bin width.}
\label{fig:mb}
\end{figure*}

The contribution from CR also changes among the different tunes and depends on the choice of PDF and its order. In particular, the amount of CR is also affected by the shape of the PDFs at small fractional momenta $x$.

{\tolerance=800 Parameters related to the overlap matter distribution function differ between the different tunes. They are strongly correlated with the other UE parameters governing the MPI and CR contributions. In general, for a given value of the matter fraction (\texttt{coreFraction}), MPI contributions increase for decreasing values of the core radius (\texttt{coreRadius}). The inclusion of the rapidity ordering for ISR in tune CP5 impacts the UE observables by reducing the number of charged particles, and needs to be compensated by a larger amount of MPI contributions. \par}

The $\chi^2$ per degree of freedom (dof) listed in Tables~\ref{tab:lo_tunes} and~\ref{tab:nlo_tunes} refers to the quantity $\chi^2(p)$ in Eq.~(\ref{chi}), divided by dof in the fit. The eigentunes (Appendix~A) correspond to the tunes in which the changes in the $\chi^2$ ($\Delta\chi^2$) of the fit relative to the best fit value equals the $\chi^2$ value obtained in the tune, \ie, $\Delta\chi^2_\text{min}=\chi^2$. Such a variation of the $\chi^2$ produces a tune whose uncertainty bands are roughly the same as the uncertainties in the fitted data points. This is the main motivation why this choice of variation was considered. For all tunes in Tables~\ref{tab:lo_tunes} and~\ref{tab:nlo_tunes}, the fit quality is good, with $\chi^2$/dof values very close to 1.

Figures~\ref{fig:1-3}--\ref{fig:6-3} show comparisons of the UE observables measured at various collision energies to predictions from the new tunes. Figures~\ref{fig:1-3} and \ref{fig:2-3} compare the charged-particle and \ptsum\ densities measured at $\sqrt{s}=13\TeV$ by the CMS experiment~\cite{CMS:2015zev} in the \tmin\ and \tmax\ regions to predictions from the LO-PDF-based tunes and the higher-order-PDF-based tunes.
Figures~\ref{fig:5-3} and \ref{fig:6-3} compare the charged-particle and \ptsum\ densities measured at $\sqrt{s}=7\TeV$ by the CMS experiment~\cite{CMS:2012zxa} in the \tmin\ and \tmax\ regions to predictions from the LO-PDF-based tunes and the higher-order-PDF-based tunes.
In Figs.~\ref{fig:3-3} and \ref{fig:4-3} similar comparisons are shown for the observables measured at $\sqrt{s}=1.96\TeV$ by the CDF experiment~\cite{Aaltonen:2015aoa} in the \tmin\ and \tmax\ regions.
All predictions reproduce well the UE observables at $\sqrt{s}=1.96$, 7, and 13\TeV.
Predictions from LO tunes are slightly better than the higher-order tunes in describing the energy dependence of the considered UE measurements.

In the region of small $\ptmax$ values ($\ptmax<3\GeV$), where contributions from diffractive processes are relevant, the predictions do not always reproduce the measurements and exhibit discrepancies up to 20\%.
Predictions from all of the new tunes cannot reproduce the UE data measured at $\sqrt{s} = 300$ and 900\GeV~\cite{Aaltonen:2015aoa}.

Figure~\ref{fig:mb} shows the charged-particle multiplicity as a function of pseudorapidity for charged particles in $\abs{\eta} < 2$ measured by the CMS experiment at $\sqrt{s}=13\TeV$~\cite{Khachatryan:2015jna} in MB events. These events were recorded with no magnetic field, so all particles irrespective of their \pt are measured. Data are compared with the predictions of the new \PYTHIAviii tunes. All of them are able to reproduce the measurement at the same level of agreement, independently of the PDF used for the UE simulation. We could not find any MB or UE observable where the level of agreement between data and predictions from the different tunes is significantly different.

\section{Validation of the new \texorpdfstring{\PYTHIAviii}{PYTHIA8} tunes}
\label{ValidationAndPerformanceOfTheNewTune}
{\tolerance=800 In this section, comparisons of the predictions obtained with the new tunes to various experimental measurements performed by the CMS experiment are provided. Unless otherwise stated, the comparisons are made at $\sqrt{s}=13\TeV$. We compare the CMS UE tunes with MB and UE data measured at central and forward pseudorapidities that are not used in the fits. We examine how well multijet, Drell--Yan, and top quark observables are predicted by MC simulations using higher-order ME generators merged with \PYTHIAviii with the various new tunes. \par}

\subsection{Comparisons using event-shape observables}
In this subsection, predictions of the new tunes are compared to event-shape observables measured at LEP, in electron-positron collisions. These observables are particularly sensitive to the value of $\alpS^\mathrm{FSR}(m_\cPZ)$. Given the leptonic initial state, there is no effect coming from the values of the MPI, color reconnection, and ISR parameters.

When predictions with \PYTHIA~8 are used, an optimal value of $\alpS^\mathrm{FSR}(m_\cPZ)\sim0.13$ is found, which best describes these observables, independent of the PDF used for the modeling of the PS evolution.

\begin{figure*}[ht!]
\centering
\includegraphics[width=0.49\textwidth]{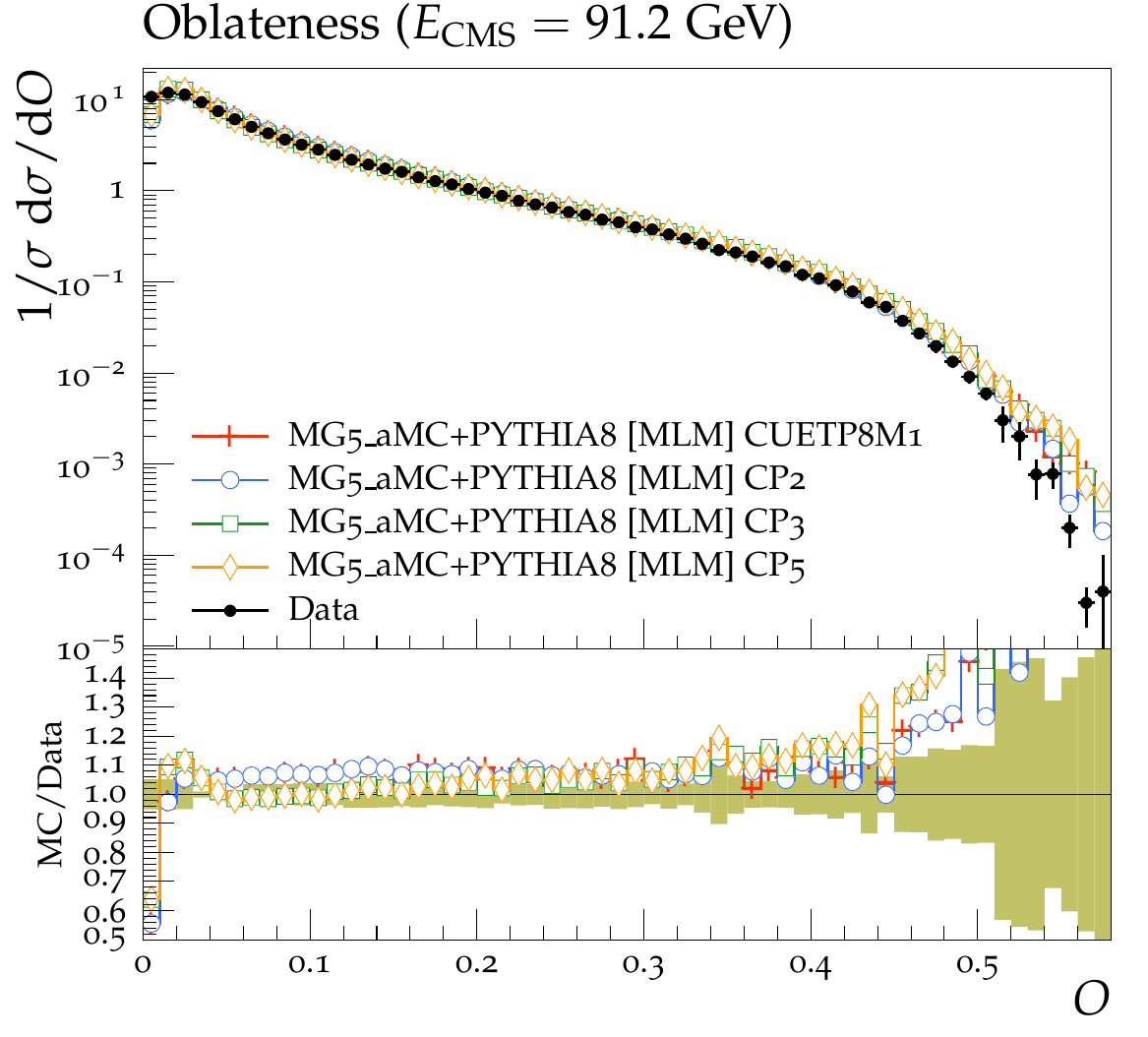}
\includegraphics[width=0.49\textwidth]{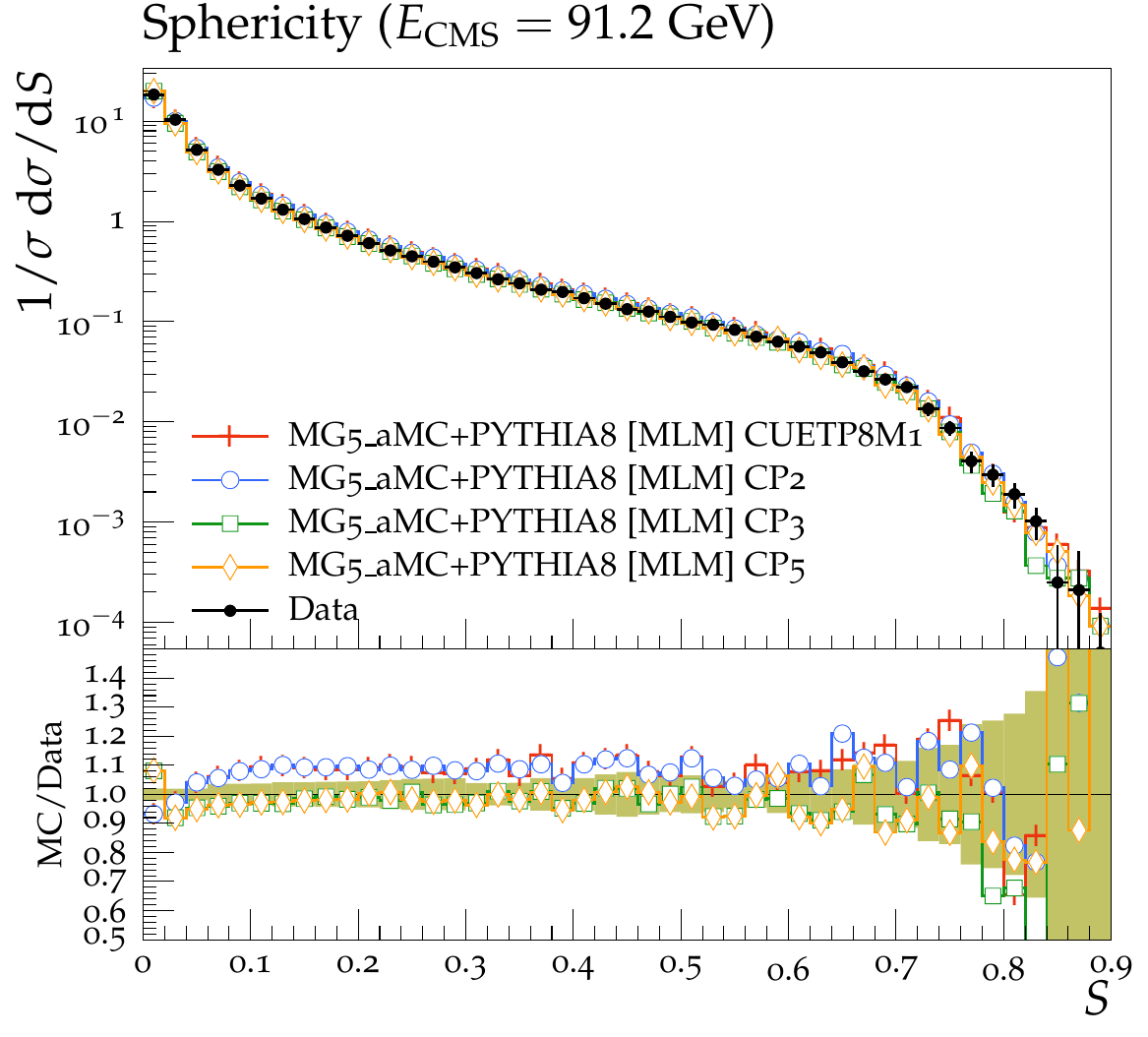}\\
\includegraphics[width=0.49\textwidth]{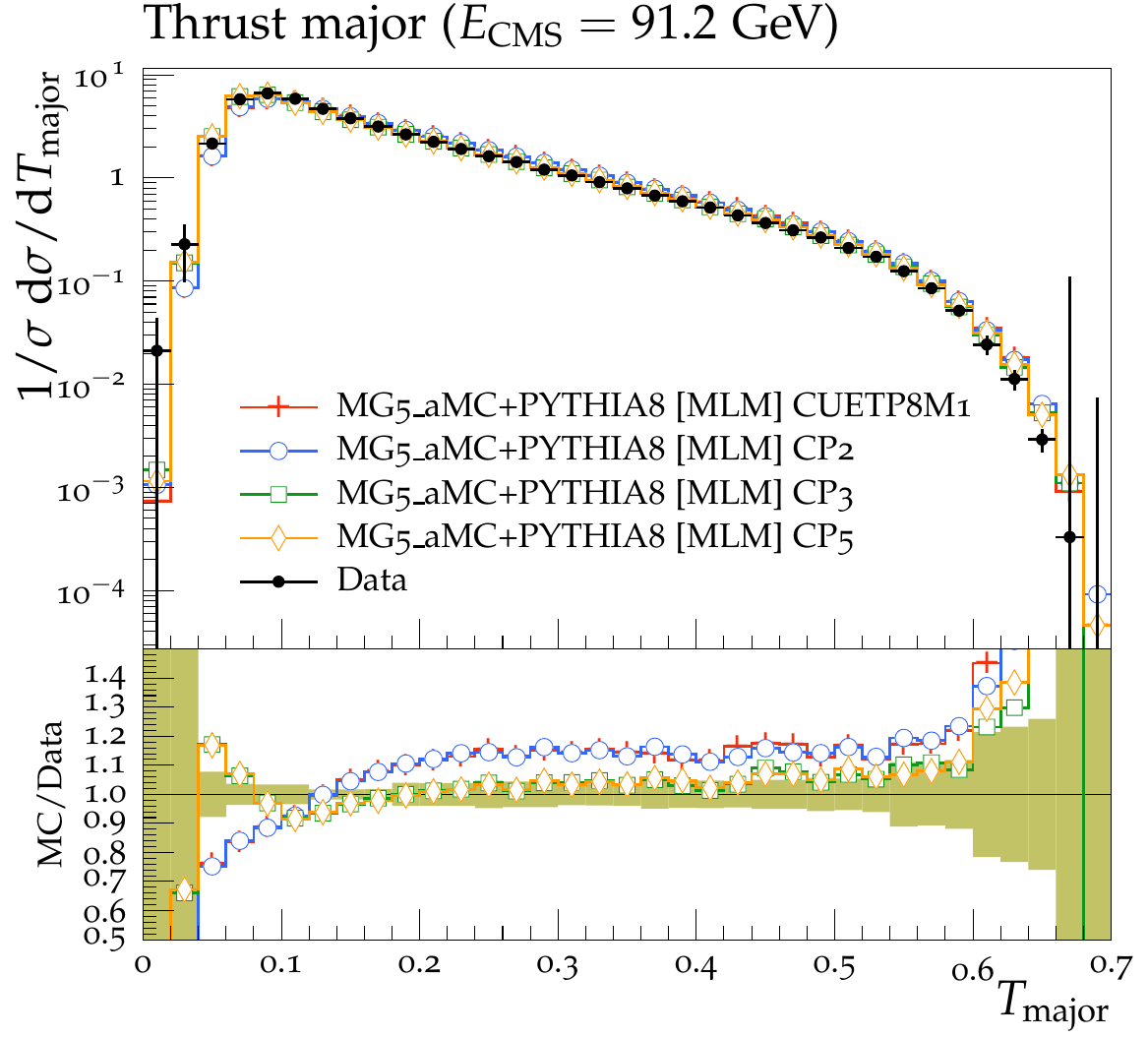}
\includegraphics[width=0.49\textwidth]{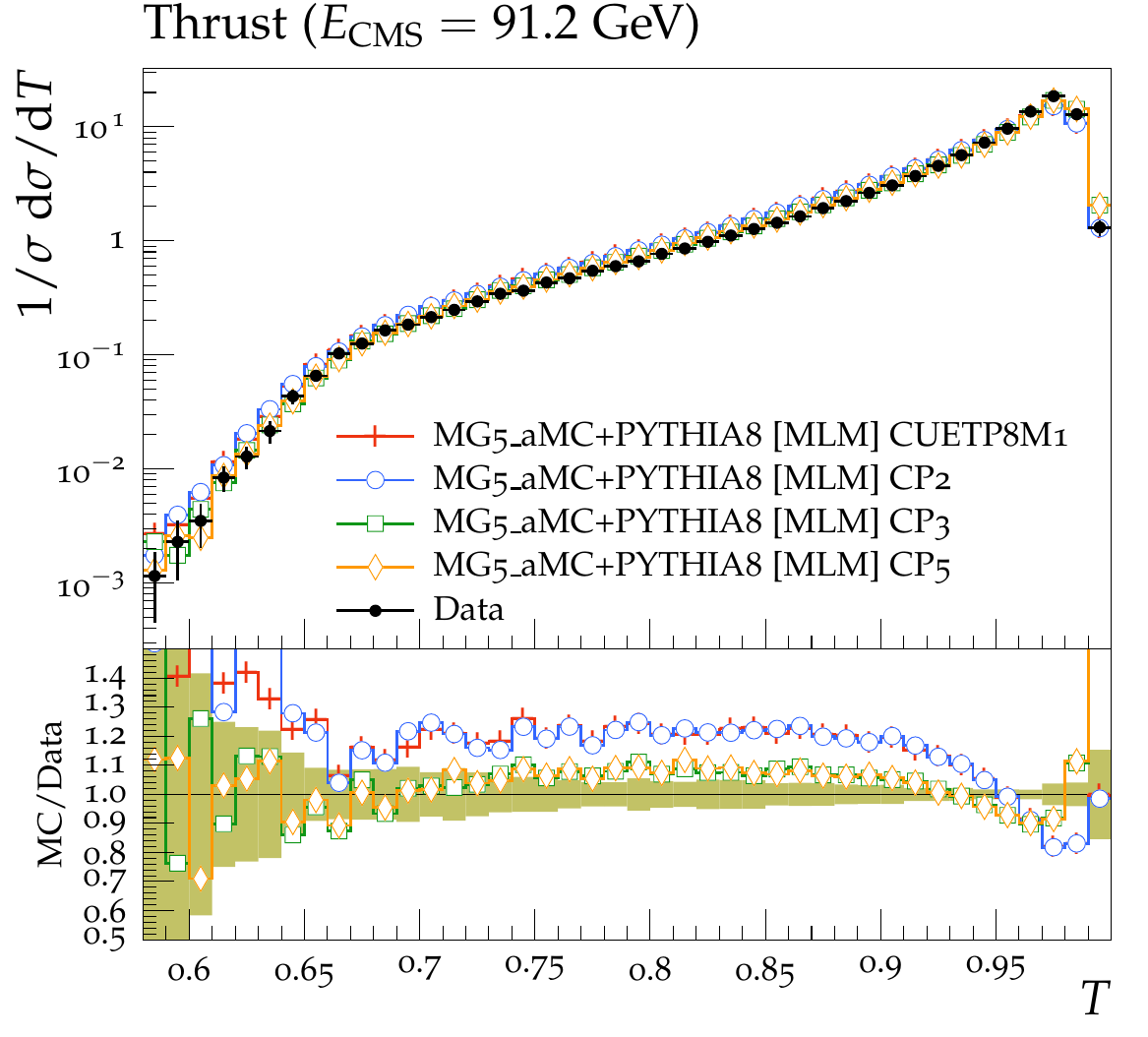}
\caption{The normalized cross sections as a function of event-shape variables, oblateness (upper left), sphericity (upper right), thrust (lower left), and thrust major (lower right) from the ALEPH $\sqrt{s} =91.2\GeV$ analysis~\cite{Heister:2003aj}, compared with the predictions by \MG~+~\PYTHIAviii with \ktMLM merging, for tunes CP2, CP3, and CP5. The ratio of the simulations to the data (MC/Data) is also shown, where the shaded band indicates the total experimental uncertainty in the data. Vertical lines drawn on the data points refer to the total uncertainty in the data. Vertical lines drawn on the MC points refer to the statistical uncertainty in the predictions. Horizontal bars indicate the associated bin width.}
\label{fig:perf_LEP}
\end{figure*}

\begin{figure*}[ht!]
\centering
\includegraphics[width=0.49\textwidth]{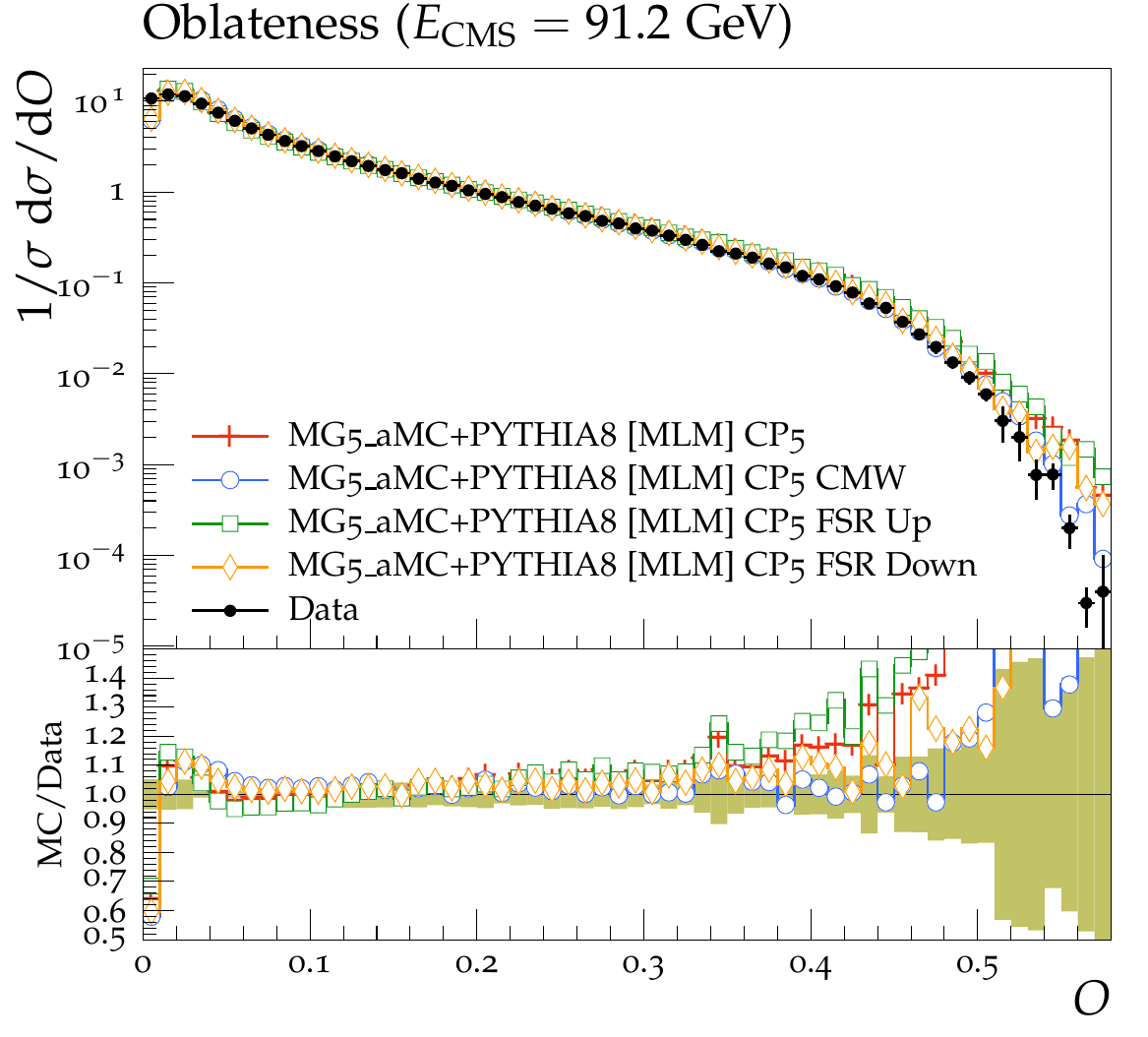}
\includegraphics[width=0.49\textwidth]{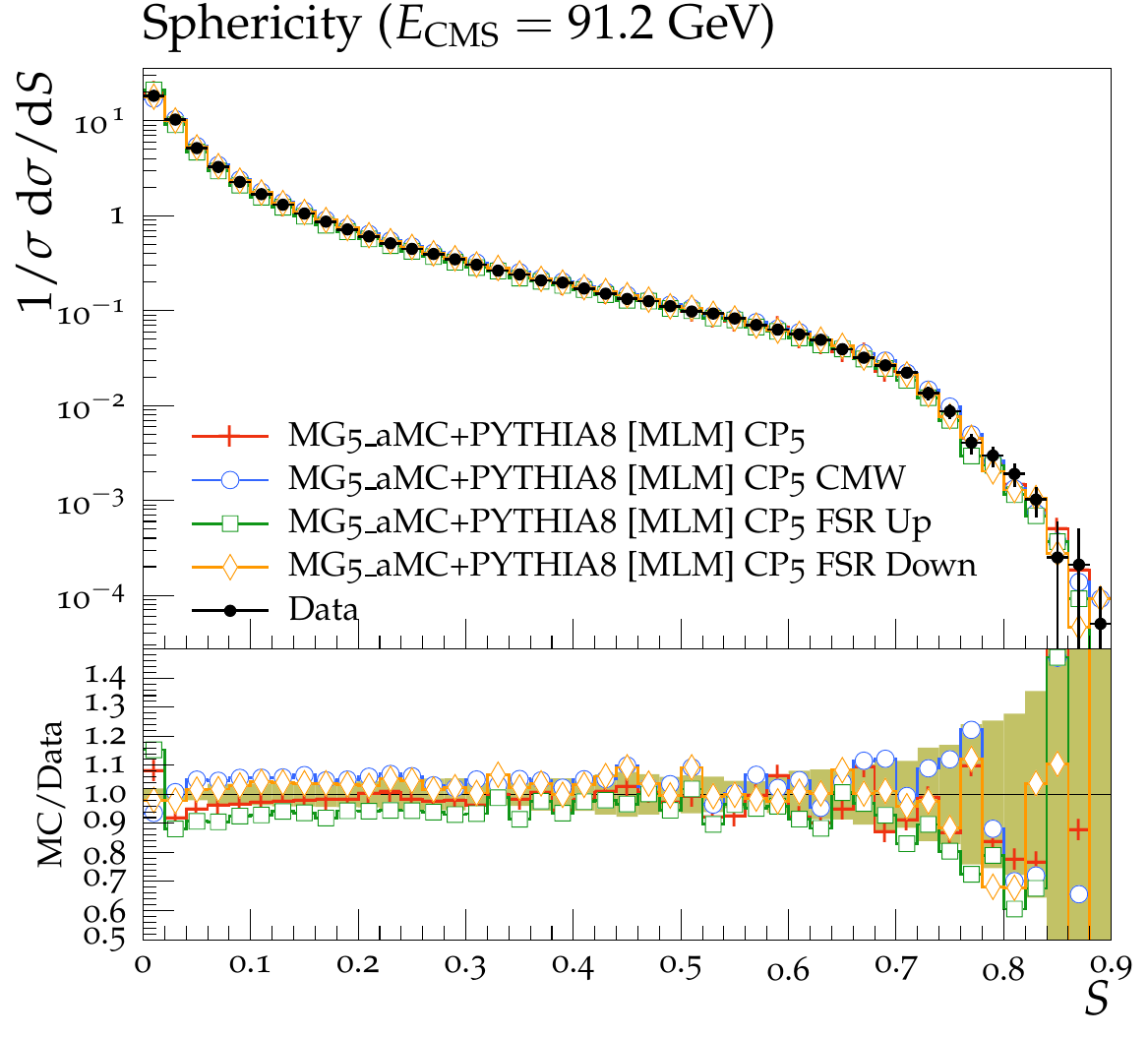}\\
\includegraphics[width=0.49\textwidth]{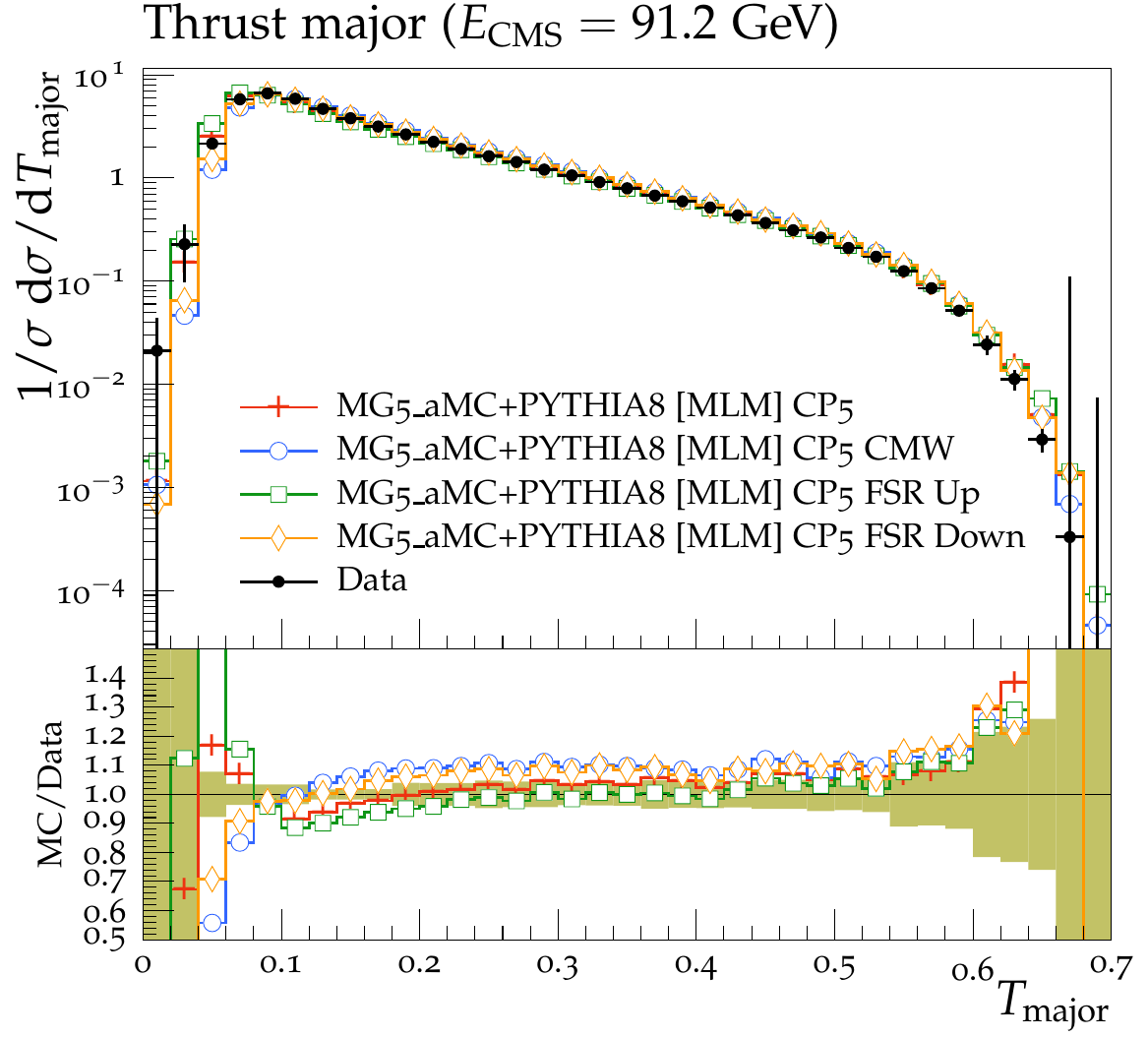}
\includegraphics[width=0.49\textwidth]{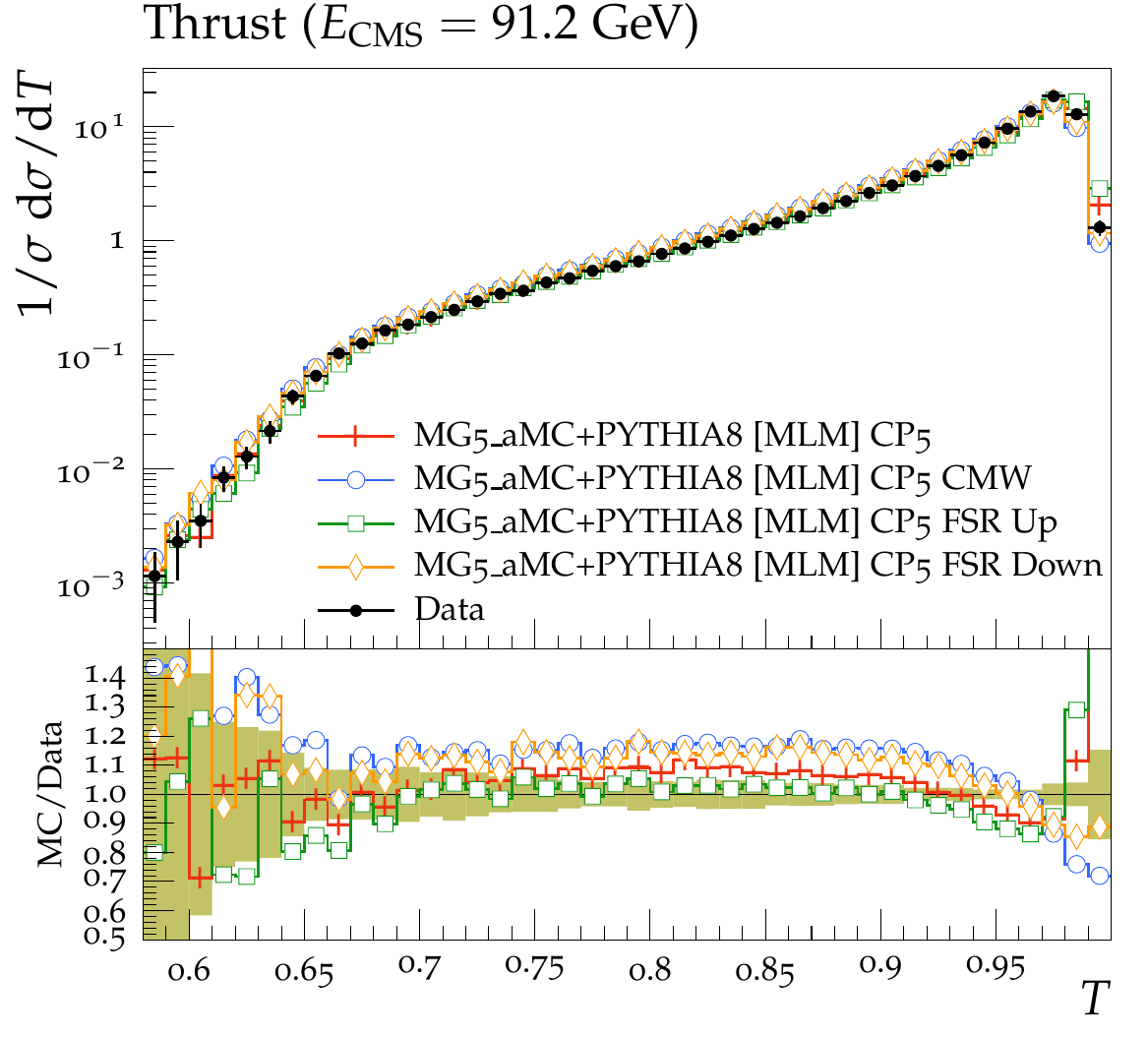}
\caption{The normalized cross sections as a function of event-shape variables, oblateness (upper left), sphericity (upper right), thrust (lower left), and thrust major (lower right) from the ALEPH $\sqrt{s} =91.2\GeV$ analysis~\cite{Heister:2003aj}, compared with the predictions by \MG~+~\PYTHIAviii with \ktMLM merging, for tune CP5, CP5 with CMW rescaling, CP5 FSR up, and CP5 FSR down. The ratio of the simulations to the data (MC/Data) is also shown, where the shaded band indicates the total experimental uncertainty in the data. Vertical lines drawn on the data points refer to the total uncertainty in the data. Vertical lines drawn on the MC points refer to the statistical uncertainty in the predictions. Horizontal bars indicate the associated bin width.}
\label{fig:perf_LEPCMW}
\end{figure*}

Figures~\ref{fig:perf_LEP} and~\ref{fig:perf_LEPCMW} display the oblateness ($O$), sphericity ($S$), thrust ($T$), and thrust major ($T_\text{major}$), measured in $\Pep\Pem \rightarrow \PZ\PGg^* \rightarrow \PQq\PAQq$ final states at $\sqrt{s} = 91.2$ \GeV by the ALEPH experiment~\cite{Heister:2003aj}. These observables measure the topology of the event. An isotropic event would have a value of $T$ close to 0.5, while values of $T$ close to 1 correspond to 2-jet events.

Predictions obtained with \MG~with up to 4 partons in the final state, and interfaced with the UE from the tune CUETP8M1 and the new  \PYTHIA~8 tunes CP2, CP3, and CP5 are considered (Figure~\ref{fig:perf_LEP}). Predictions using the tune CP2 do not describe the event-shape observables very well, with discrepancies with the data up to 30\% in the $T$ and $T_\text{major}$. In particular, tune CP2 predicts too many isotropical events. A similar description is obtained for predictions of \MG+\PYTHIA~8 with the tune CUETP8M1. A better agreement in the event-shape variables is observed for predictions using tune CP3 and CP5. A correct description of event-shape observables strongly depends on the value of the FSR strong coupling.
The observations above indicate that when merged configurations are considered, \ie, \MG~+~\PYTHIAviii,
where partons at higher multiplicities in the final state are simulated at the ME level, the description of event-shape observables degrades.
A value of $\alpS^\mathrm{FSR}(m_\cPZ)\sim0.13$ generally overestimates the number of final-state partons, while a lower $\alpS^\mathrm{FSR}(m_\cPZ)\sim0.12$ performs better.

At large values of $T$, where the hadronization effects become relevant, we observe a large difference between predictions from tunes using a small $\alpS^\mathrm{FSR}$ (CP3 and CP5) and tunes using a large $\alpS^\mathrm{FSR}$ (CP2 and CUETP8M1). These differences may be due to the interplay between the value of the strong coupling and the hadronization. Analyses particularly sensitive to hadronization should carefully evaluate the corresponding systematic uncertainties. In some cases retuning hadronization parameters may be desired.

We also compared \MG+\PYTHIA~8 with CP5, and CP5 with CMW rescaling \cite{Catani:1990rr} (Figure~\ref{fig:perf_LEPCMW}). Apart from $T$, for all shape variables considered, CP5 without CMW rescaling describes the data better.

\subsection{Comparisons using MB and other UE observables}
\label{PerformanceSoftEvents}

\begin{figure*}[ht!]
\centering
\includegraphics[width=0.49\textwidth]{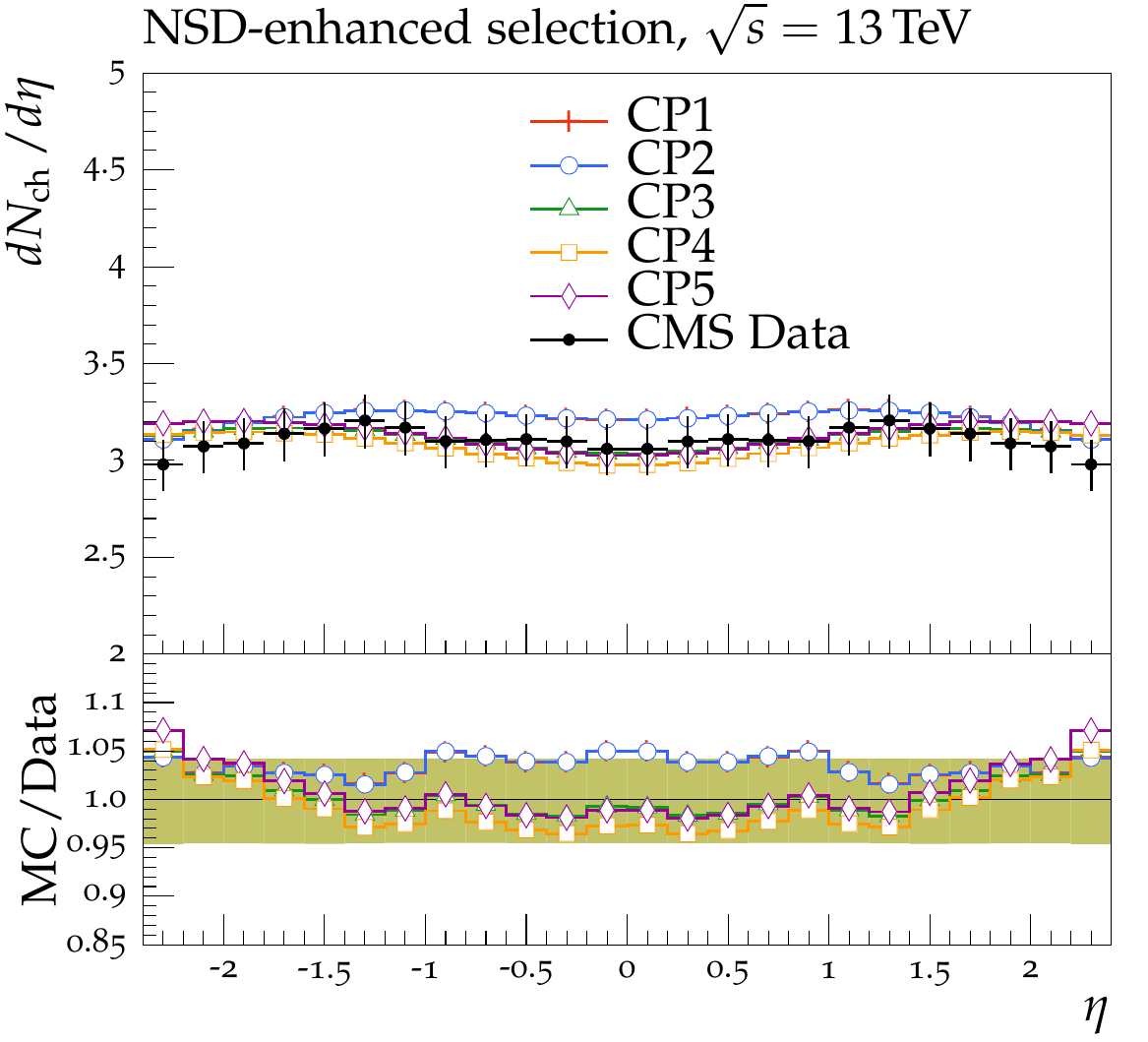}
\includegraphics[width=0.49\textwidth]{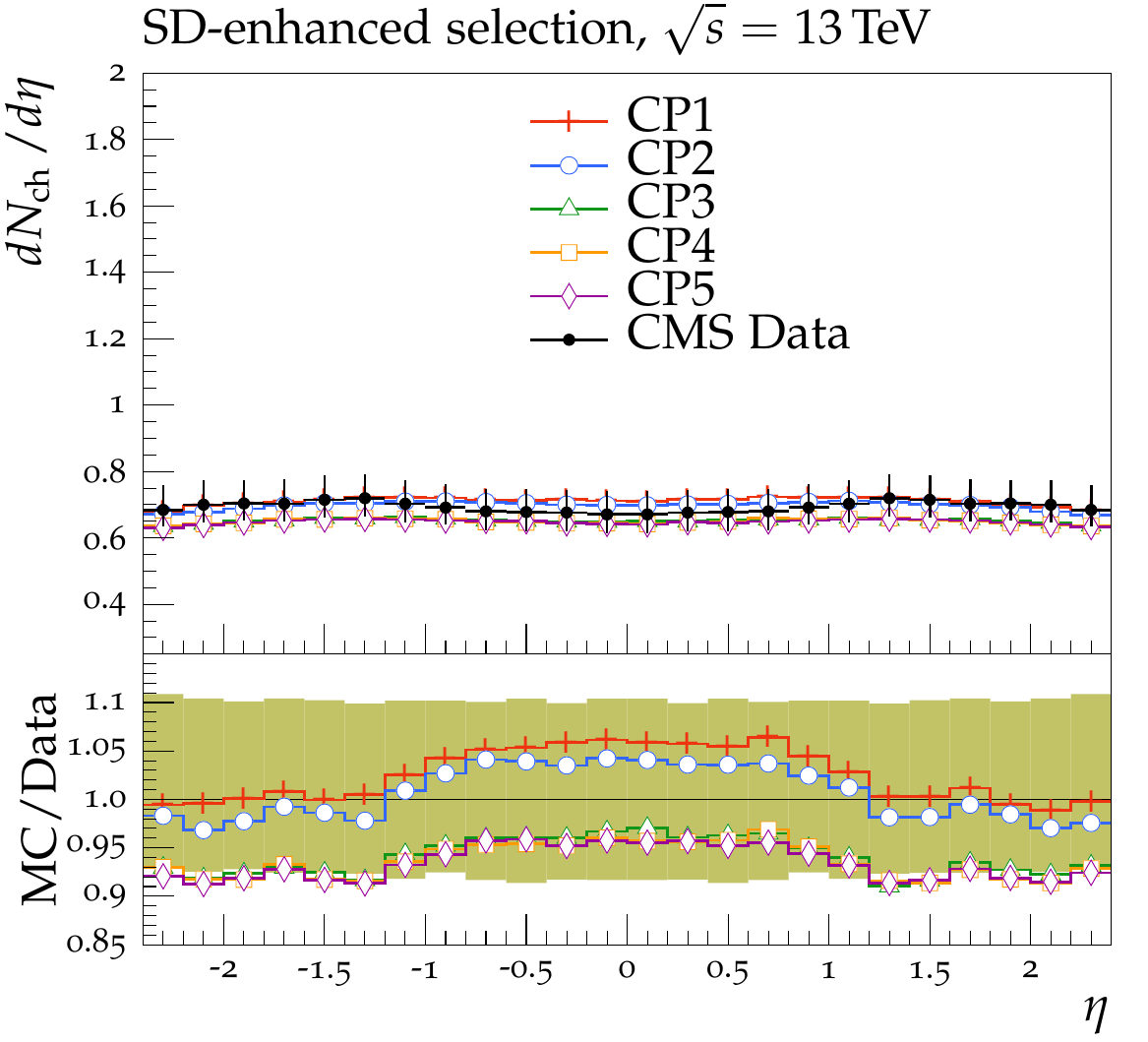}
\caption{The pseudorapidity distribution ($\pt>0.5\GeV$, $\abs{\eta}<2.4$) for (left) the NSD-enhanced and (right) the SD-enhanced event selection of charged particles in inelastic proton-proton collisions, from the CMS $\sqrt{s}=13\TeV$ analysis~\cite{Sirunyan:2018zdc}.  The data are compared with the CMS \PYTHIAviii LO-PDF tunes CP1 and CP2, the CMS \PYTHIAviii NLO-PDF tune CP3, and the CMS
\PYTHIAviii NNLO-PDF tunes CP4 and CP5.  The ratio of the simulations to the data (MC/Data) is also shown, where the shaded band indicates the total experimental uncertainty in the data. Vertical lines drawn on the data points refer to the total uncertainty in the data. Vertical lines drawn on the MC points refer to the statistical uncertainty in the predictions. Horizontal bars indicate the associated bin width.}
\label{fig:perf_13TeV_II}
\end{figure*}

{\tolerance=800 In this subsection, predictions of the new tunes are compared to the observables measured in MB collisions that are sensitive to contributions from soft emissions and MPI. Figure~\ref{fig:perf_13TeV_II} shows the charged-particle multiplicity as a function of pseudorapidity~\cite{Sirunyan:2018zdc} in NSD-enhanced and SD-enhanced event samples. The details of the selections can be found in Ref.~\cite{Sirunyan:2018zdc}. These observables are sensitive to SD, CD, and DD dissociation. It is observed that predictions from all of the tunes are similar to each other and describe well the measurements for both considered selections. This shows that the number of charged particles produced in diffractive processes and inelastic collisions is simultaneously described by the new CMS tunes.
Figure~\ref{fig:perf_13TeV_II} also demonstrates that tunes based on NNPDF3.1 PDF sets at orders higher than LO adequately describe the MB data. \par}

\begin{figure*}[ht!]
\centering
\includegraphics[width=0.49\textwidth]{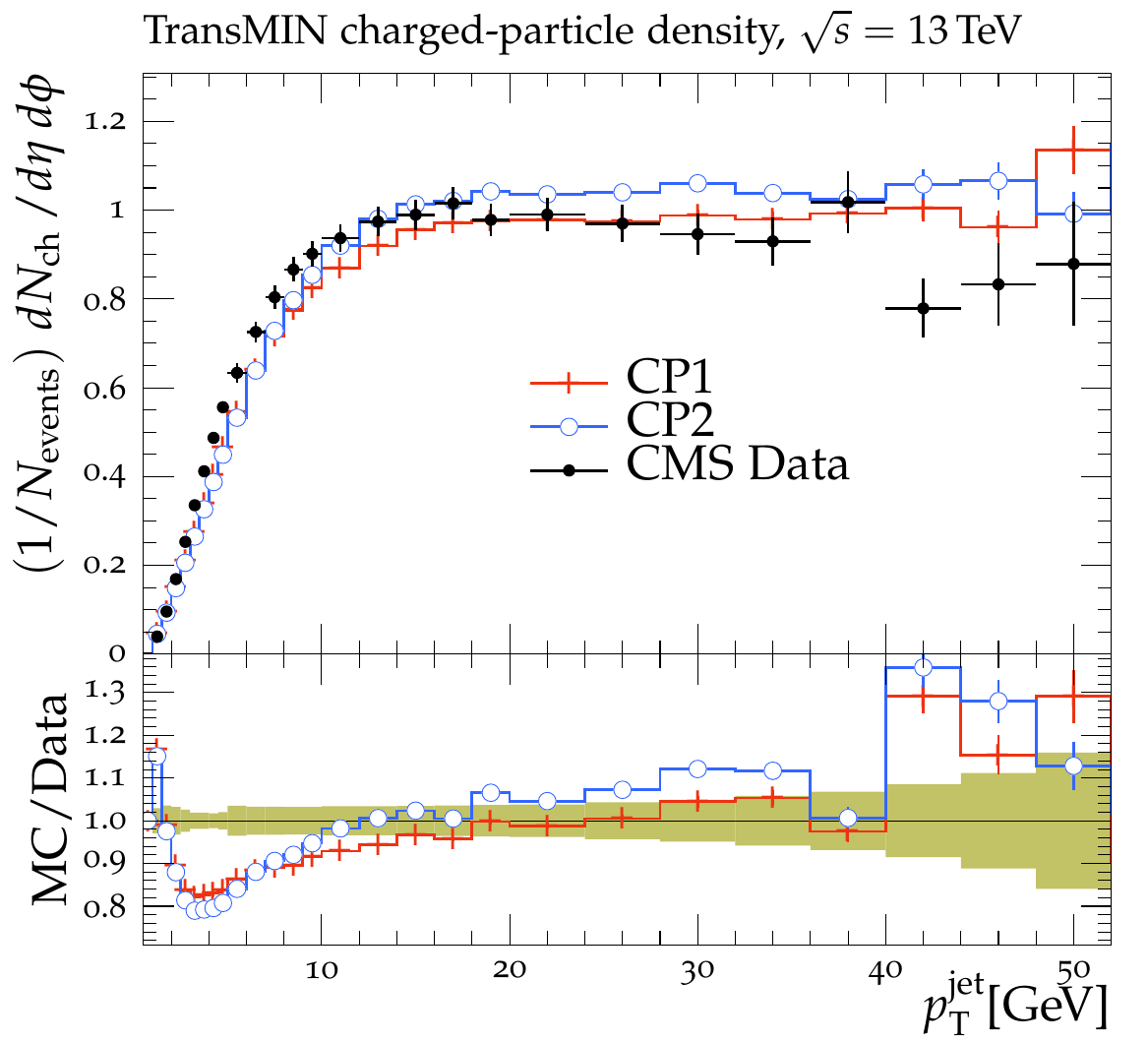}
\includegraphics[width=0.49\textwidth]{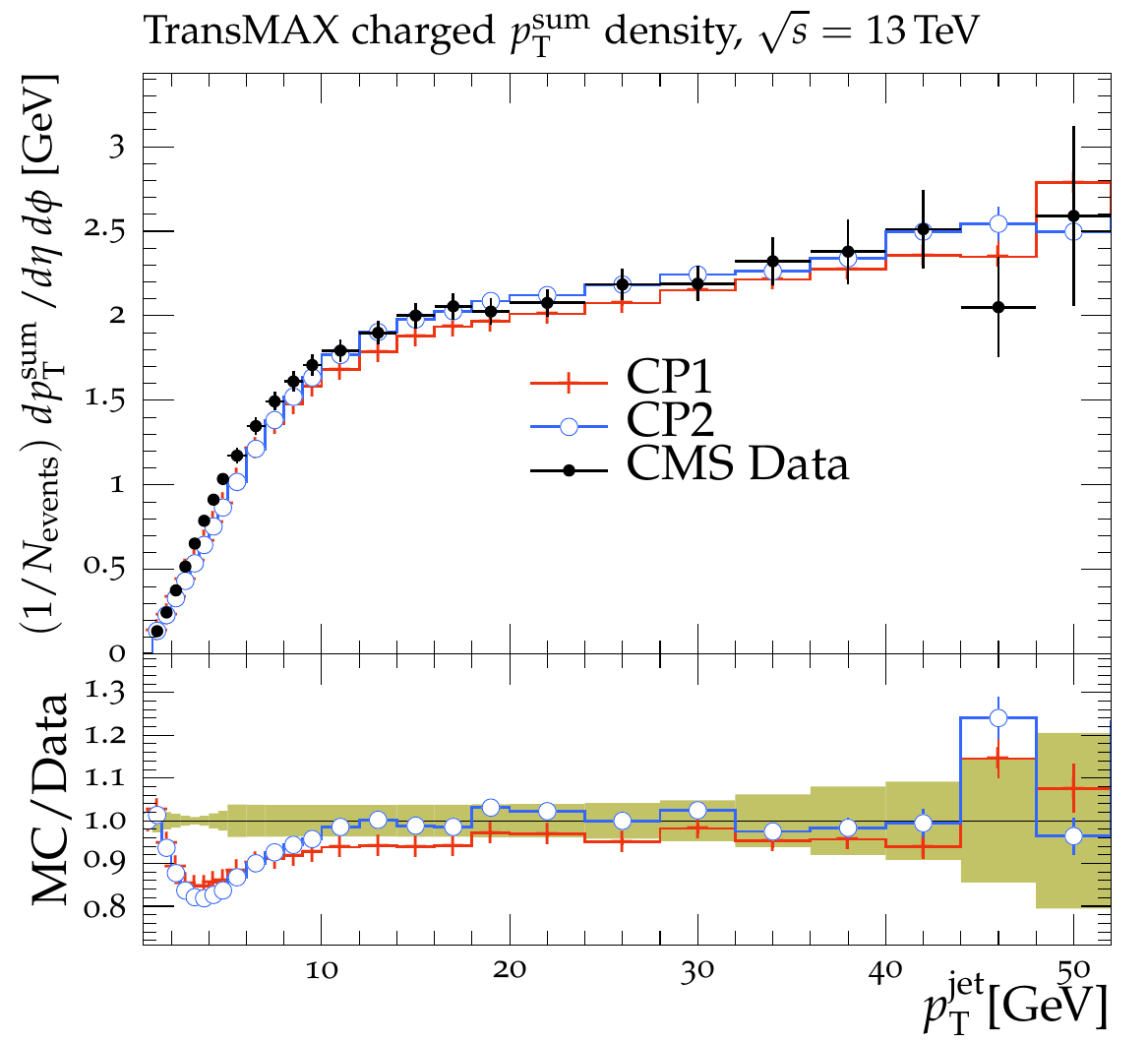}\\
\includegraphics[width=0.49\textwidth]{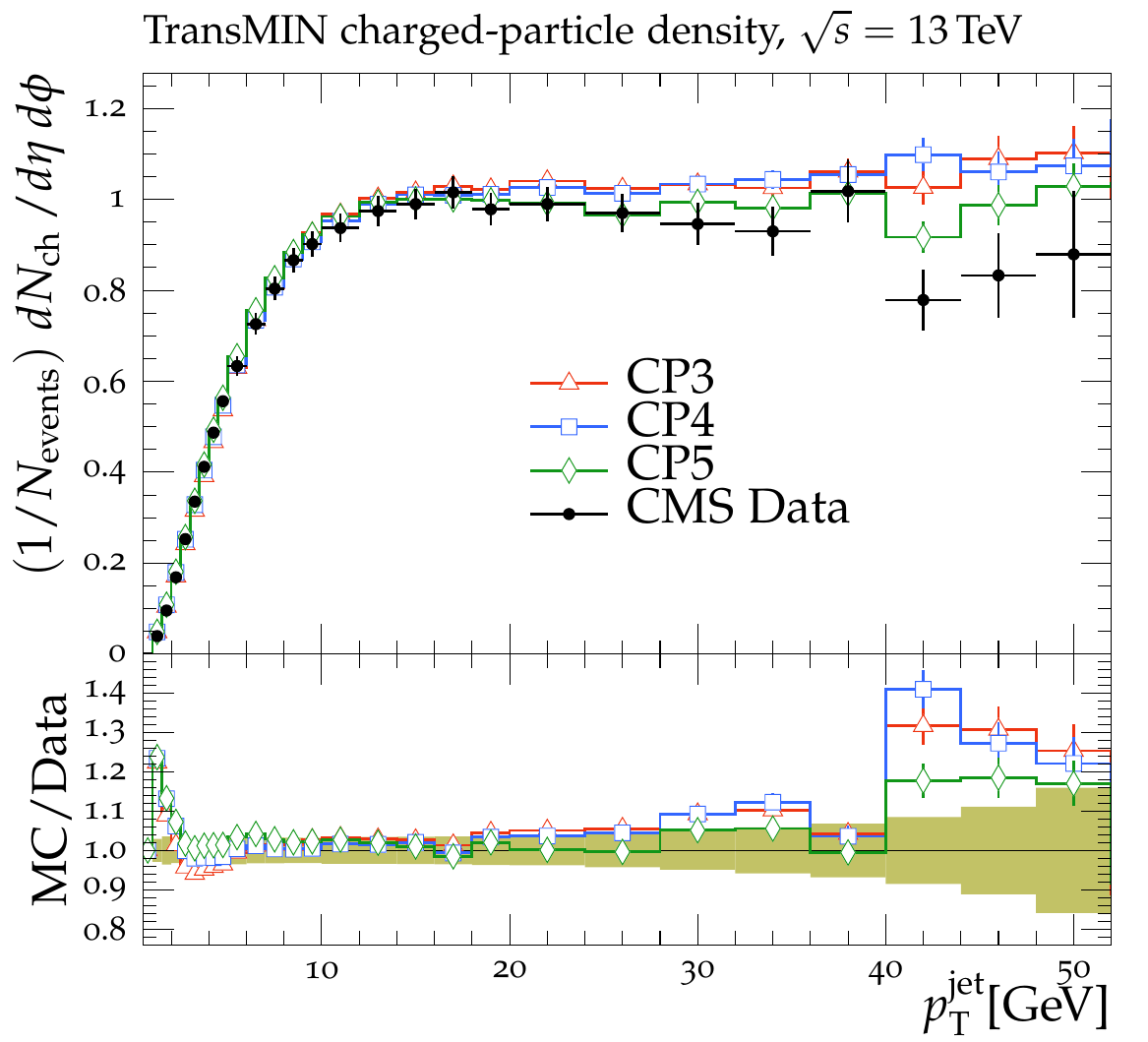}
\includegraphics[width=0.49\textwidth]{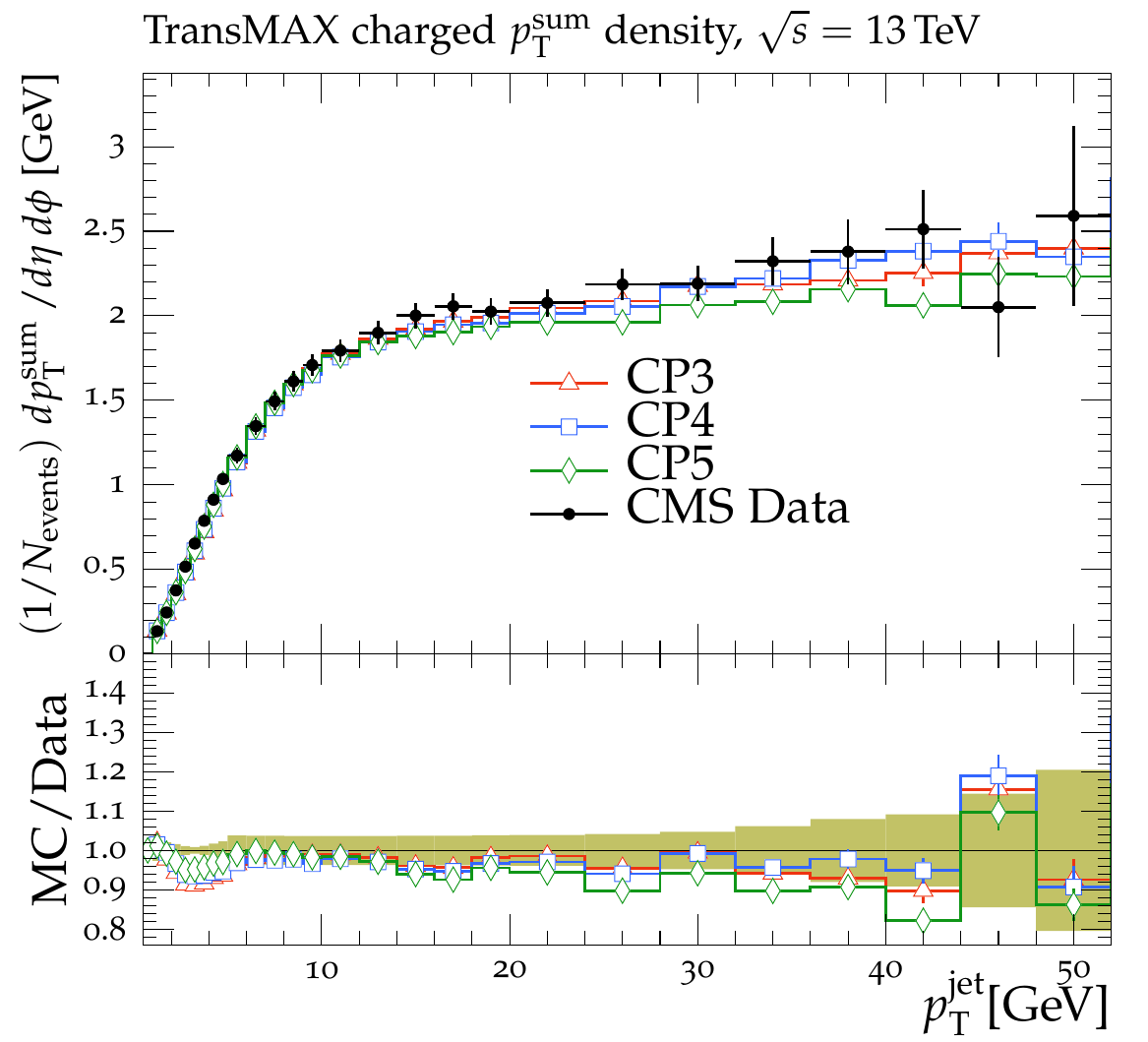}
\caption{The \tmin\ charged-particle multiplicity (left column) and \pt\ sum densities (right column) for particles with $\pt>0.5\GeV$ in $\abs{\eta}< 2.0$ as a function of the transverse momentum of the leading charged-particle jet, $\pt^\text{jet}$, from the CMS $\sqrt{s}=13\TeV$ analysis~\cite{CMS:2015zev}. The upper-row plots show the LO tunes, while the lower-row plots show the higher-order tunes. The ratio of the simulations to the data (MC/Data) is also shown, where the shaded band indicates the total experimental uncertainty in the data. Vertical lines drawn on the data points refer to the total uncertainty in the data. Vertical lines drawn on the MC points refer to the statistical uncertainty in the predictions. Horizontal bars indicate the associated bin width.}
\label{fig:perf_13TeV_III}
\end{figure*}

{\tolerance=800 Figure \ref{fig:perf_13TeV_III} shows the UE observables, \ie, charged-particle multiplicity and \ptsum\ densities~\cite{CMS:2015zev}, as a function of the \pt\ of the leading jet reconstructed using just charged particles. The observables shown in Fig.~\ref{fig:perf_13TeV_III} are from events selected without requiring any NSD- or SD-enhanced selections. The CMS UE tunes describe well UE-sensitive data measured using the leading charged-particle jet for \pt$^{\text{jet}}>10\GeV$. Tunes based on NLO or NNLO PDF sets, \ie, CP3, CP4, and CP5, describe the region at lower \pt$^{\text{jet}}$ better than CP1 and CP2, which are based on LO PDF sets. Predictions obtained with CP1 and CP2 underestimate the UE observables by about $\approx$15--20\%. Predictions obtained with CP3, CP4, and CP5 describe the UE in events characterized using the leading charged particle, as well as those characterized by the leading charged-particle jet, quite well. \par}

\begin{figure*}[ht!]
\centering
\includegraphics[width=0.49\textwidth]{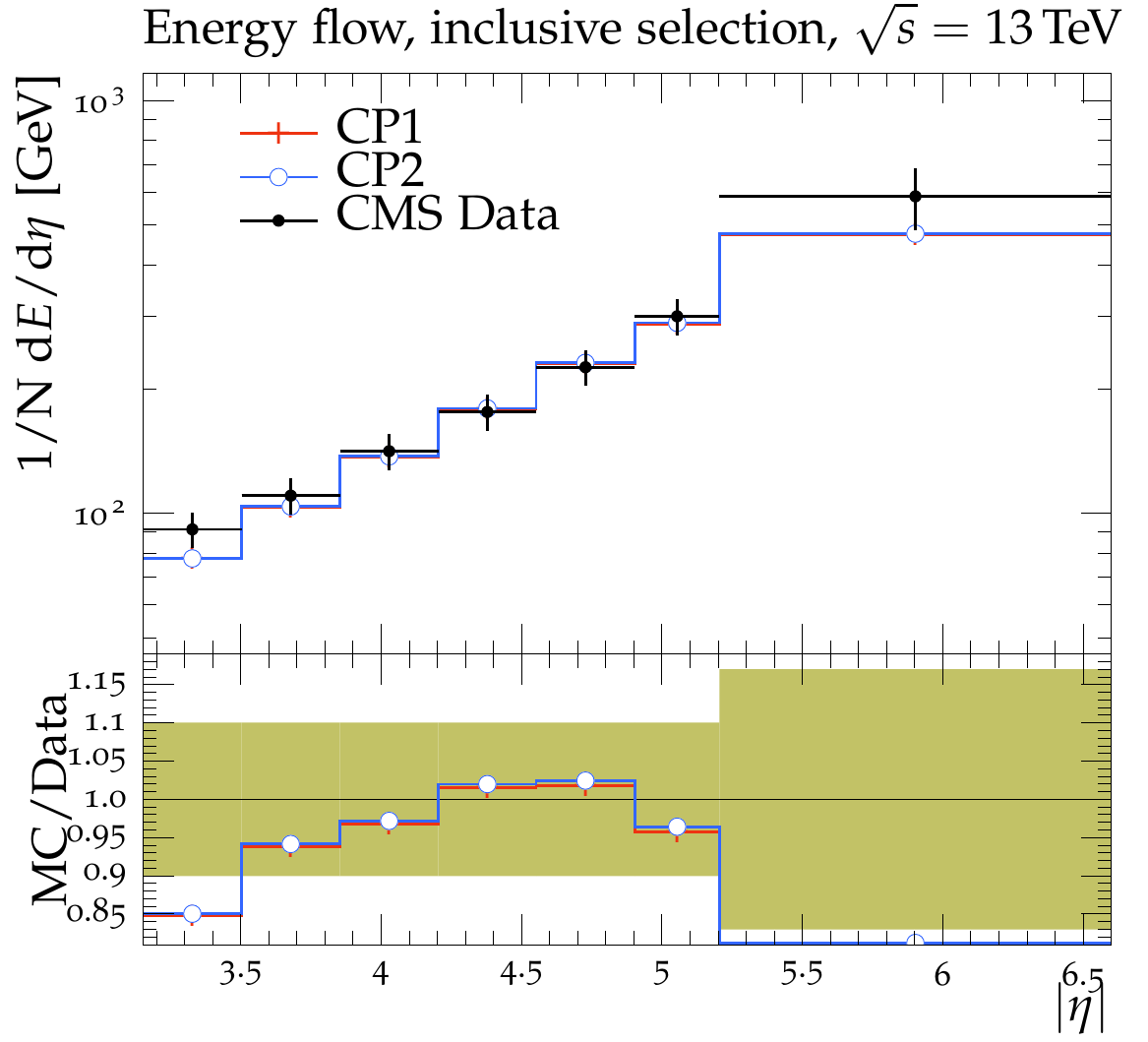}
\includegraphics[width=0.49\textwidth]{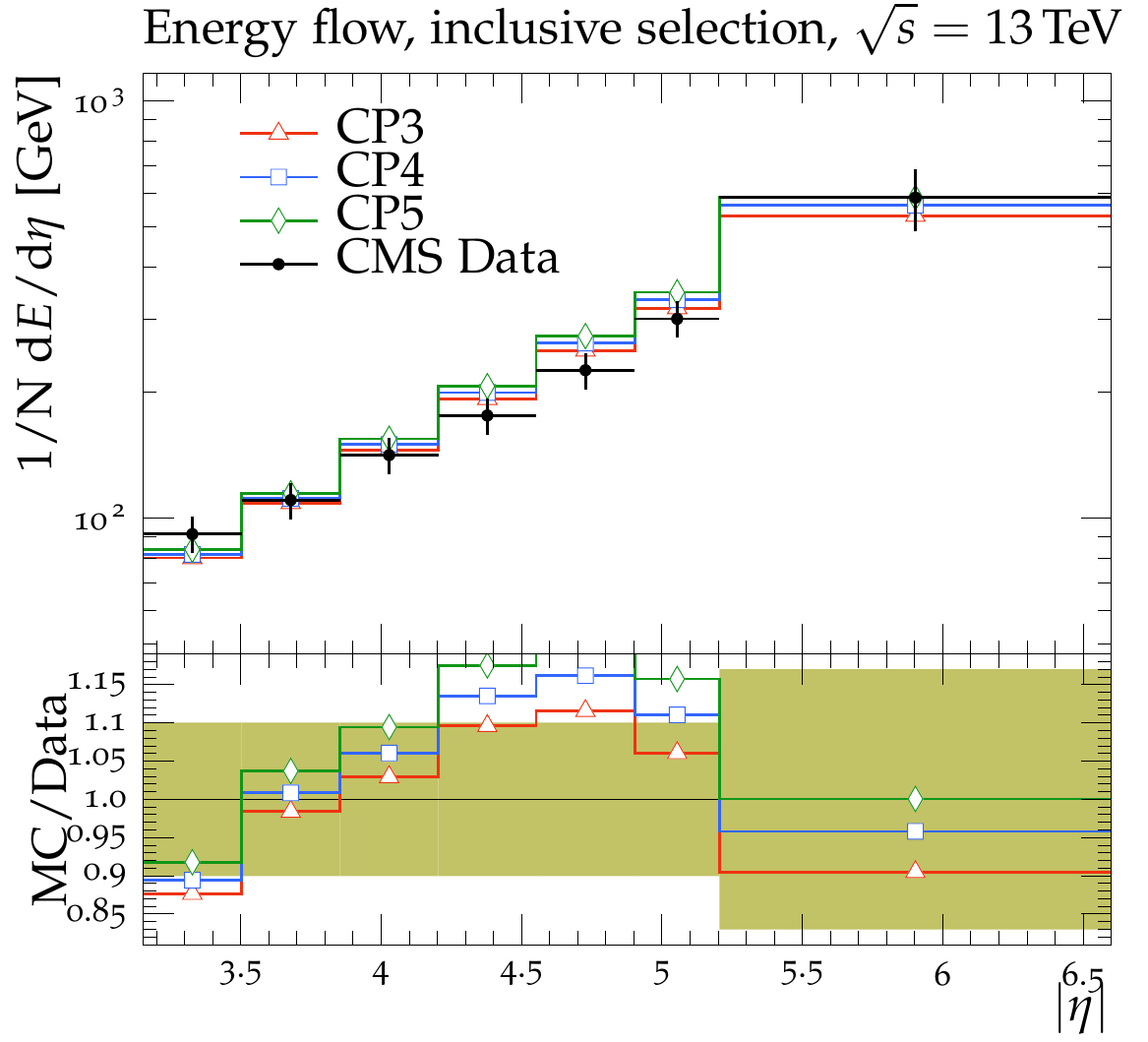}\\
\caption{The energy flow measured in an inclusive selection as a function of pseudorapidity, from the CMS $\sqrt{s}=13\TeV$ analysis~\cite{CMS:2016mxs}. The data are compared with (left) the CMS \PYTHIAviii LO-PDF tunes CP1 and CP2, and with (right) the CMS \PYTHIAviii NLO-PDF tune CP3 and the CMS \PYTHIAviii NNLO-PDF tunes CP4 and CP5. The ratio of the simulations to the data (MC/Data) is also shown, where the shaded band indicates the total experimental uncertainty in the data. Vertical lines drawn on the data points refer to the total uncertainty in the data. Vertical lines drawn on the MC points refer to the statistical uncertainty in the predictions. Horizontal bars indicate the associated bin width.}
\label{fig:fwdregion}
\end{figure*}

\begin{figure*}[ht!]
\centering
\includegraphics[width=0.49\textwidth]{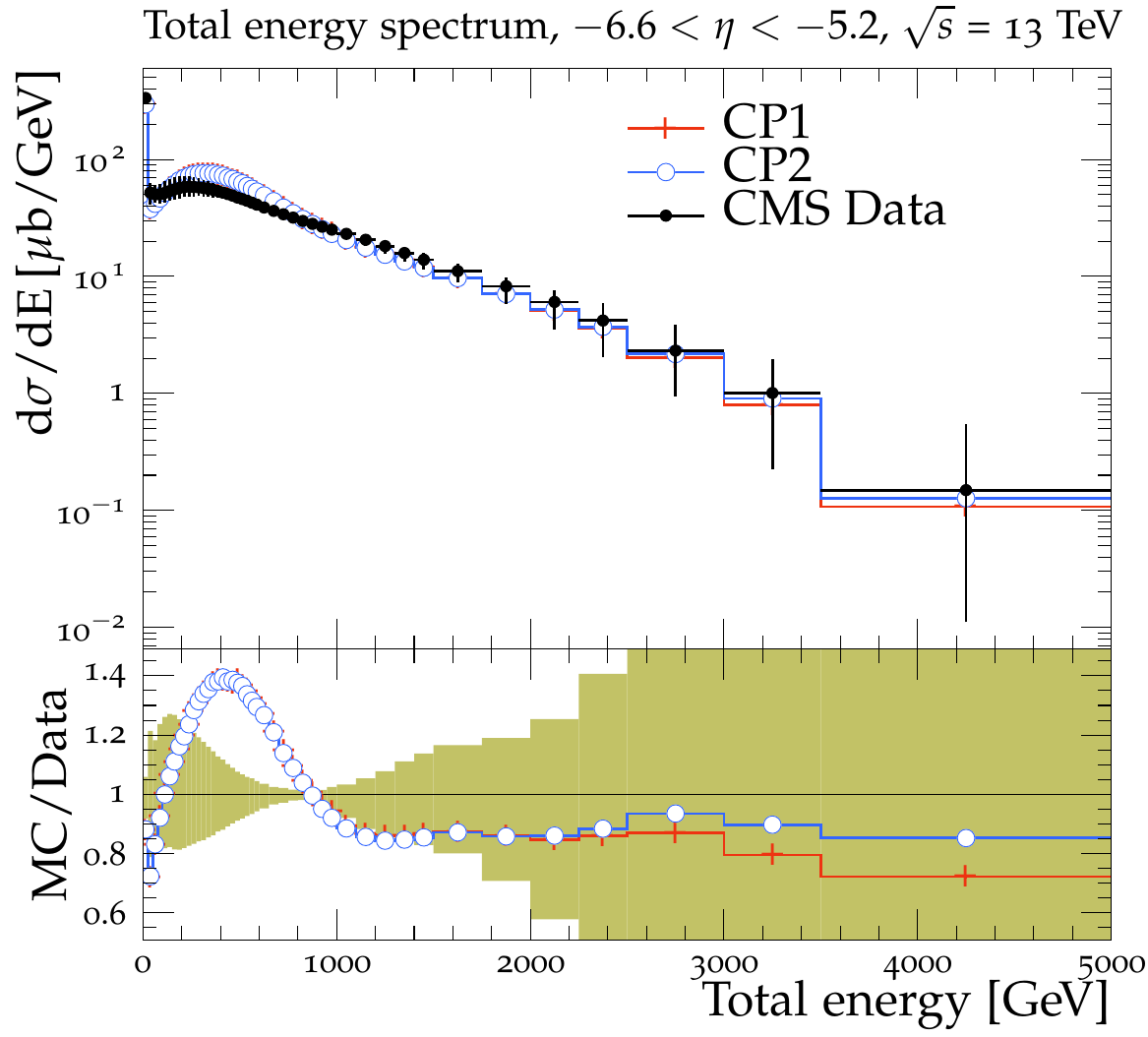}
\includegraphics[width=0.49\textwidth]{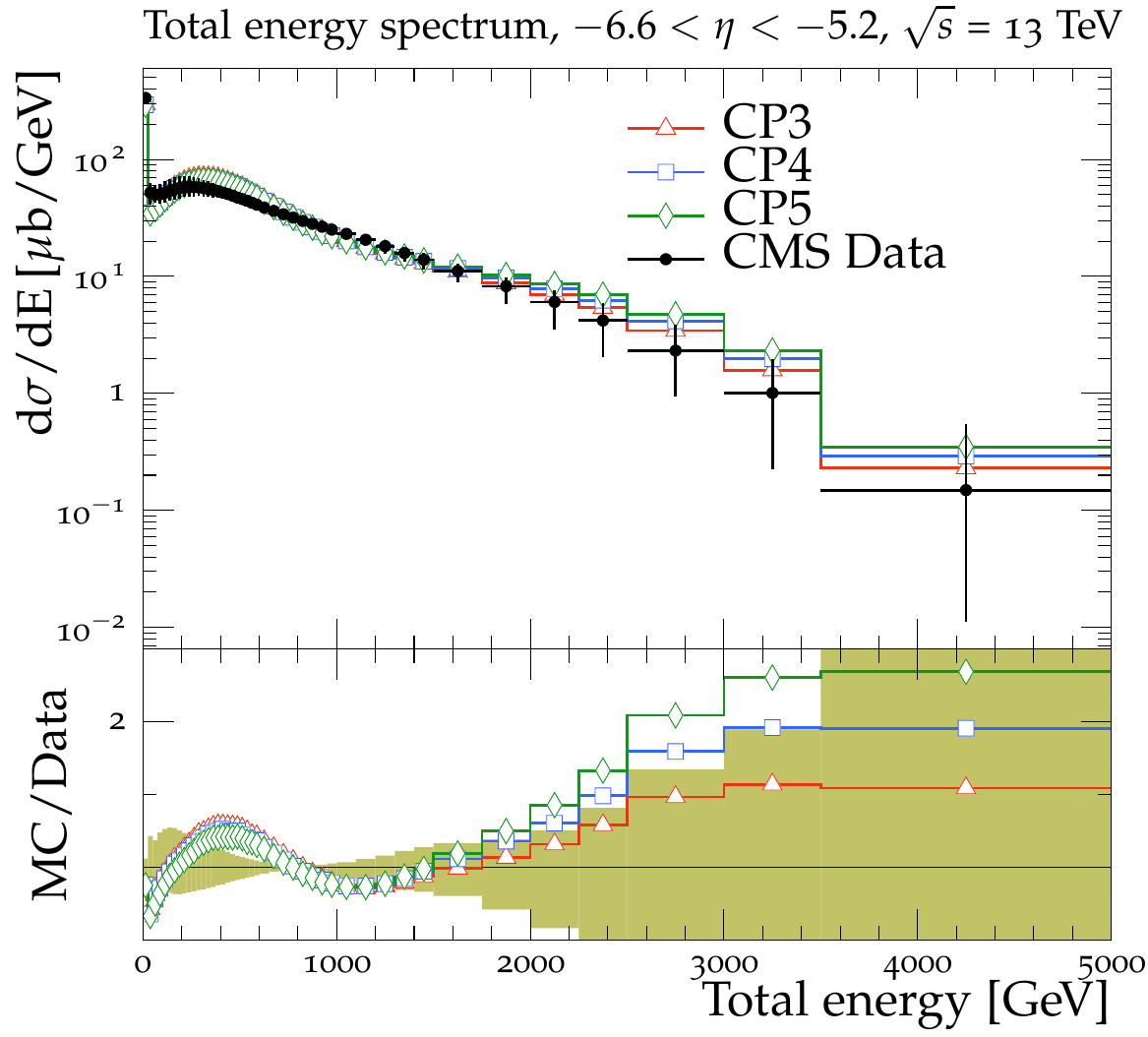}\\
\caption{The total energy spectrum measured in the pseudorapidity interval $-6.6<\eta<-5.2$, from the CMS $\sqrt{s}=13\TeV$ analysis~\cite{Sirunyan:2017nsj}. The data are compared with (left) the CMS \PYTHIAviii LO-PDF tunes CP1 and CP2, and with (right) the CMS \PYTHIAviii NLO-PDF tune CP3 and the CMS \PYTHIAviii NNLO-PDF tunes CP4 and CP5. The ratio of the simulations to the data (MC/Data) is also shown, where the shaded band indicates the total experimental uncertainty in the data. Vertical lines drawn on the data points refer to the total uncertainty in the data. Vertical lines drawn on the MC points refer to the statistical uncertainty in the predictions. Horizontal bars indicate the associated bin width.}
\label{fig:fwdregion2}
\end{figure*}

Predictions for observables measured in the forward region are compared with data and shown in Figs.~\ref{fig:fwdregion} and~\ref{fig:fwdregion2}. The energy flow, defined as the average energy per event~\cite{CMS:2016mxs}, as a function of $\eta$ with the Hadron Forward (HF) calorimeter~\cite{Chatrchyan:2008aa} covering $3.15 <\abs{\eta}< 5.20$ and the CASTOR calorimeter~\cite{Chatrchyan:2008aa} covering $-6.6<\eta<-5.2$, is well reproduced by all tunes. A different level of agreement is achieved for predictions from the new CMS tunes for the spectrum of the total energy E measured in the CASTOR calorimeter at $\sqrt{s}=13\TeV$~\cite{Sirunyan:2017nsj}, displayed in Fig.~\ref{fig:fwdregion2}. In particular, the tunes based on LO PDF sets reproduce the energy spectrum well at large values ($\text{E}>2000\GeV$), but have differences of up to 40\% at low values ($\text{E}<800\GeV$). The tunes using higher-order PDF sets are closer to the data at low energy values, with differences up to 20\%, but tend to overestimate the energy at large values. This dissimilar behaviour is driven by the different  \texttt{pT0Ref} values of the tunes. The fiducial inelastic cross sections~\cite{CMS:2016ael}, when two different selections are applied in the forward region, are not well reproduced by any of the new tunes or by CUETP8M1, with differences up to 10\%. This might be because of the Sch\"uler--Sj\"ostrand~\cite{Schuler:1993wr}  diffraction model used in the simulation, which might have a suboptimal description of the low-mass  diffractive components. A better description might be provided by tunes using the Donnachie--Landshoff~\cite{Donnachie:1984xq} or minimum-bias Rockefeller~\cite{Ciesielski:2012mc} diffractive models.

\subsection{Comparisons using observables in multijet final states}
\label{SoftHardQCD}

In this subsection, we present comparisons of observables measured in multijet final states. For these studies, the NLO dijet MEs implemented in the \POWHEG event generator merged with the \PYTHIAviii simulation of the PS and UE are used. The merging between the \POWHEG ME calculations and the \PYTHIAviii UE simulation is performed using the shower-veto procedure, which rejects showers if their transverse momentum is greater than the minimal \pt of all final-state partons simulated in the ME (parameter \pt$^\text{hard}= 2\GeV$~\cite{Alioli:2010xa}). Variables in multijet events, such as jet transverse momenta or azimuthal dijet correlations, are expected to be less affected by MPI contributions, since jets at high \pt ($>$100\GeV) mainly originate from the hard scattering or additional hard emissions, which are simulated in the \POWHEG calculation by the ME formalism.
However, the MPI contribution still has some impact because it adds an average energy offset to the event, which is then included in the jet reconstruction~\cite{1748-0221-6-11-P11002,Khachatryan:2016kdb}. The predictions reproduce well inclusive jet cross sections as a function of jet \pt at both central and forward jet rapidities, irrespective of the cone size (0.4 or 0.7) used for the jet clustering algorithm~\cite{Khachatryan:2016wdh}.

\begin{figure*}[ht!]
\centering
\includegraphics[width=0.48\textwidth]{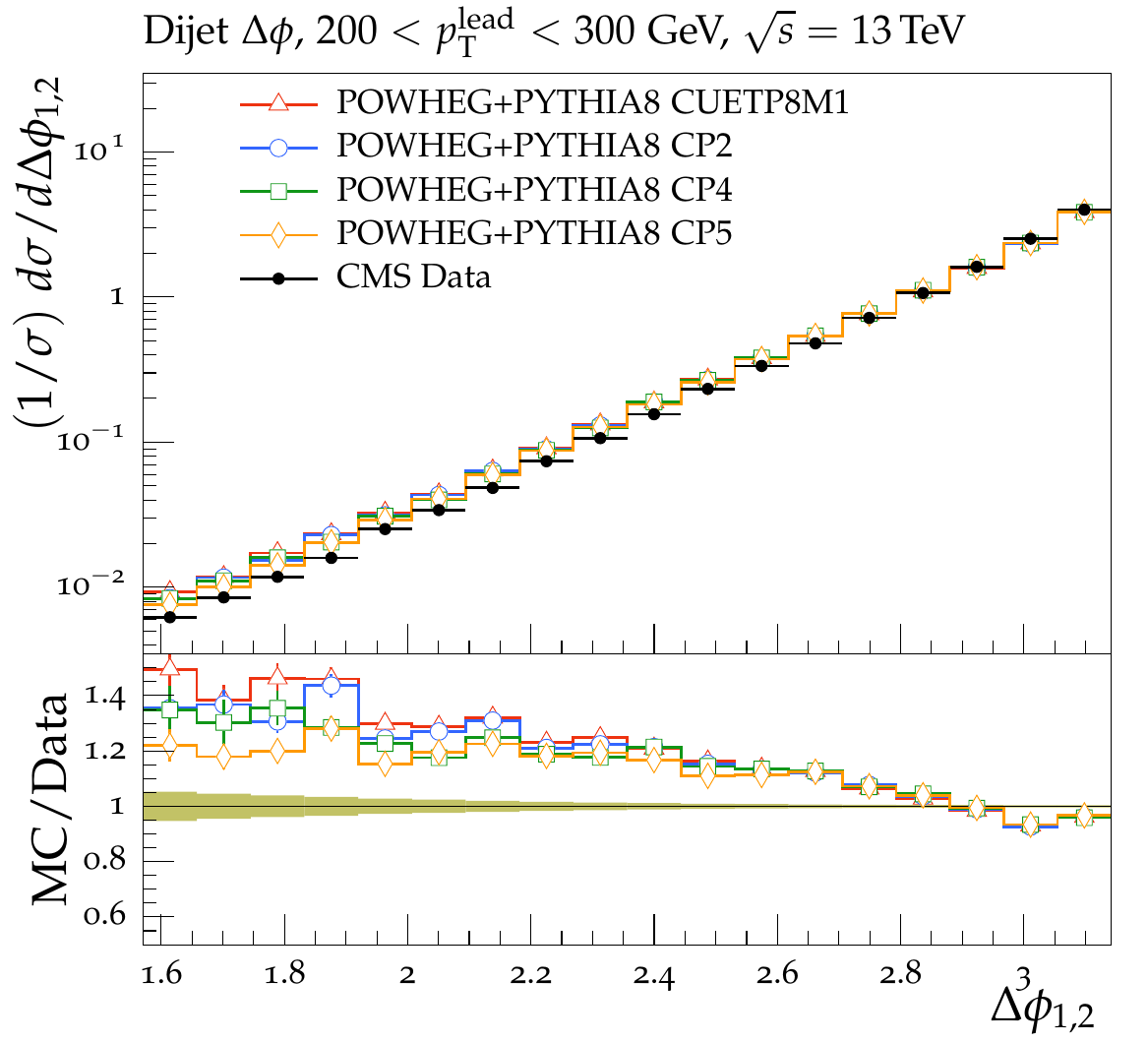}
\includegraphics[width=0.48\textwidth]{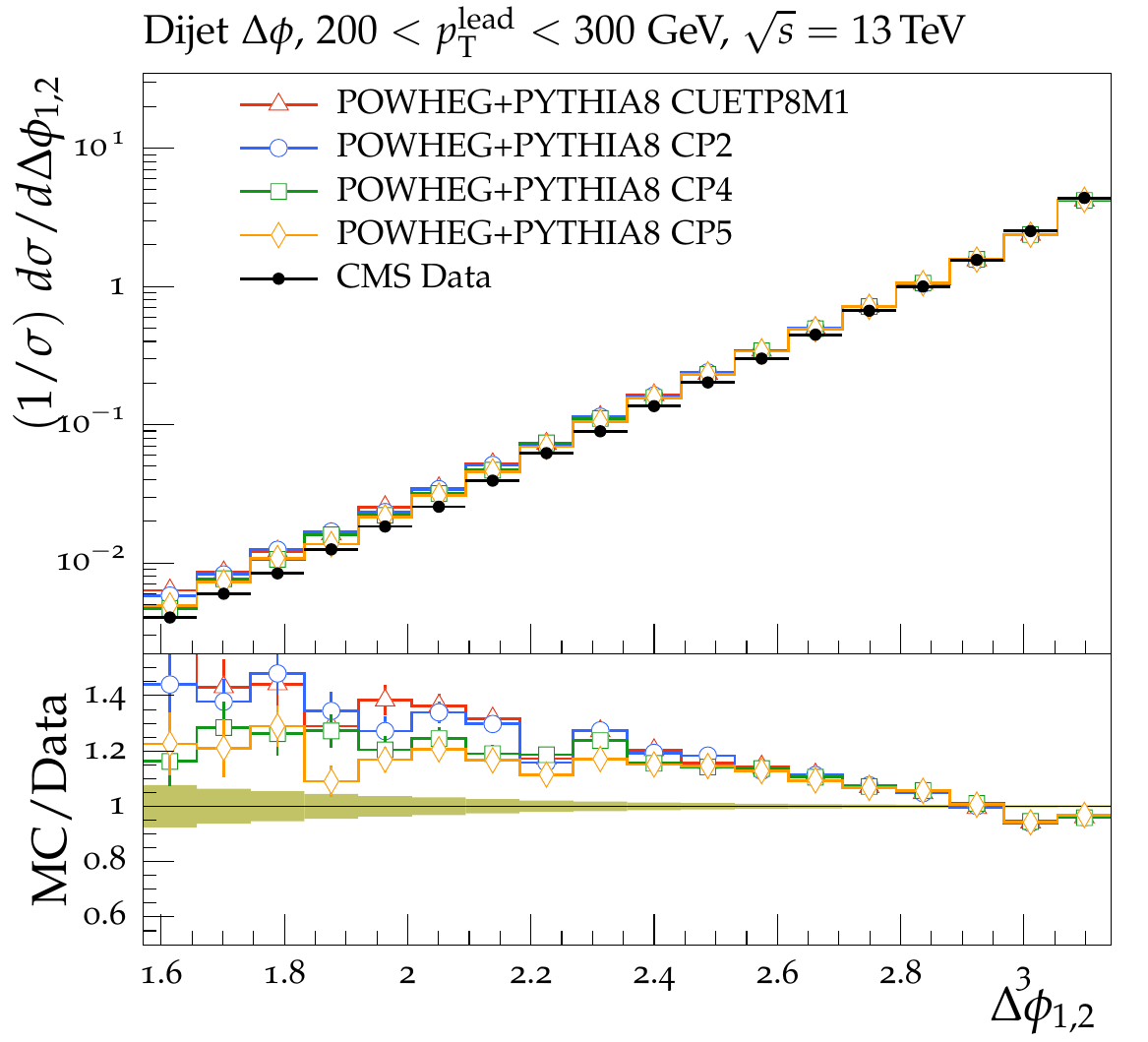}\\
\caption{The azimuthal difference $\delphi_{1,2}$ between the leading two jets with $\abs{\eta}<2.4$ in dijet events with leading-jet transverse momentum in the range (left) $200<\pt^{\text{lead}}< 300\GeV$ and (right) $300<\pt^{\text{lead}}<400\GeV$, from the
CMS $\sqrt{s}=13\TeV$ analysis~\cite{Sirunyan:2017jnl}. The jets are reconstructed using the anti-\kt jet finding algorithm~\cite{Cacciari:2008gp,Cacciari:2011ma} with a distance parameter of 0.4. The data are compared with predictions of the NLO dijet ME calculation from \POWHEG, interfaced to the \PYTHIAviii tunes CUETP8M1, CP2, CP4, and CP5. Tunes CP1 and CP3 are not shown in the plot but present a similar behavior as tunes CP2 and CP4. The ratios of simulations to the data (MC/Data) are also shown, where the shaded band indicates the total experimental uncertainty in the data. Vertical lines drawn on the data points refer to the total uncertainty in the data. Vertical lines drawn on the MC points refer to the statistical uncertainty in the predictions. Horizontal bars indicate the associated bin width.}
\label{fig:deltaphijj}
\end{figure*}

Figure \ref{fig:deltaphijj} shows the normalized cross section~\cite{Sirunyan:2017jnl} as a function of the azimuthal difference $\delphi_{1,2}$ between the two leading jets for two different selections on the leading jet \pt: $200<\pt<300\GeV$ and $300<\pt<400\GeV$. The results indicate that UE tunes based on an NLO evaluation of $\alpS(m_\cPZ)$ describe the data better than UE tunes based on LO evolution.
In particular, the better agreement is driven by the lower value of $\alpS^\mathrm{ISR}(m_\cPZ)$. In fact, predictions obtained with \POWHEG merged with \PYTHIAviii with the CUETP8M1 or CP2 tune exhibit a strong jet decorrelation, due to a large contribution from emissions simulated from the PS, and they overestimate the cross sections at small and medium $\delphi_{1,2}$ values ($\delphi_{1,2}< 2.4$). The PS component is reduced by the lower value of $\alpS(m_\cPZ)$, which increases the degree of correlation between the selected jets, resulting in a better description of the data by predictions of the CP4 and CP5 tunes. A similar outcome was also observed for an analogous measurement performed at the D0 experiment at $\sqrt{s}=1.96\TeV$~\cite{PhysRevLett.94.221801}. In general, predictions obtained from \POWHEG~+ \PYTHIAviii tend to differ from the data at low and intermediate $\delphi_{1,2}$ values ($\delphi_{1,2}< 2.7$) by 10--40\%.

\subsection{Comparisons using observables sensitive to double-parton scattering}
\label{DPS}
 In this subsection, we present comparisons of predictions of the new tunes to DPS-sensitive observables measured by the CMS experiment at $\sqrt{s}=7\TeV$ in final states with four jets (4j)~\cite{Chatrchyan:2013qza}, and with two jets originating from bottom quarks ({\cPqb} jets)~\cite{Aaboud:2016mmw} and two other jets (2b2j)~\cite{Khachatryan:2016rjt}.

The topology in the transverse plane of the physics objects measured in the final state is sensitive to contributions from DPS. In particular, the 4j analysis performed by the CMS experiment requires two jets at high \pt (hard jets) and two jets at low \pt (soft jets); the 2b2j measurement selects two jets originating from {\cPqb}  quarks and two other jets (light-flavor jets). Both of them measured the $\Delta$S observable, defined as:
\begin{equation}\label{deltas}
\Delta S=\arccos\left(\frac{\vec{p}_{\text{T,1}}\cdot \vec{p}_{\text{T,2}}}{\abs{\vec{p}_{\text{T,1}}} \abs{\vec{p}_{\text{T,2}}}}\right),\\
\end{equation}
where $\vec{p}_\mathrm{T,1}$ refers to the momentum of the hard-jet or bottom jet pair system and $\vec{p}_\mathrm{T,2}$ to that of the soft-jet or light-flavor jet pair system. This variable relates the production planes of the hard (bottom) jet and soft- (light-flavor) jet pairs. Details of the event selection and of the specific analyses can be found in Refs.~\cite{Chatrchyan:2013qza} and~\cite{Khachatryan:2016rjt}.

Assuming that the two hard scatterings occurring within the same collision are completely independent of each other, the DPS cross section for a given process can be expressed through the inclusive partonic cross sections of the two single scatterings and an effective cross section, \eff. In a geometrical approach, this cross section is related to the transverse size of the proton and to the total inelastic proton-proton ($\Pp\Pp$) cross section \cite{HUMPERT1985211, Mangano1989}. When no correlations among the partons inside the proton are present, \eff\ is similar to the inelastic $\Pp\Pp$ cross section. In this simple factorized approach, one expects \eff\ to be independent of the partonic final states of the two hard processes occurring within the same collision.
In \PYTHIAviii, the value of \eff~is calculated by dividing the ND cross section by the so-called ``enhancement factor", which depends on the parameters of the overlap matter distribution function and on \texttt{pT0Ref}~\cite{Sjostrand:1987su}. For central $\Pp\Pp$ collisions, the enhancement factor tends to be large, translating to a lower value of \eff\ and a larger DPS contribution. For peripheral interactions, enhancement factors are small, giving large values of \eff\ and a small DPS contribution.\\

Table~\ref{DPSreference} shows the values of \eff\ published by the CMS Collaboration for the 4j and the 2b2j measurements. A previous study~\cite{Khachatryan:2015pea} concluded that observables sensitive to semi-hard MPI and those sensitive to DPS cannot be described by a single set of parameters. Table~\ref{DPStable} displays the \eff\ values obtained from the new CMS UE tunes. The central values of \eff\ are consistent among the new tunes and are slightly larger than the values of the DPS-based tunes~\cite{Khachatryan:2015pea}.

\begin{figure*}[ht!]
\centering
\includegraphics[width=0.49\textwidth]{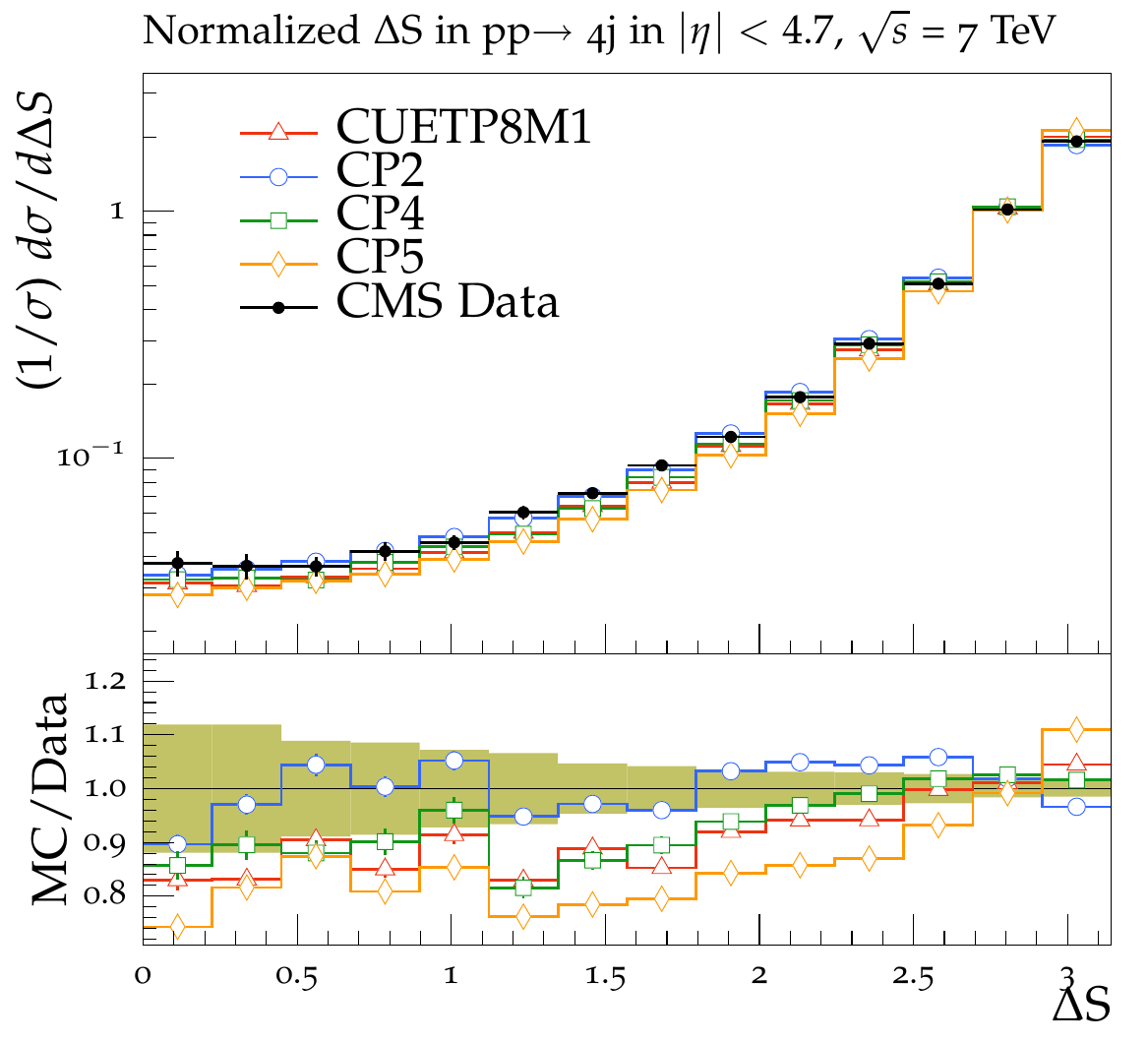}
\includegraphics[width=0.49\textwidth]{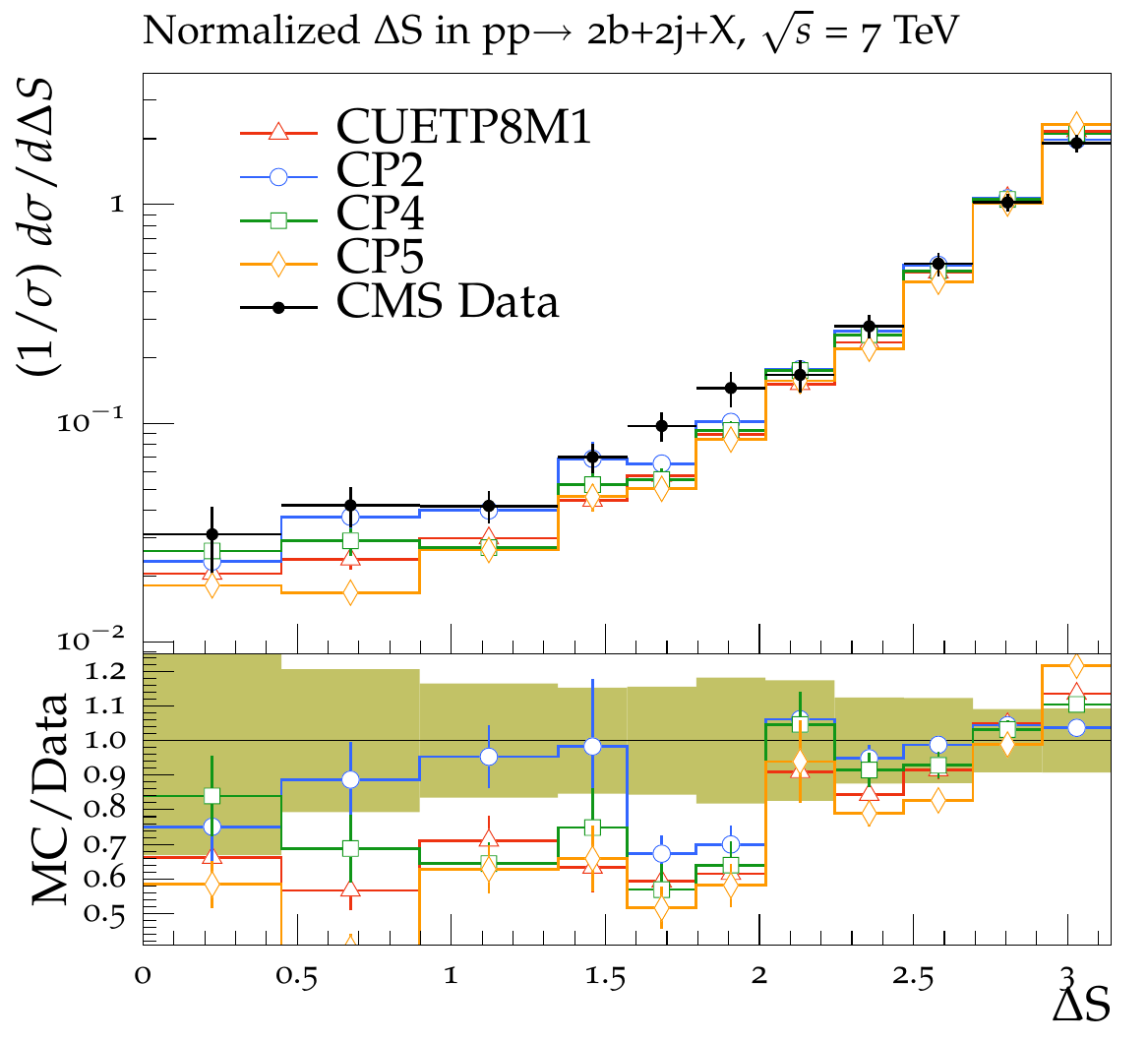}
\caption{The correlation observable $\Delta$S measured in 4j (left) and 2b2j (right) production, compared to predictions of \PYTHIAviii tunes CUETP8M1, CP2, CP4, and CP5, from the
CMS $\sqrt{s}=7\TeV$ analyses~\cite{Chatrchyan:2013qza,Khachatryan:2016rjt}. Tunes CP1 and CP3 are not shown in the plot but show\
 a similar behaviour as, respectively, tunes CP2 and CP4. The ratios of simulations to the data (MC/Data) are also shown, where the shaded band indicates the total experimental uncertainty in the data. Vertical lines drawn on the data points refer to the total uncertainty in the data. Vertical lines drawn on the MC points refer to the statistical uncertainty in the predictions. Horizontal bars indicate the associated bin width.}
\label{fig:dps_results_4j_2b2j}
\end{figure*}

Figure~\ref{fig:dps_results_4j_2b2j}  shows the comparisons of predictions obtained from \PYTHIAviii with tunes CUETP8M1, CP2, CP4, and CP5 to the DPS observables measured in the 4j and 2b2j final states. Predictions from the CP2 tune based on a LO PDF set describe the central values better than the CP4 and CP5 tunes based on an NNLO PDF set or the old tune CUETP8M1. This is due to the different \texttt{pT0Ref} value used by CP2, CP4, and CP5, which determines the amount of simulated MPI. The value of the \texttt{pT0Ref} parameter is driven by the distribution of the gluon distribution function at low $x$, which is very different in LO and NNLO PDF sets. Additionally, predictions obtained with CP4 describe the DPS-sensitive observables better than CP5. This is due to the different rapidity ordering used for the PS emissions in the two tunes. By removing the rapidity ordering for the PS emissions (CP4), the simulation produces more radiation and decreases the correlation between the selected jet pairs compared to CP5. This reduced jet correlation tends to mimic a DPS event by producing low values of $\Delta$S. We have checked that the observables sensitive to color coherence, which were measured by the CMS experiment at $\sqrt{s}=7\TeV$~\cite{Chatrchyan:2013fha}, are well described by predictions from both CP4 and CP5 tunes, despite the difference in the rapidity ordering of the PS simulation between the two tunes. 

\begin{table}
\centering
\topcaption{Values of \eff\ at $\sqrt{s}=7\TeV$ published by the CMS Collaboration for the four-jet final states, obtained by fitting predictions of the \PYTHIAviii MC event generator to DPS-sensitive measured observables.}
\label{DPSreference}
\begin{tabular}{ccc}
Final state  & Generator & \eff\ [mb] ($\sqrt{s}=7\TeV$)  \\
\hline
4j & \PYTHIAviii & 19.0$^{+4.7}_{-3.0}$ \cite{Khachatryan:2015pea} \\
2b2j & \PYTHIAviii & 23.2$^{+3.3}_{-2.5}$ \cite{Gunnellini:2014kwa} \\
\end{tabular}
\end{table}

\begin{table}
\centering
\topcaption{Values of \eff\ at $\sqrt{s}=7$ and 13\TeV obtained with the new CMS UE tunes.}
\label{DPStable}
\begin{tabular}{c c c}
          & $\sqrt{s}=7\TeV$ & $\sqrt{s}=13\TeV$ \\
          & \eff\ [mb]  & \eff\ [mb] \\
\hline
CP1       & 26.3$^{+1.0}_{-1.7}$ & 27.8$^{+1.1}_{-1.4}$ \\
CP2       & 24.7$^{+1.0}_{-1.6}$ & 26.0$^{+1.0}_{-1.3}$ \\
CP3       & 24.1$^{+1.0}_{-1.5}$ & 25.2$^{+1.0}_{-1.3}$ \\
CP4       & 23.9$^{+1.0}_{-1.5}$ & 25.3$^{+1.1}_{-1.4}$ \\
CP5       & 24.0$^{+1.0}_{-1.6}$ & 25.3$^{+1.0}_{-1.3}$ \\
\end{tabular}
\end{table}

\subsection{Comparisons using observables in top quark production}
\label{TopSector}
In the following, we investigate how the new \PYTHIAviii tunes describe the CMS \ttbar data when different ME generators, namely \POWHEG and \MG, are employed.
Both ME configurations use the NNPDF3.1 NNLO PDF with $\alpS(m_\cPZ)=0.118$ and assume a top quark mass
($m_\PQt$) value of 172.5\GeV.

{\tolerance=800 In the \POWHEG configuration, the ME heavy quark production mode~\cite{powhegv21,powhegv22,powhegv23} is used.
In this configuration, \POWHEG simulates inclusive \ttbar production at NLO, where the first additional jet is computed at LO, while \MG~performs the calculation with up to 2 additional jets at NLO, with a third jet simulated at LO.
The \POWHEG generator scales the real emission cross section by a damping function that controls the ME-PS merging and that regulates the high-\pt\ radiation.
The damping variable used in the \POWHEG simulation is set to 1.379 times $m_\PQt$, a value derived from data at $\sqrt{s}=8\TeV$ in the dilepton channel using a similar ME calculation and assuming the CP5 tune.
The factorization and renormalization scales are assumed equal to the transverse mass of the top quark,
$\mT^\PQt=\sqrt{\smash[b]{m_\PQt^2+\pt^2}}$.
The minimum \pt for the emission of light quarks in \POWHEG is  0.8\GeV.
The \texttt{pThard} parameter is set to 0 and
the \POWHEG hardness criterion, defined by the \texttt{pTdef} option, is set to 1.
The merging scale in \MG~ is set to 40\GeV, and the threshold applied to regulate multijet MEs in the \MG~ FxFx merging procedure, is 20\GeV. \par}

\begin{figure*}[ht!]
\centering
\includegraphics[width=0.4\textwidth]{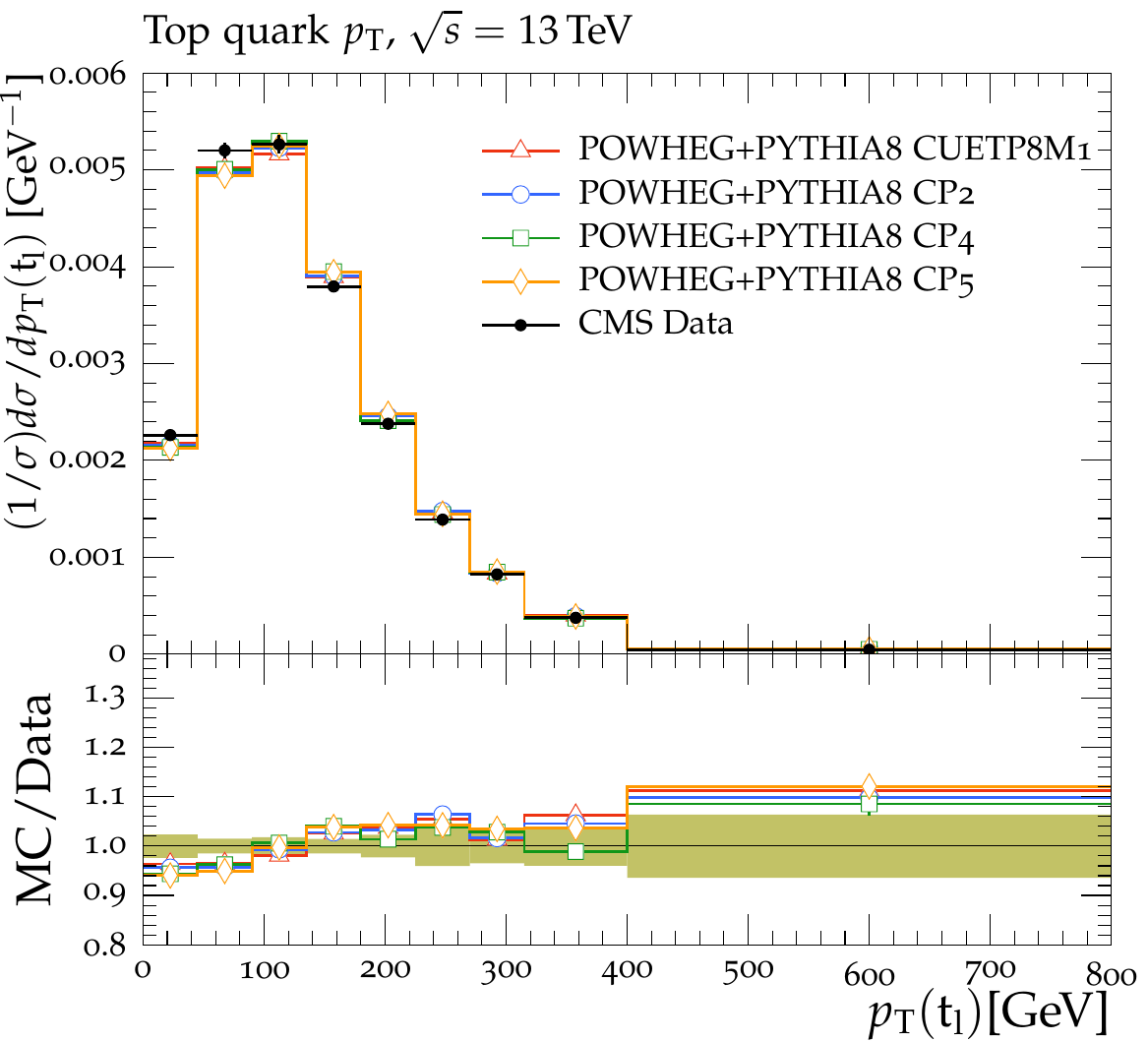}
\includegraphics[width=0.4\textwidth]{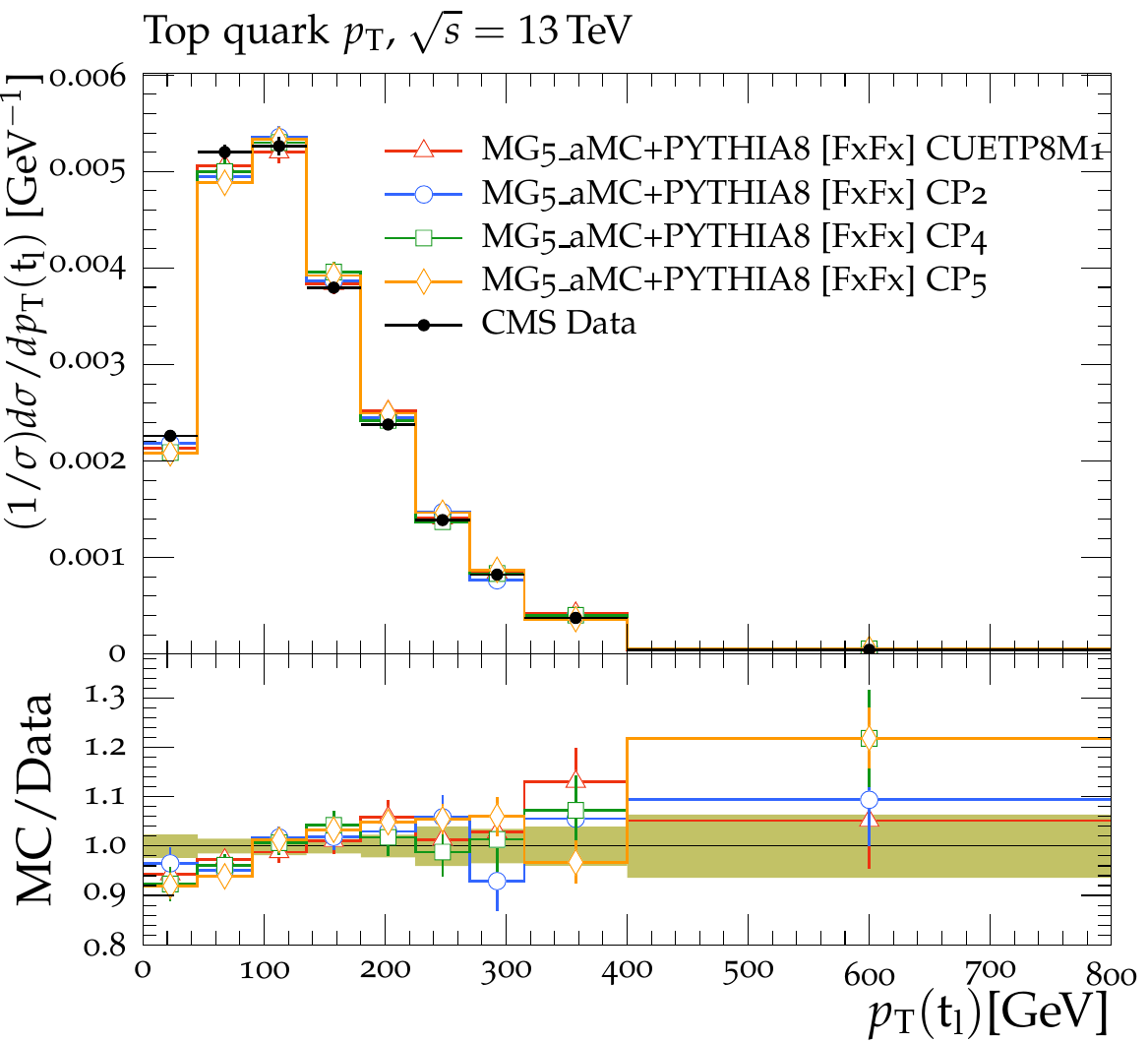} \\
\includegraphics[width=0.4\textwidth]{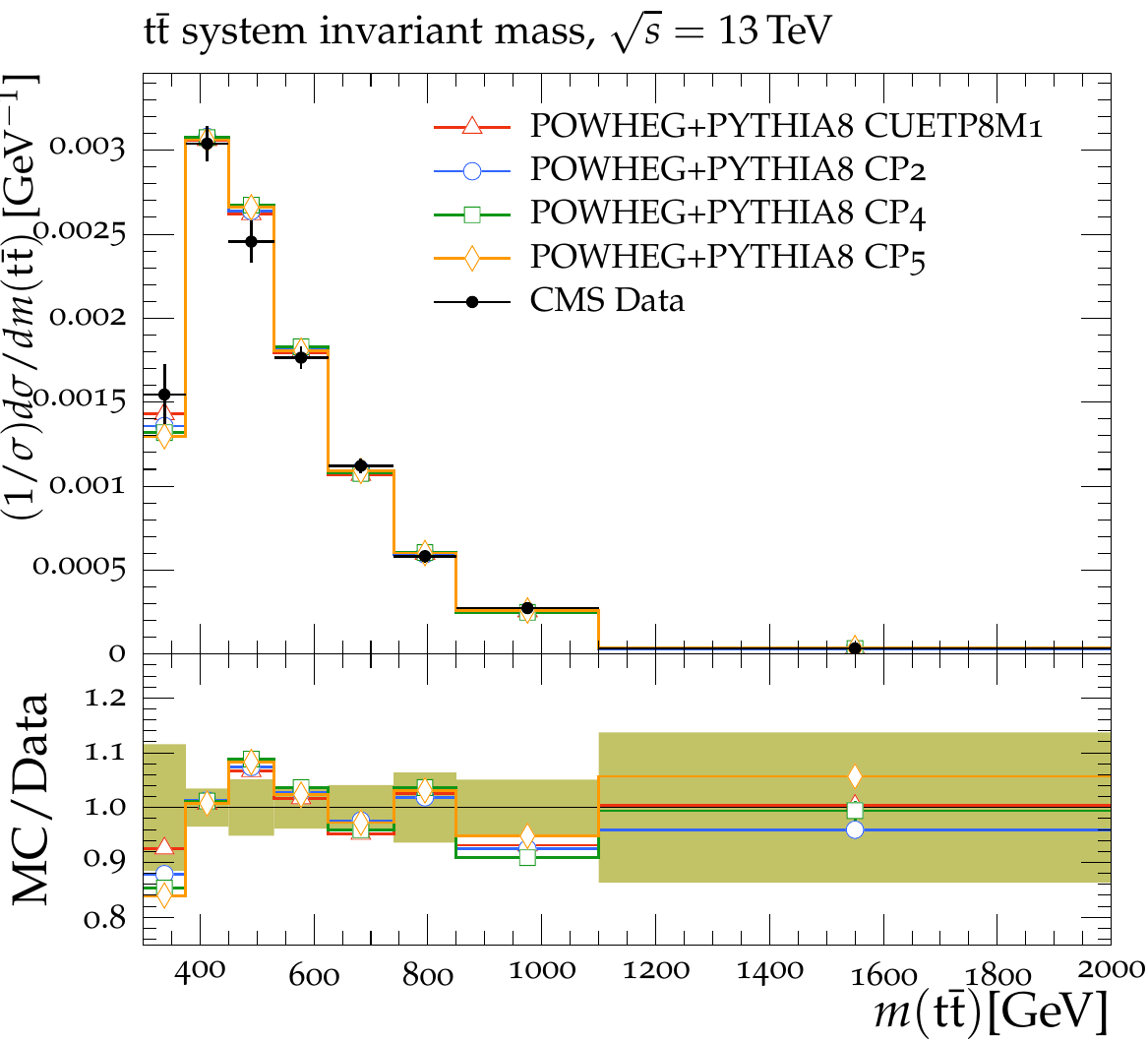}
\includegraphics[width=0.4\textwidth]{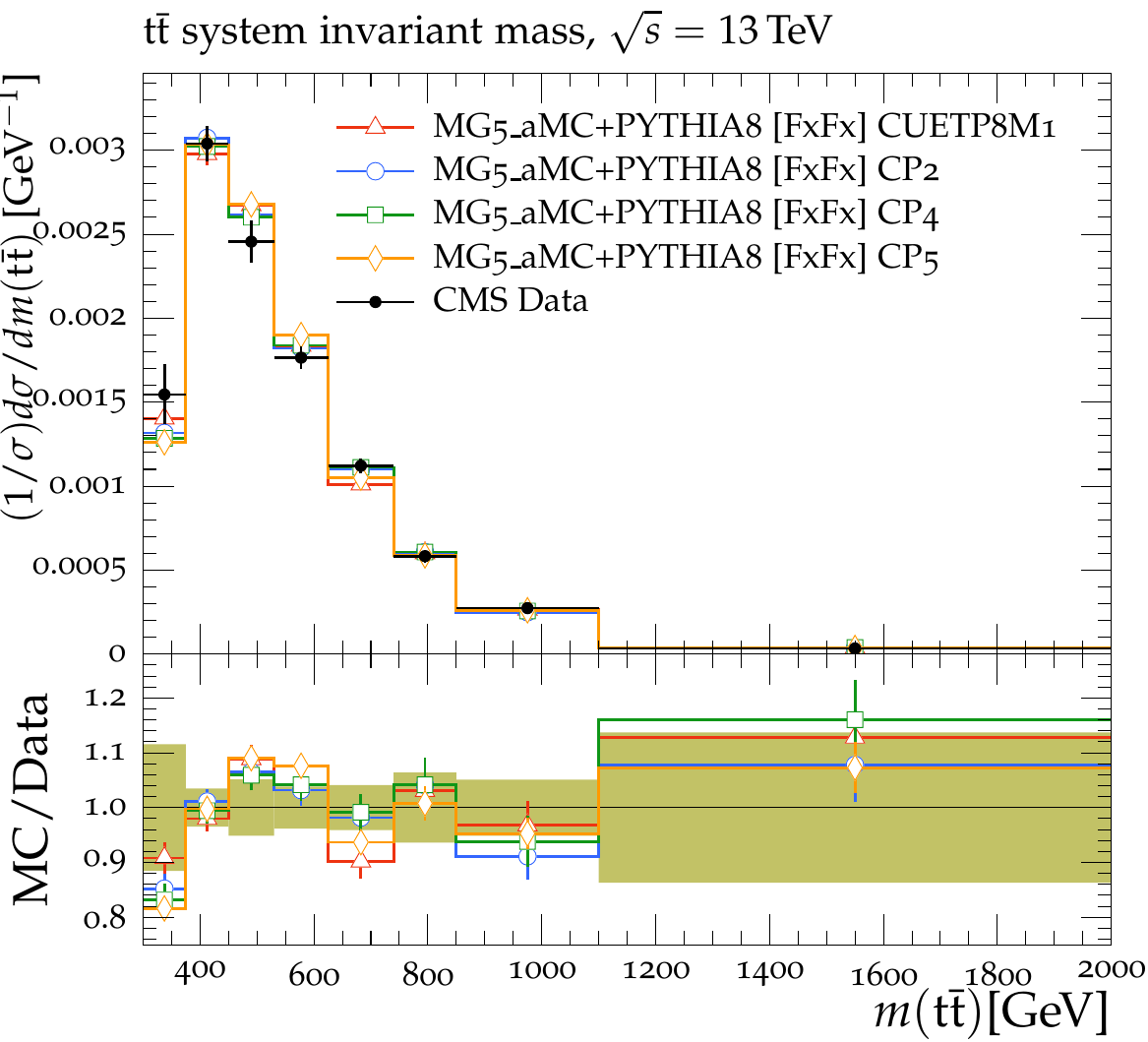} \\
\includegraphics[width=0.4\textwidth]{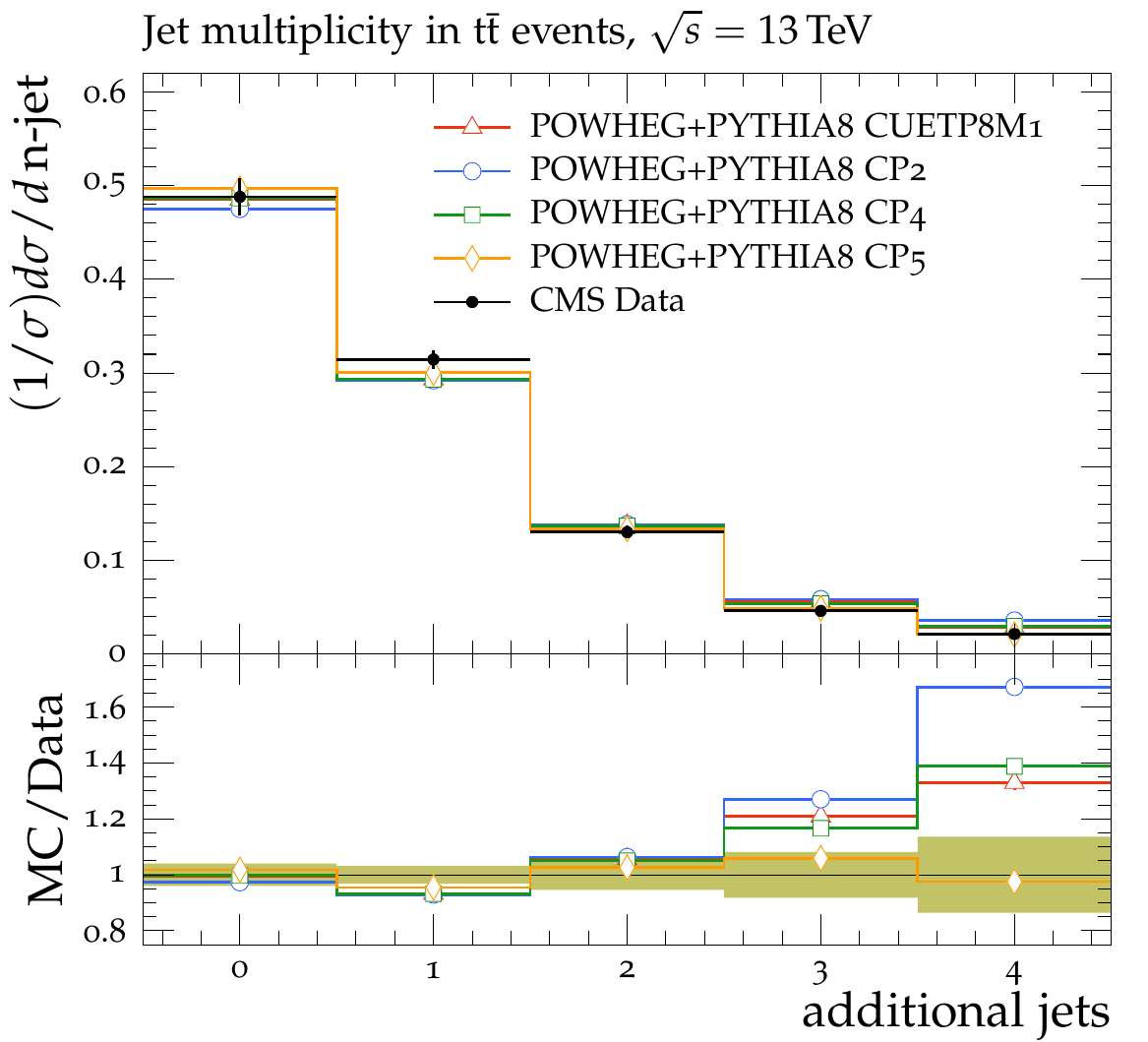}
\includegraphics[width=0.4\textwidth]{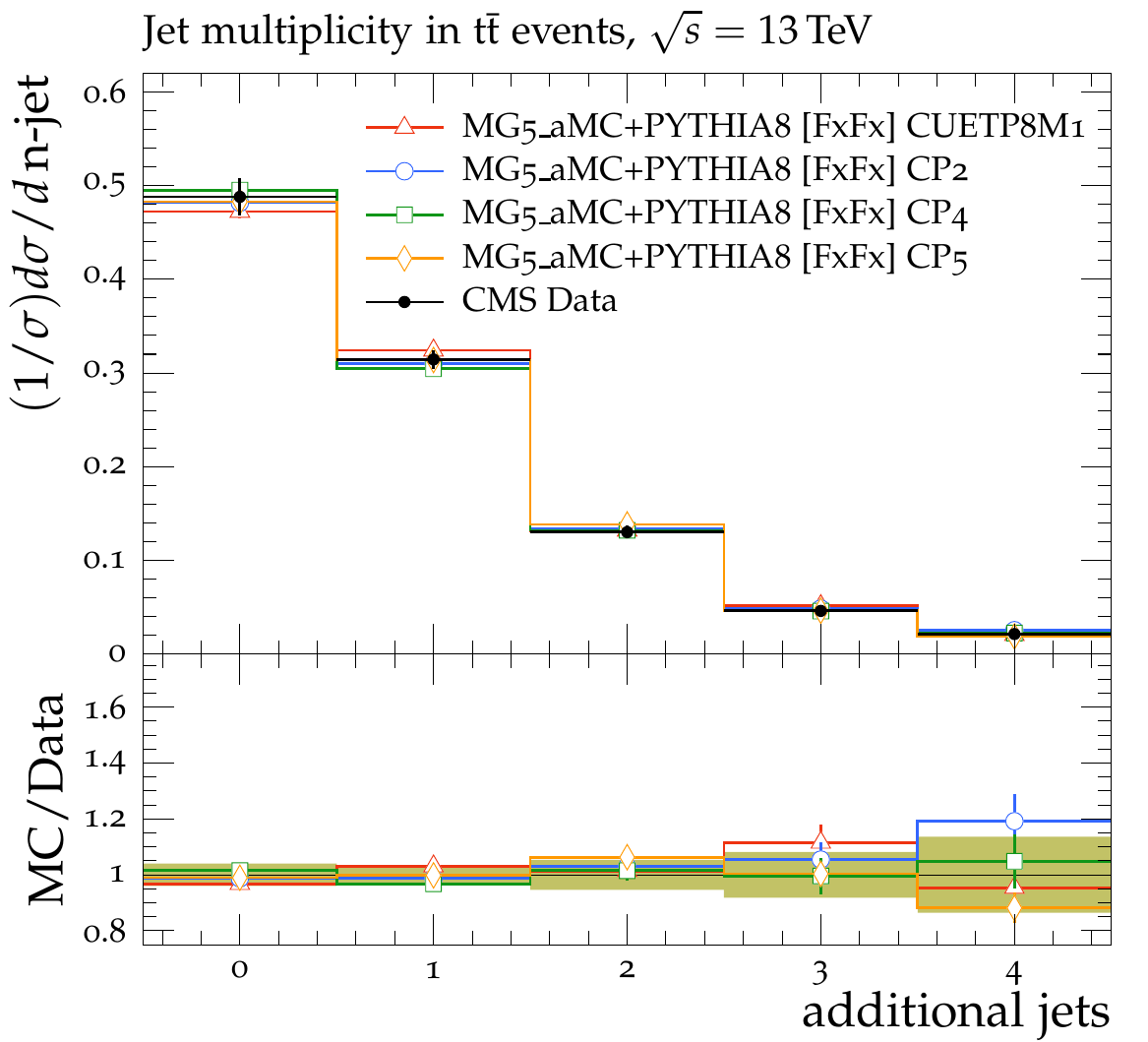}
\caption{The normalized \ttbar cross section in the lepton+jets channel, as a function of the transverse momentum of the top quark for leptonically decaying top quarks ($t_\ell$) (upper), the invariant mass of the \ttbar system, $m$(\ttbar) (middle), and in bins of number of additional jets (lower) from
CMS $\sqrt{s}=13\TeV$ analysis~\cite{TOP-16-008}. The data are compared with the predictions of \POWHEG (left) and \MG~[FxFx] (right). In both  cases, the PS simulation is done with the \PYTHIAviii tunes CUETP8M1, CP2, CP4, or CP5. Tunes CP1 and CP3 are not shown in the plot but present a similar behaviour as, respectively, tunes CP2 and CP4. The ratios of simulations to the data (MC/Data) are also shown, where the shaded band indicates the total experimental uncertainty in the data. Vertical lines drawn on the data points refer to the total uncertainty in the data. Vertical lines drawn on the MC points refer to the statistical uncertainty in the predictions. Horizontal bars indicate the associated bin width.}
\label{fig:top}
\end{figure*}

{\tolerance=3000 Distributions~\cite{TOP-16-008} in the lepton+jets channel are compared to predictions from different tunes using various settings, namely, \POWHEG~+\ \PYTHIAviii, and \MG+\PYTHIAviii with FxFx merging~\cite{fxfx}, referred to as \MG~[FxFx] hereafter, with the CUETP8M1, CP2, CP4, and CP5 tunes.  Figure \ref{fig:top} (upper panel) displays the normalized \ttbar cross section in bins of \pt of the top quark decaying leptonically ($\PQt_\ell$), in data and simulation. For all tunes, \POWHEG~+~\PYTHIAviii predictions have deviations below 10\% with respect to the central values of the data. The central values of predictions from \MG~[FxFx] and data agree within $\approx$10\% for $\pt(\PQt_\ell)<400\GeV$ and within $\approx$20\% for higher \pt. \par}

{\tolerance=800 Figure \ref{fig:top} (middle panel) shows the normalized \ttbar cross section in bins of $m$(\ttbar) in data and simulation. Predictions from \POWHEG and \MG~[FxFx] with the new tunes describe the central values of the data reasonably well. Normalized \ttbar cross sections in bins of number of additional jets in data and simulation in the lepton+jets channel at $\sqrt{s}=13\TeV$ are shown in Fig.~\ref{fig:top} (lower panel).
The cross sections are compared with the predictions of \POWHEG and of \MG~[FxFx]. The central values predicted by \POWHEG~+~\PYTHIAviii are in good agreement with data when CP5 tune is used.
The value of $\alpS^\mathrm{ISR}(m_\cPZ)$ in combination with the rapidity ordering for ISR in the \PYTHIAviii simulation affects the additional jet distribution in \ttbar events.
Predictions obtained from \POWHEG~+~\PYTHIAviii overestimate the data when a high value of $\alpS^\mathrm{ISR}(m_\cPZ)\approx0.13$ is used (CUETP8M1 and CP2 tunes) irrespective of rapidity ordering for ISR.
It is observed that even when $\alpS^\mathrm{ISR}(m_\cPZ)=0.118$ is used, predictions from the CP4 tune overshoot the data at high jet multiplicities. A much better agreement of central values is obtained only when rapidity ordering for ISR is switched on in the \PYTHIAviii simulation and $\alpS^\mathrm{ISR}(m_\cPZ)=0.118$ is used as in the CP5 tune.
Predictions from \MG~[FxFx]+\PYTHIAviii with CUETP8M1, CP2, CP4, and CP5 tunes describe the central values of the data reasonably well. \par}

\begin{figure*}[ht!]
\centering
\includegraphics[width=0.49\textwidth]{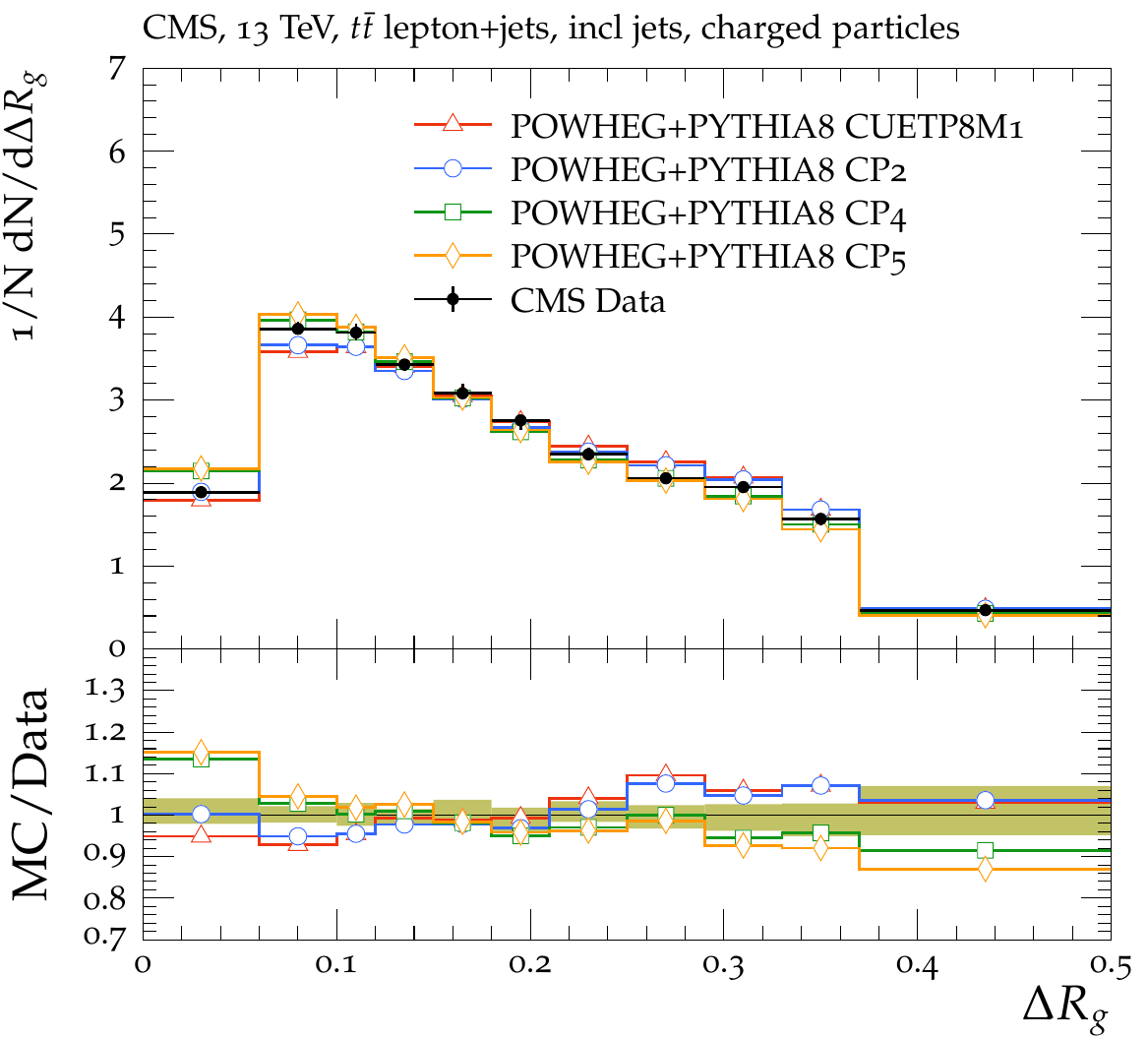}
\includegraphics[width=0.49\textwidth]{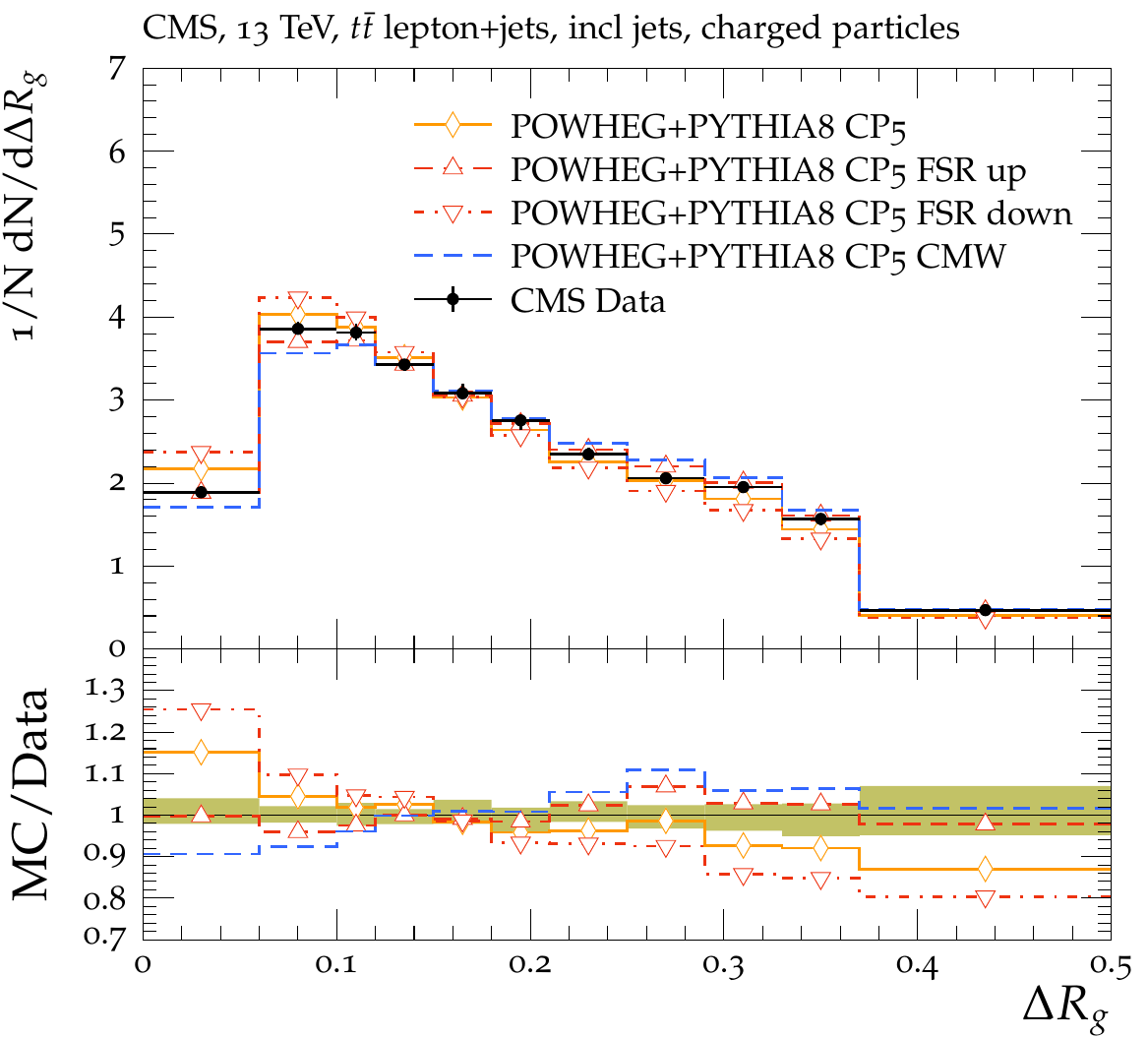}
\caption{Comparison with the measurement~\cite{Sirunyan:2018asm} of the angle between two groomed subjets, $\Delta R_g$ in \ttbar events predicted by \POWHEG~+~\PYTHIAviii for the different tunes. The data are compared to tunes CUETP8M1, CP2, CP4, and CP5 (left). Tunes CP1 and CP3 are not displayed but they present a similar behavior as tunes CP2 and CP4, respectively. The data are also compared to CP5, CP5 FSR up, CP5 FSR down, and CP5 with CMW rescaling (right).
The ratios of simulations to the data (MC/Data) are also shown, where the shaded band indicates the total experimental uncertainty in data. Vertical lines drawn on the data points refer to the total uncertainty in the data. Horizontal bars indicate the associated bin width.}
\label{fig:ttbar_jet_substructure}
\end{figure*}

{\tolerance=800
Comparisons are also made using jet substructure observables in \ttbar events in the lepton+jets channel using measurements by CMS at $\sqrt{s}=13\TeV$~\cite{Sirunyan:2018asm}.
Figure~\ref{fig:ttbar_jet_substructure} displays the comparisons using the distribution of the angle between two groomed subjets, $\Delta R_g$, which is found to be the most sensitive to $\alpS^\mathrm{FSR}(m_\cPZ)$~\cite{Sirunyan:2018asm}.
The data are compared to simulations with the tunes CUETP8M1, CP2, CP4, and CP5, as well as CP5 FSR up ($\alpS^\mathrm{FSR}(m_\cPZ)=0.122$), CP5 FSR down ($\alpS^\mathrm{FSR}(m_\cPZ)=0.115$), and CP5 with CMW rescaling.
It is observed that tunes with higher $\alpS^\mathrm{FSR}(m_\cPZ)$ (CUETP8M1, CP2, and CP5 FSR up) describe the data better.
Tune CP5 with CMW rescaling resolves the discrepancy of CP5 at high $\Delta R_g$, but worsens the description at $\Delta R_g\sim0.27$ compared to CP5.
It should be noted that a fit to the $\Delta R_g$ distribution using a b-enriched sample yields $\alpS^\mathrm{FSR}(m_\cPZ)=0.130^{+0.016}_{-0.020}$ ~\cite{Sirunyan:2018asm} without CMW rescaling, while a fit to the distirubtion of the UE observable $\overline{\pt}$ measured in \ttbar events yields $\alpS^\mathrm{FSR}(m_\cPZ)=0.120\pm0.006$~\cite{Sirunyan:2018avv}. Therefore, in \ttbar events, UE and jet substructure observables prefer different central $\alpS^\mathrm{FSR}(m_\cPZ)$ values, but they are compatible within uncertainties.
\par}

\subsection{Comparisons using observables in \texorpdfstring{\PW{} and \cPZ{}}{W and Z} boson production}
\label{DYUE}
{\tolerance=800 In this subsection, we present a validation of the new CMS UE tunes for observables measured in events with a \PW{} or \cPZ{} boson in the final state at $\sqrt{s}=13\TeV$. For the comparisons, we use predictions obtained with \MG~$+$~\PYTHIAviii at LO using the \ktMLM merging scheme, and at NLO using the \FxFx merging scheme. The \ktMLM merging scale is set to 19\GeV, while for \FxFx the corresponding scale is set to 30\GeV. In both cases the MEs include the final states with 0, 1, 2, and 3 partons, and up to 2 partons are calculated at NLO precision in the \FxFx case. To ease the comparison of the different tunes, the same PDF, NNPDF3.1 NNLO, and $\alpS(m_\cPZ) = 0.118$ are used for the ME calculation independently of the tune. \par}

\begin{figure*}[ht!]
\centering
\includegraphics[width=0.4\textwidth]{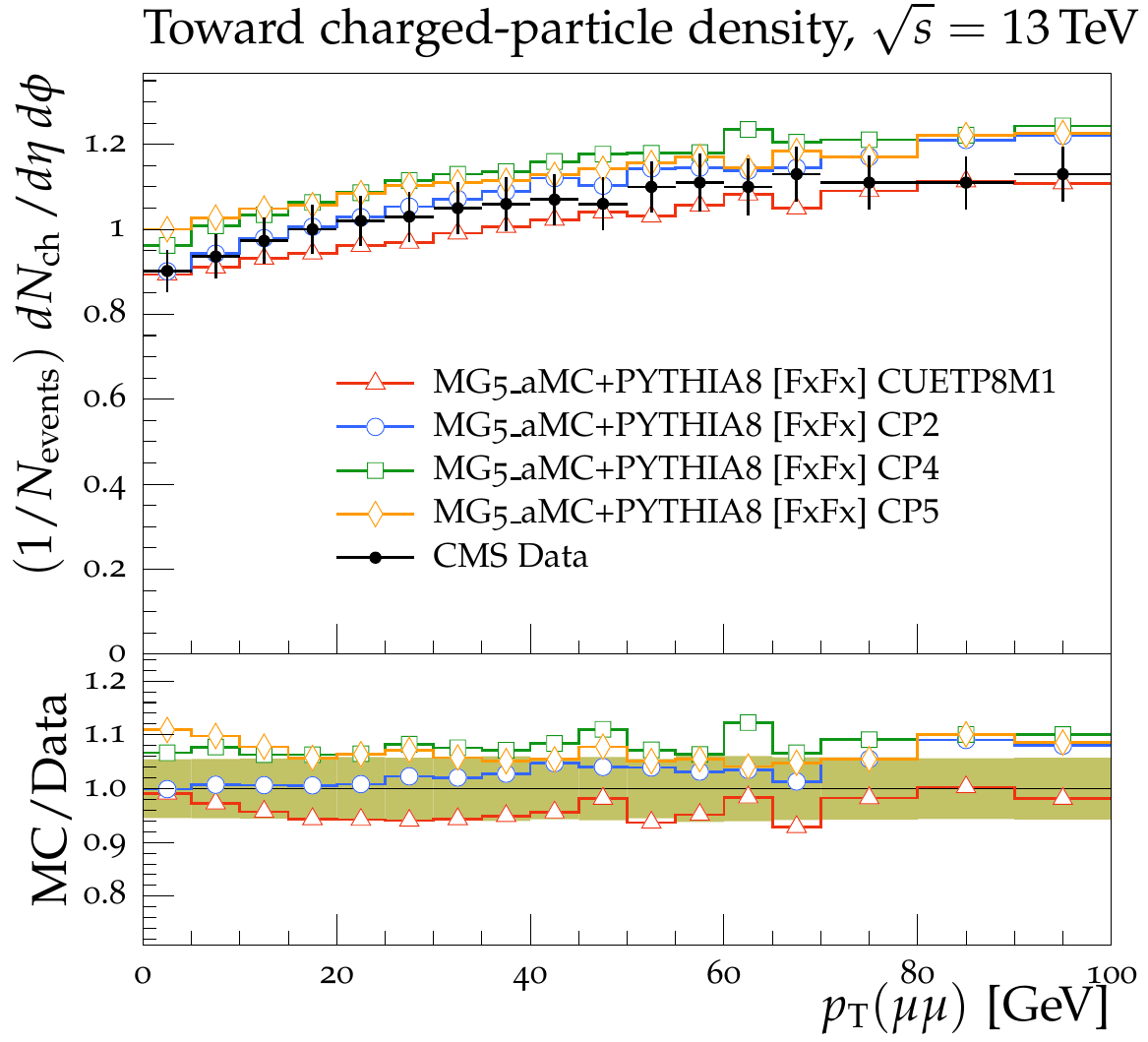}
\includegraphics[width=0.4\textwidth]{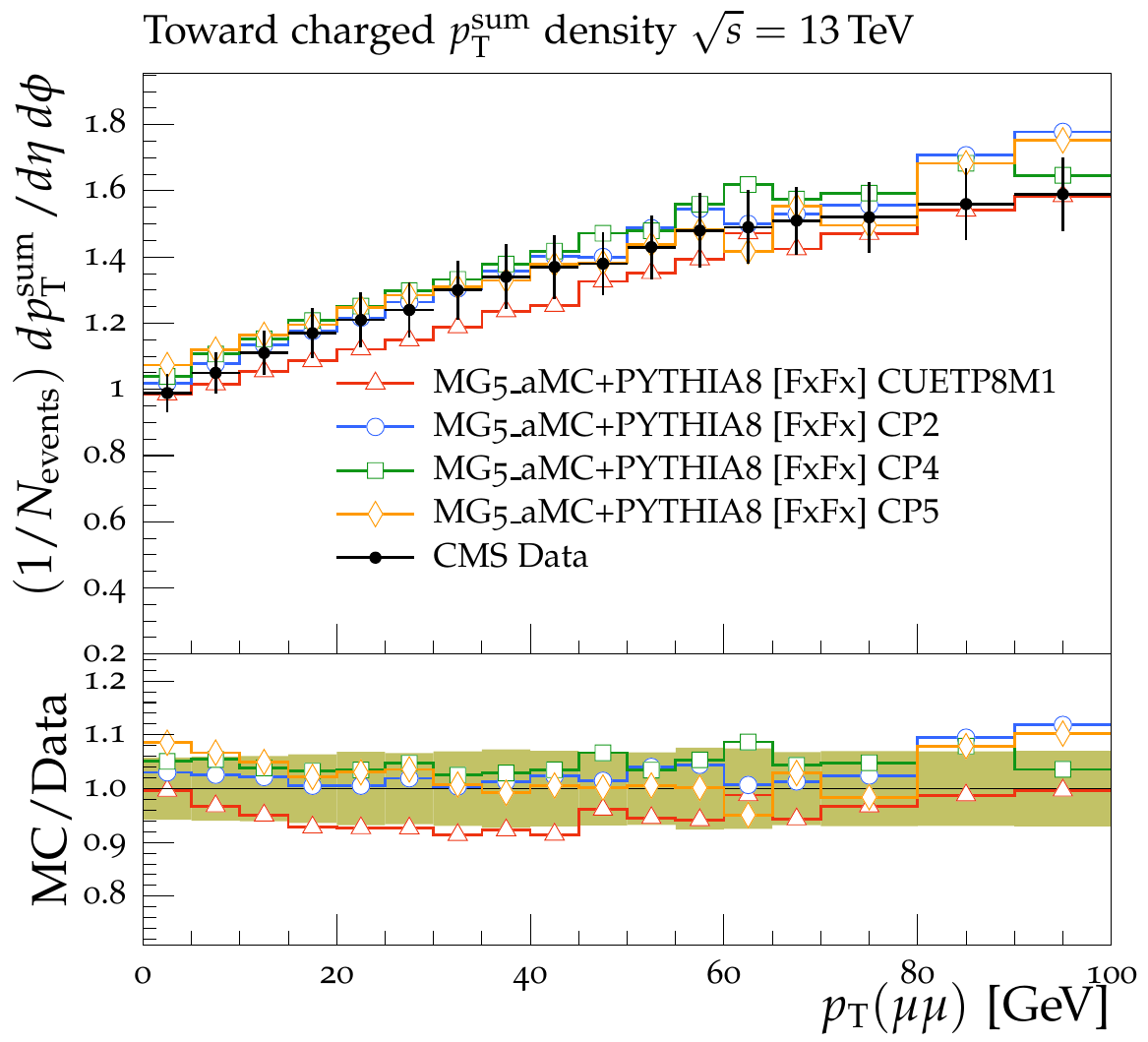}\\
\includegraphics[width=0.4\textwidth]{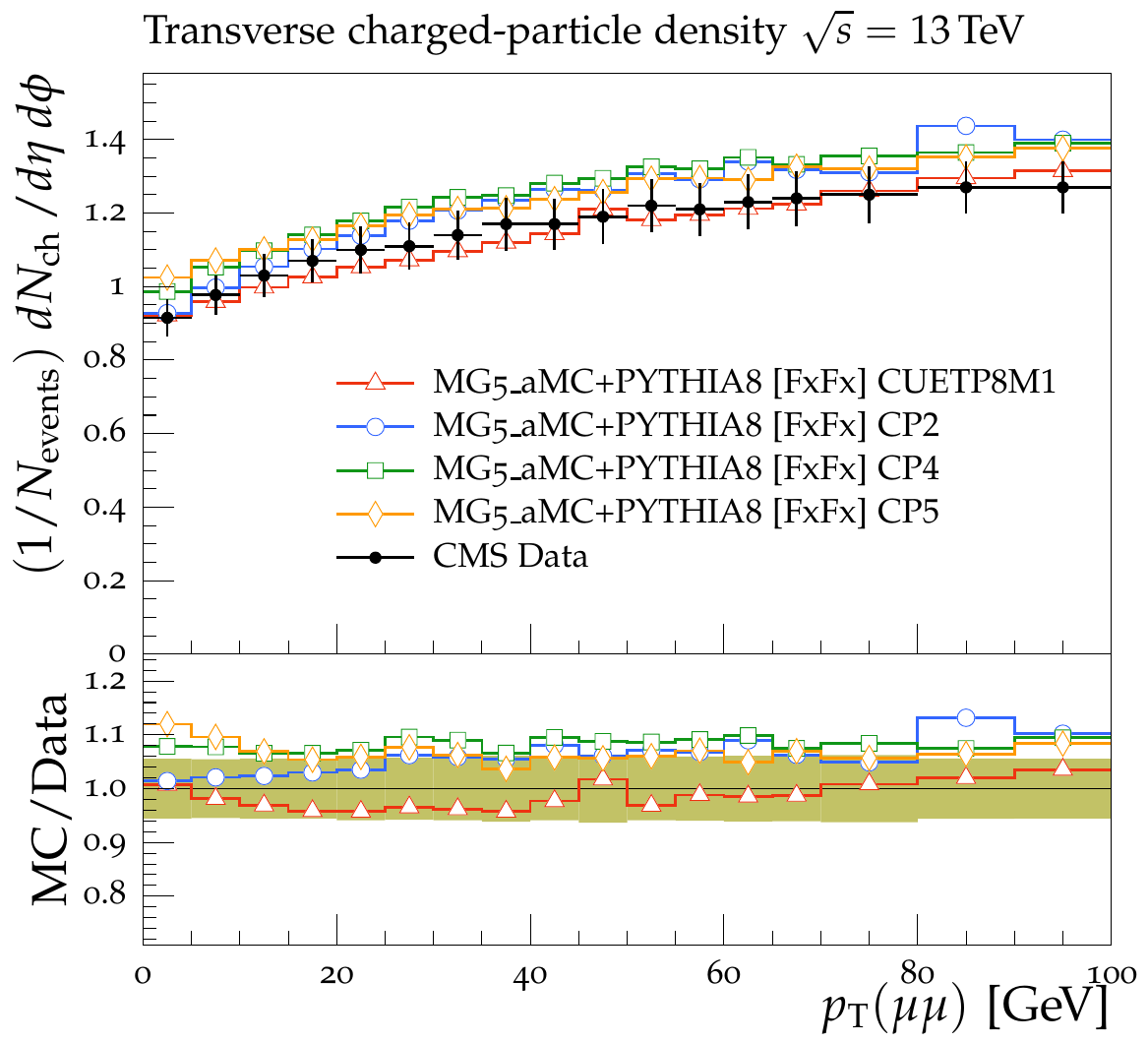}
\includegraphics[width=0.4\textwidth]{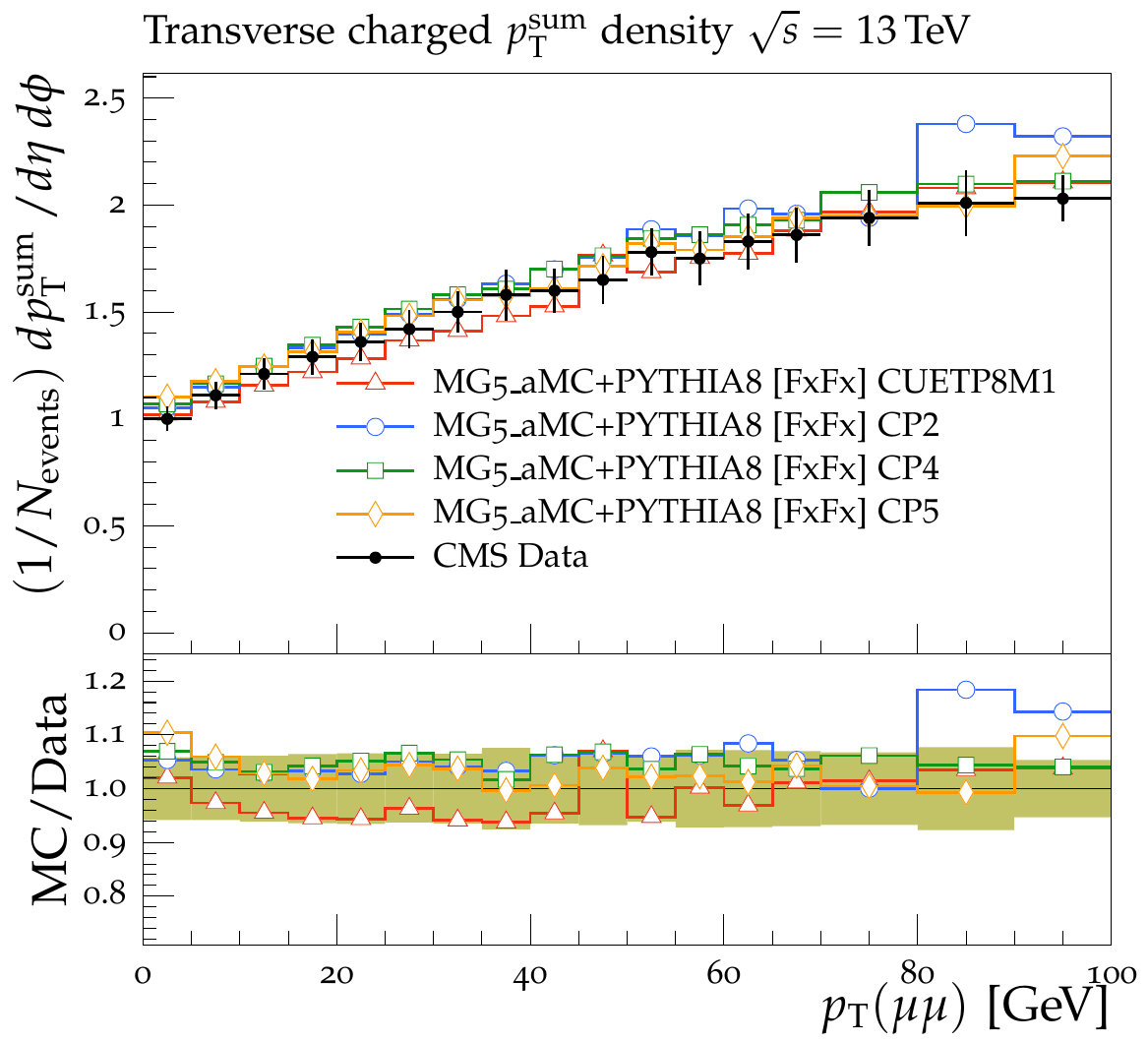}\\
\includegraphics[width=0.4\textwidth]{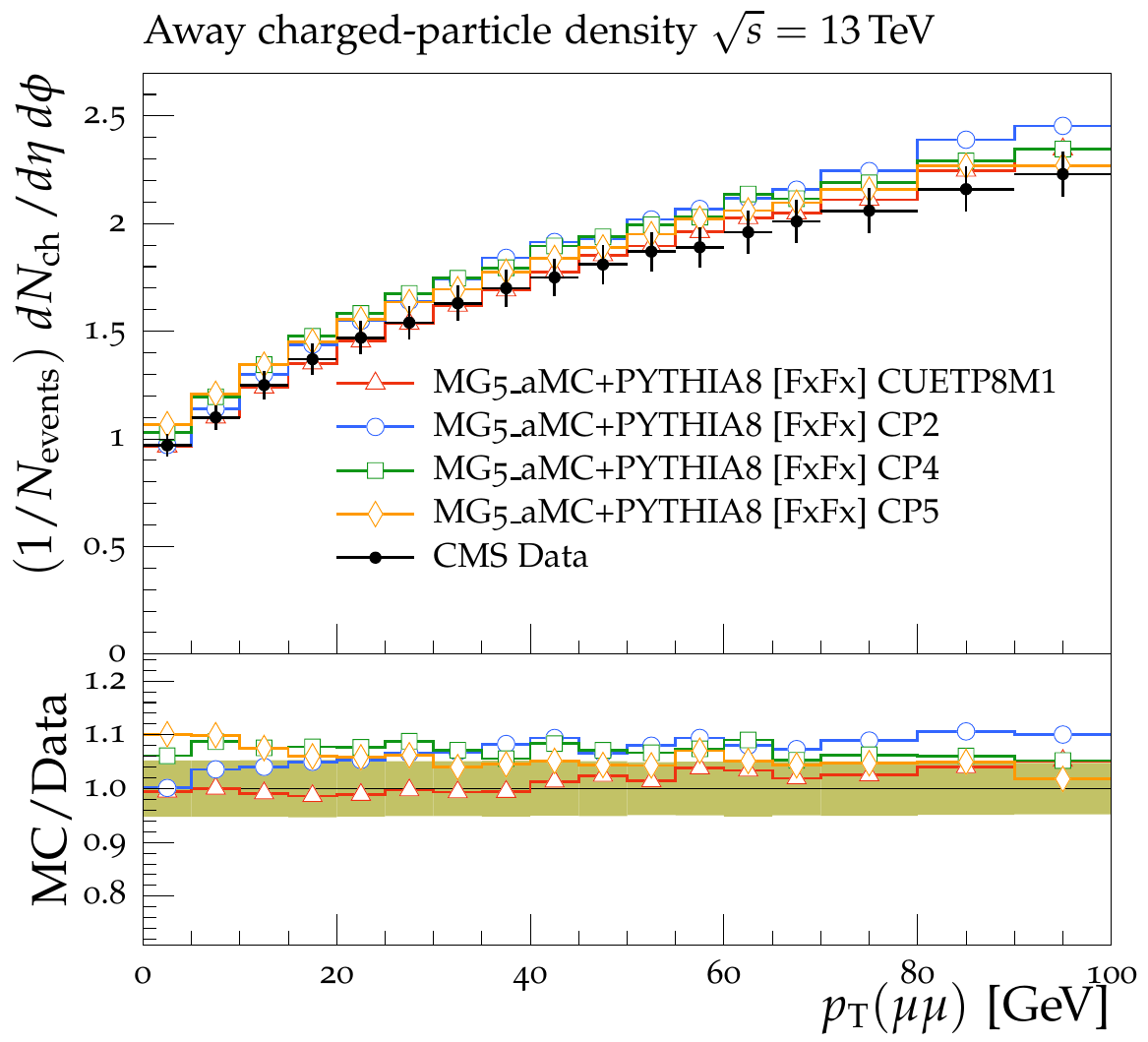}
\includegraphics[width=0.4\textwidth]{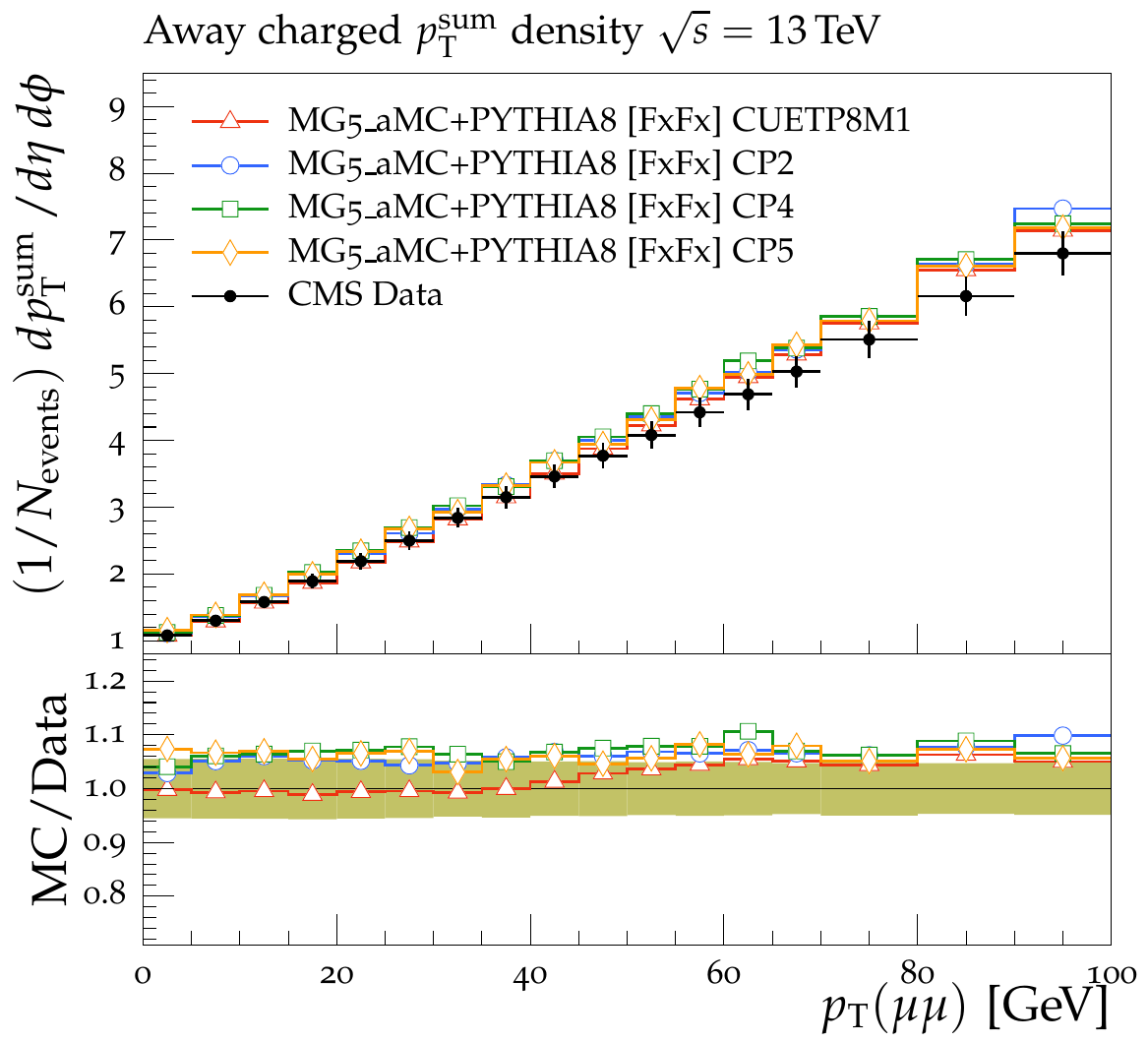}\\
\caption{The charged-particle multiplicity (left) and \ptsum\ (right) in the toward (upper), transverse (middle), and away (lower) regions measured as a function of the \cPZ{} boson \pt\ in Drell--Yan events at $\sqrt{s}=13\TeV$~\cite{Sirunyan:2017vio}, and compared with the predictions obtained by an inclusive NLO ME calculated by \MG, interfaced to the UE simulation of \PYTHIAviii with the CUETP8M1, CP2, CP4, and CP5 tunes. Tunes CP1 and CP3 are not shown in the plot but present a similar behaviour as, respectively, tunes CP2 and CP4. The ratios of simulations to the data (MC/Data) are also shown, where the shaded band indicates the total experimental uncertainty of the data. Vertical lines drawn on the data points refer to the total uncertainty in the data. Vertical lines drawn on the MC points refer to the statistical uncertainty in the predictions. Horizontal bars indicate the associated bin width.}
\label{fig13TeVDY}
\end{figure*}

First, UE observables~\cite{Sirunyan:2017vio} in Drell--Yan events in an invariant mass window of 81--101\GeV around the \cPZ{}  boson peak for muonic decays are studied. The charged-particle density and transverse momentum sum are measured as a function of the \cPZ{} boson \pt in the three regions introduced in Section~\ref{PerformanceSoftEvents}: toward, away, and transverse. The regions are defined with respect to the \cPZ{} boson direction. The measurements are compared with \FxFx predictions obtained with the CUETP8M1, CP2, CP4, and CP5 tunes in Fig.~\ref{fig13TeVDY}. The measurements are, in general, well-described by all tunes.

The description of the cross section as a function of the jet multiplicity is also investigated in  \cPZ+jets~\cite{Sirunyan:2018cpw} and \PW+jets~\cite{ref:W} final states. The \cPZ+jets measurement is restricted to the phase space where the two leptons have $\pt>20\GeV$ and $\abs{y}<2.4$ and the dilepton mass lies in a $\pm20\GeV$ window around 91\GeV. The momenta of the photons inside a cone of $\Delta\text{R}<0.1$ are added to the lepton momentum in order to partly recover the energy lost by FSR. Jets are clustered using the anti-\kt algorithm with $R = 0.4$ and must satisfy the criteria $\pt > 30\GeV$ and $\abs{y} < 2.4$. The distance between the selected leptons and the leading jet $\Delta R(\ell,\mathrm{j})$ must be greater than 0.4.
For the \PW+jets measurement, the phase space is restricted by a transverse mass requirement, $\mT> 50\GeV$, and by requirements on the muon, $\pt > 25\GeV$ and $\abs{y}<2.4$. In the \cPZ+jets measurements the same clustering algorithm, the FSR recovery prescription described above, and the lepton jet separation requirement are applied.

\begin{figure*}
\centering
\includegraphics[width=0.49\textwidth]{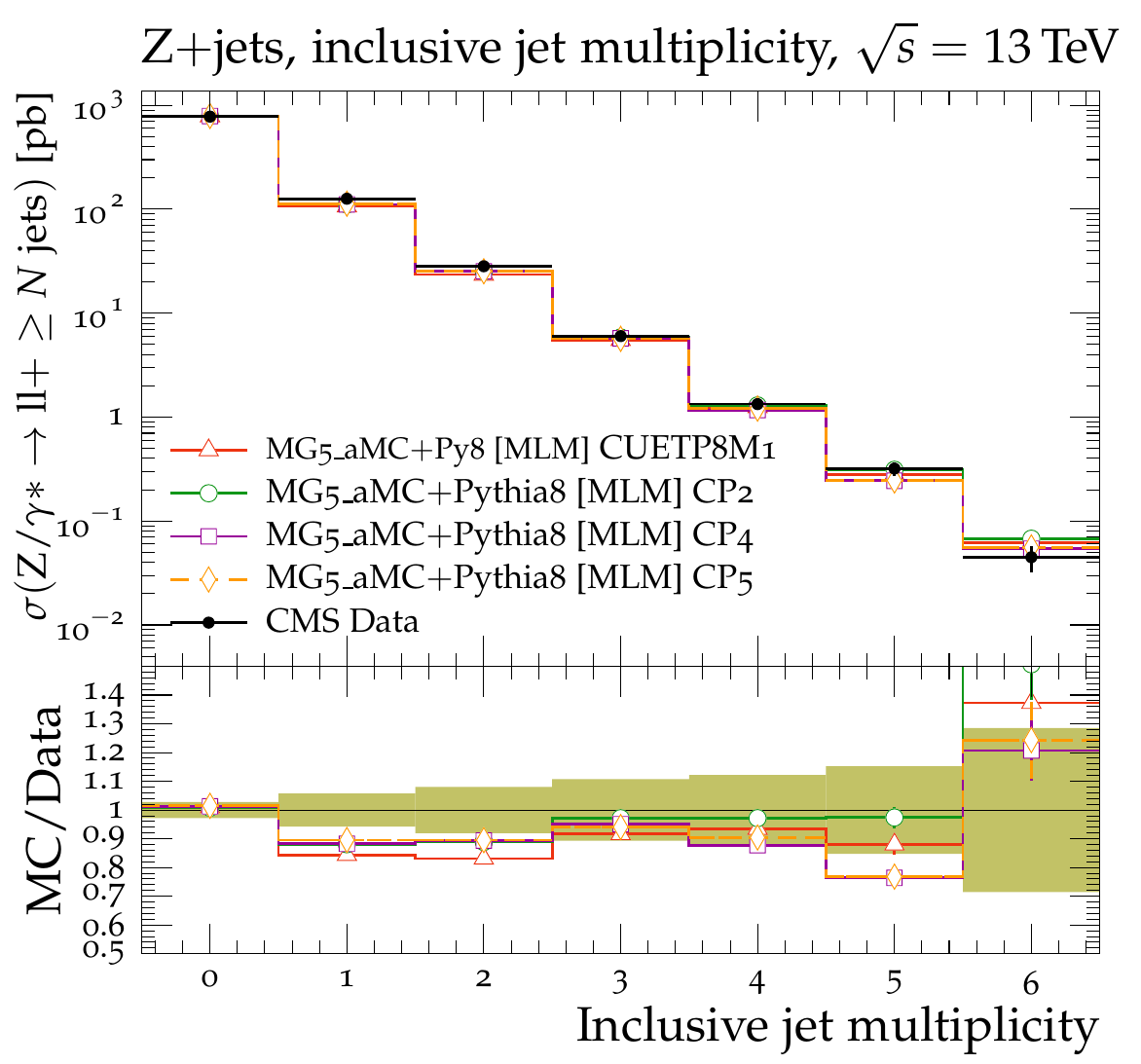}
\includegraphics[width=0.49\textwidth]{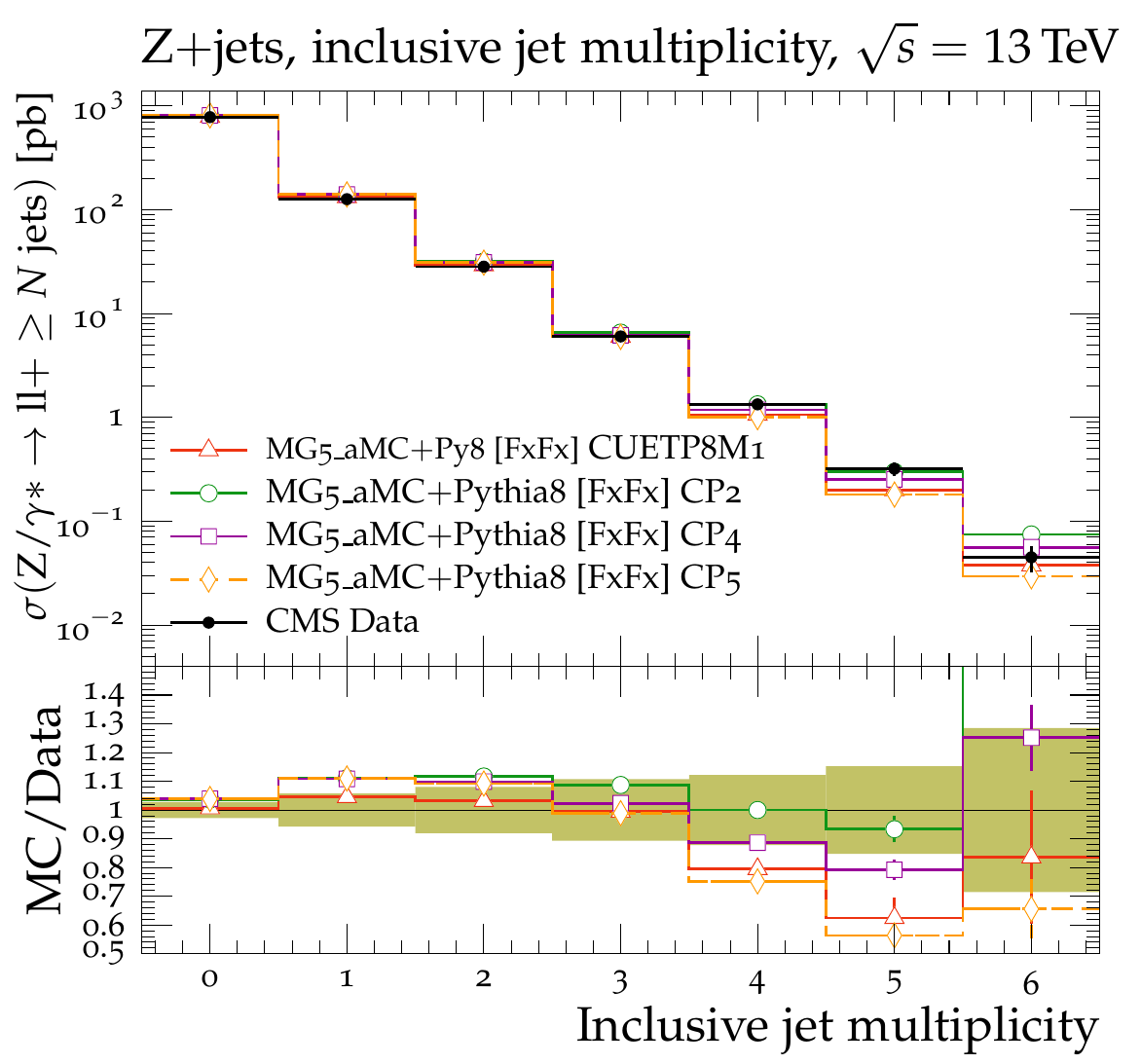} \\
\includegraphics[width=0.49\textwidth]{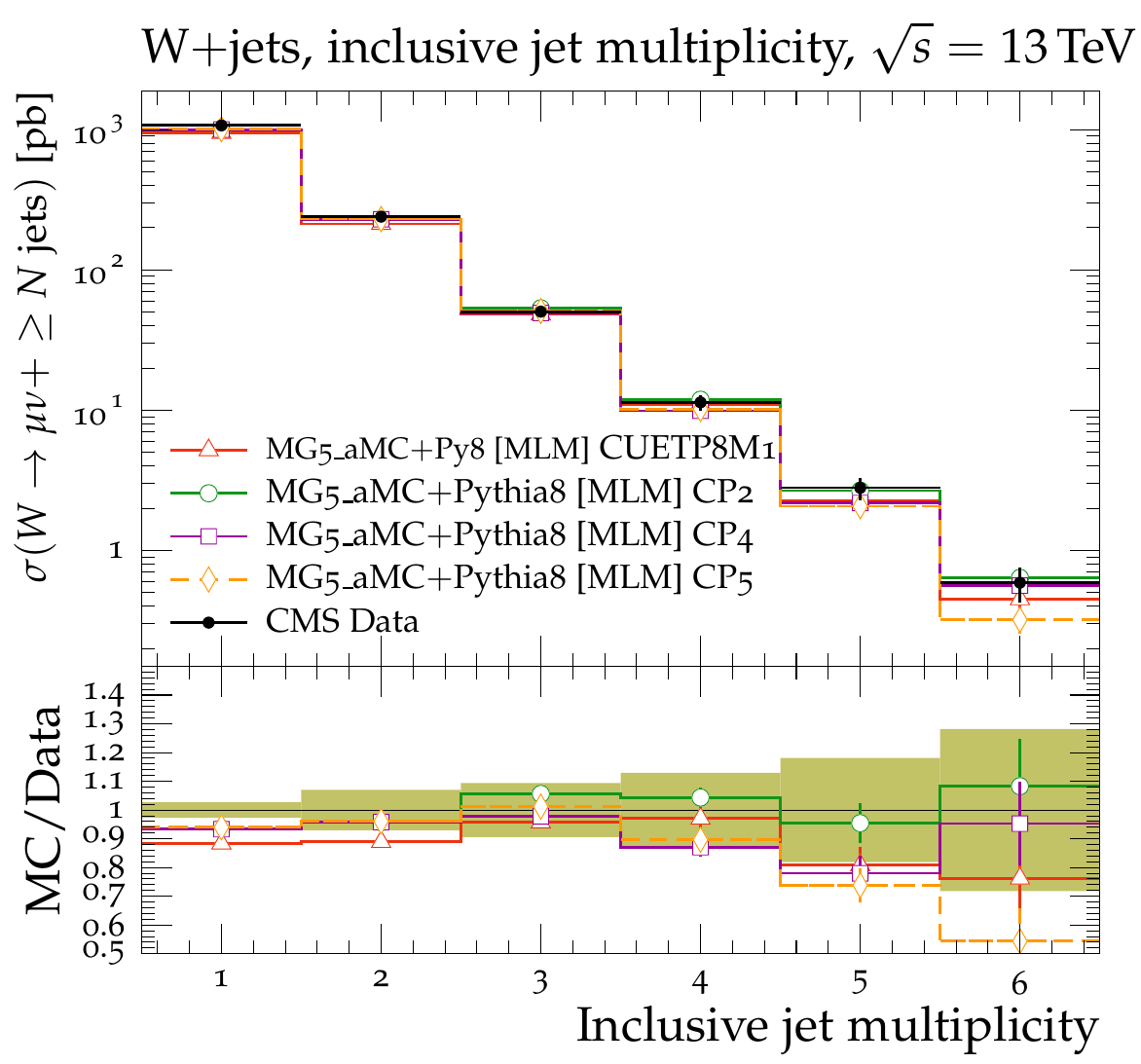}
\includegraphics[width=0.49\textwidth]{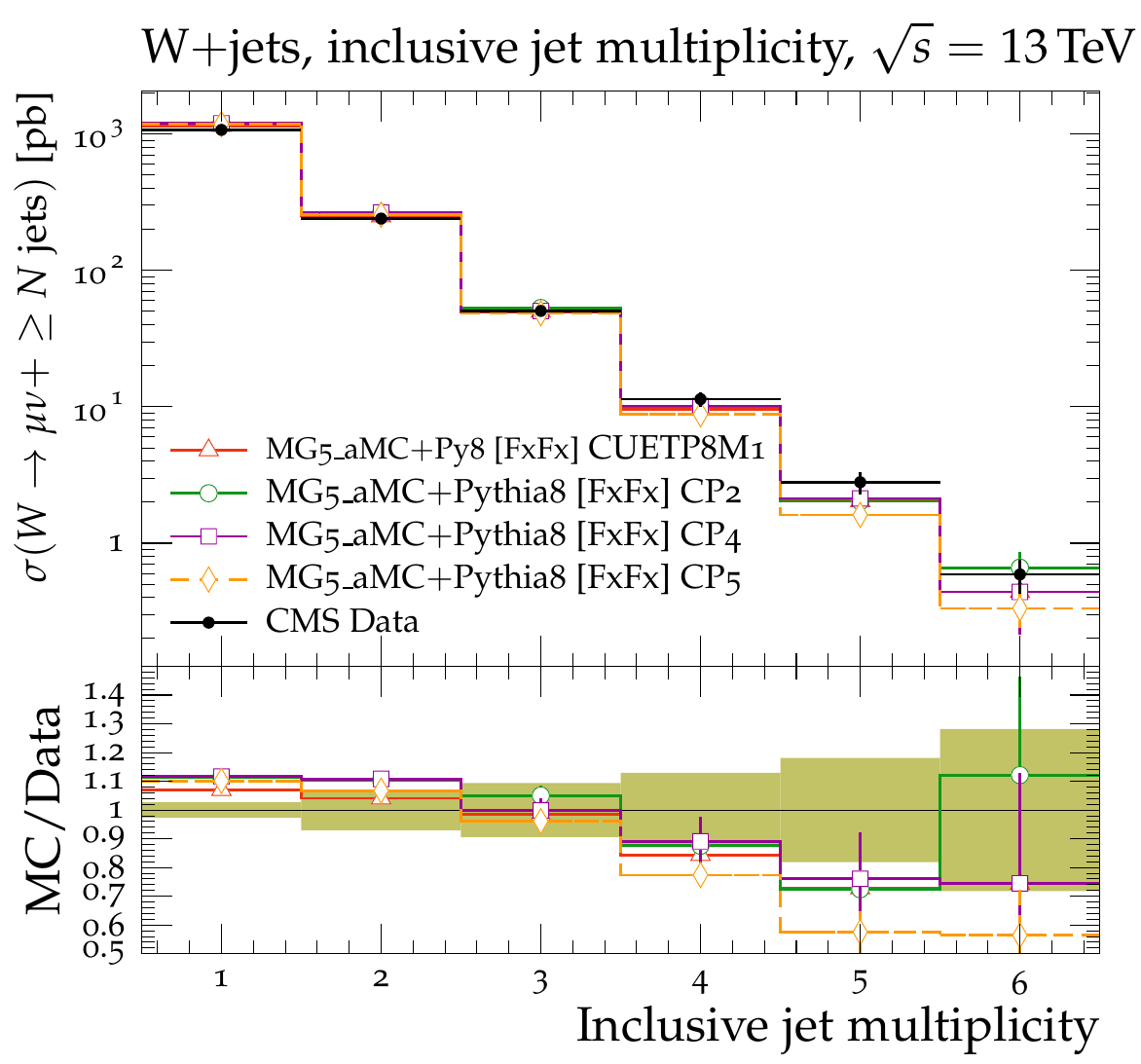}
\caption{Comparison with the measurement~\cite{Sirunyan:2018cpw,ref:W} of the inclusive jet multiplicity in \cPZ+jets (upper) and \PW+jets (lower) events predicted by \MG~+~\PYTHIAviii with \ktMLM merging (left) and \FxFx merging (right) for the different tunes. Tunes CP1 and CP3 are not shown in the plot but present a similar behaviour as, respectively, tunes CP2 and CP4. The ratios of simulations to the data (MC/Data) are also shown, where the shaded band indicates the total experimental uncertainty in the data. Vertical lines drawn on the data points refer to the total uncertainty in the data. Vertical lines drawn on the MC points refer to the statistical uncertainty in the predictions. Horizontal bars indicate the associated bin width.}

\label{fig:vj_nj}
\end{figure*}

The comparisons of the jet multiplicities to various predictions are shown in Fig.~\ref{fig:vj_nj}. The measurement of the cross section inclusive in the number of jets, $N$, is not available for the \PW+jets analysis and the lower plots start at $N=1$. The \ktMLM predictions of the jet multiplicity have little sensitivity to the UE and PS tunes, so all the tunes provide a good description this observable, with a slightly better agreement observed for the CP2 tune. In the case of the \FxFx sample, the CP5 tune predicts fewer events with a jet multiplicity of more than four with respect to the measurement. The deficit increases for increasing jet multiplicities. The CUETP8M1 tune shows a similar behaviour, though.

\begin{figure*}
\centering
\includegraphics[width=0.49\textwidth]{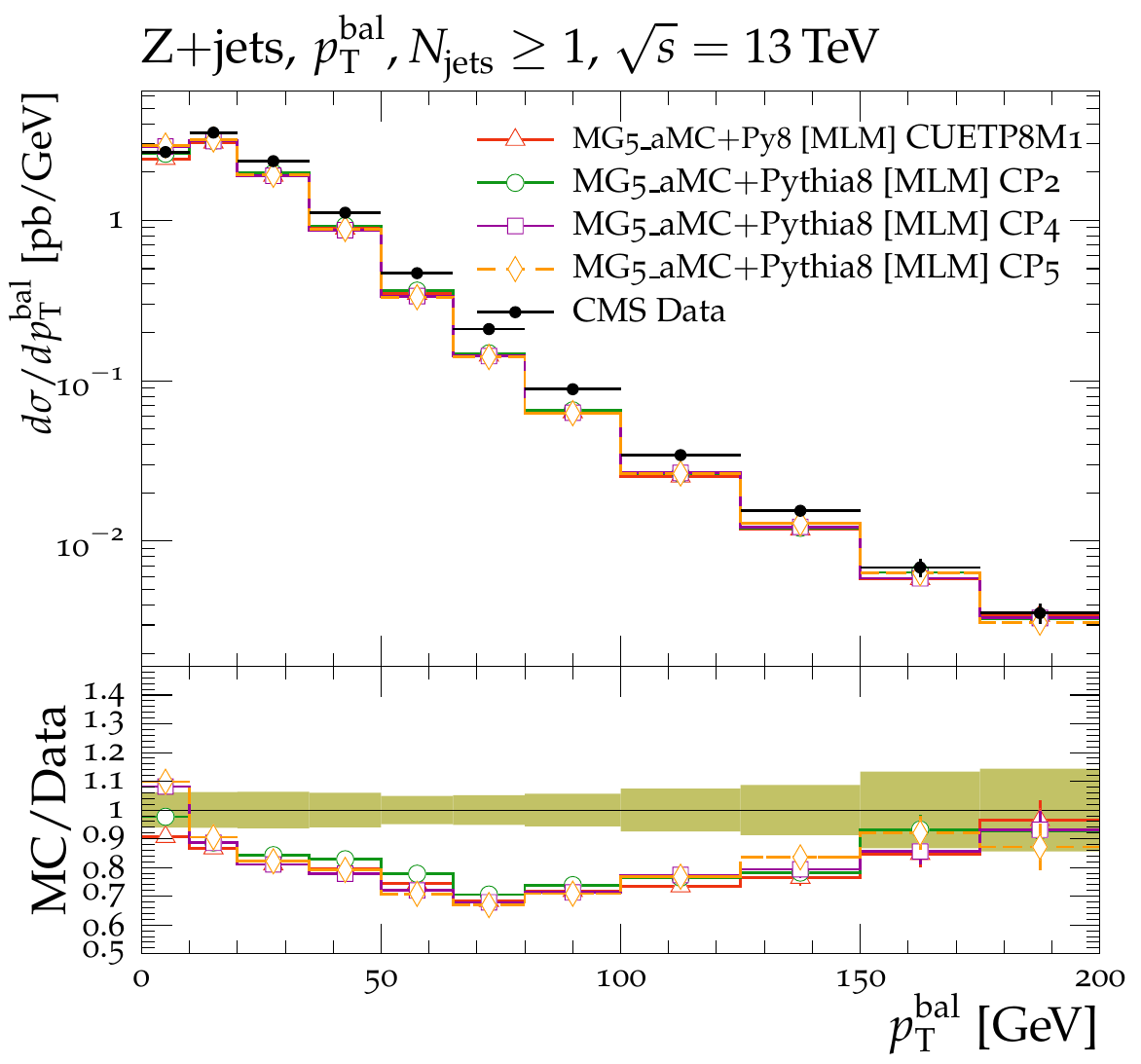}
\includegraphics[width=0.49\textwidth]{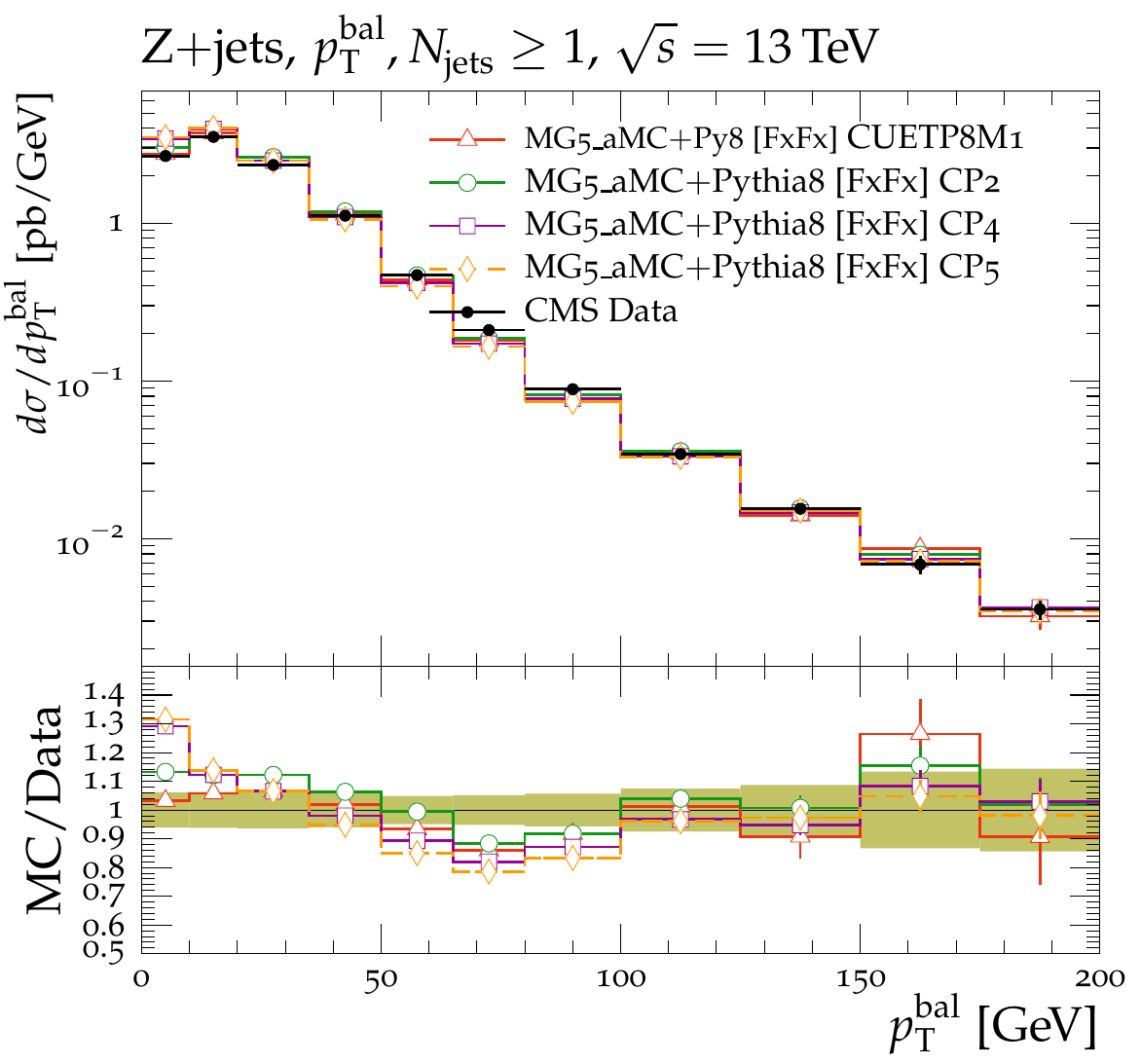}
\caption{Comparison with the measurement~\cite{Sirunyan:2018cpw} of the \pt balance predicted by \MG~+~\PYTHIAviii with \ktMLM merging (left) and \FxFx merging (right) for the different tunes for events with at least one jet. Tunes CP1 and CP3 are not shown in the plot but they present a similar behaviour as tunes CP2 and CP4, respectively.  The ratios of simulations to the data (MC/Data) are also shown, where the shaded band indicates the total experimental uncertainty in the data. Vertical lines drawn on the data points refer to the total uncertainty in the data. Vertical lines drawn on the MC points refer to the statistical uncertainty in the predictions. Horizontal bars indicate the associated bin width.}
\label{fig:ptbal}
\end{figure*}

Predictions using the new CMS UE tunes are also compared with the $\pt$ balance between the \cPZ{} boson and the jets with $\pt > 30\GeV$ and $\abs{y}<2.4$ using the variable \pt$^\text{bal} = \lvert \ensuremath{\ptvec}(\cPZ) + \sum_{\text{jets}} \ptvec(\mathrm{j}_i) \rvert$~\cite{Sirunyan:2018cpw}. This variable is sensitive to PS and UE.  The comparison is shown in Fig.~\ref{fig:ptbal} for events with at least one jet. Differences between the tunes are significant only in the region below $\approx$20\GeV. The discrepancy in this region for the \FxFx samples indicates that the distribution peaks at lower values for CP4 and CP5 than in data.

\begin{figure*}
\centering
\includegraphics[width=0.49\textwidth]{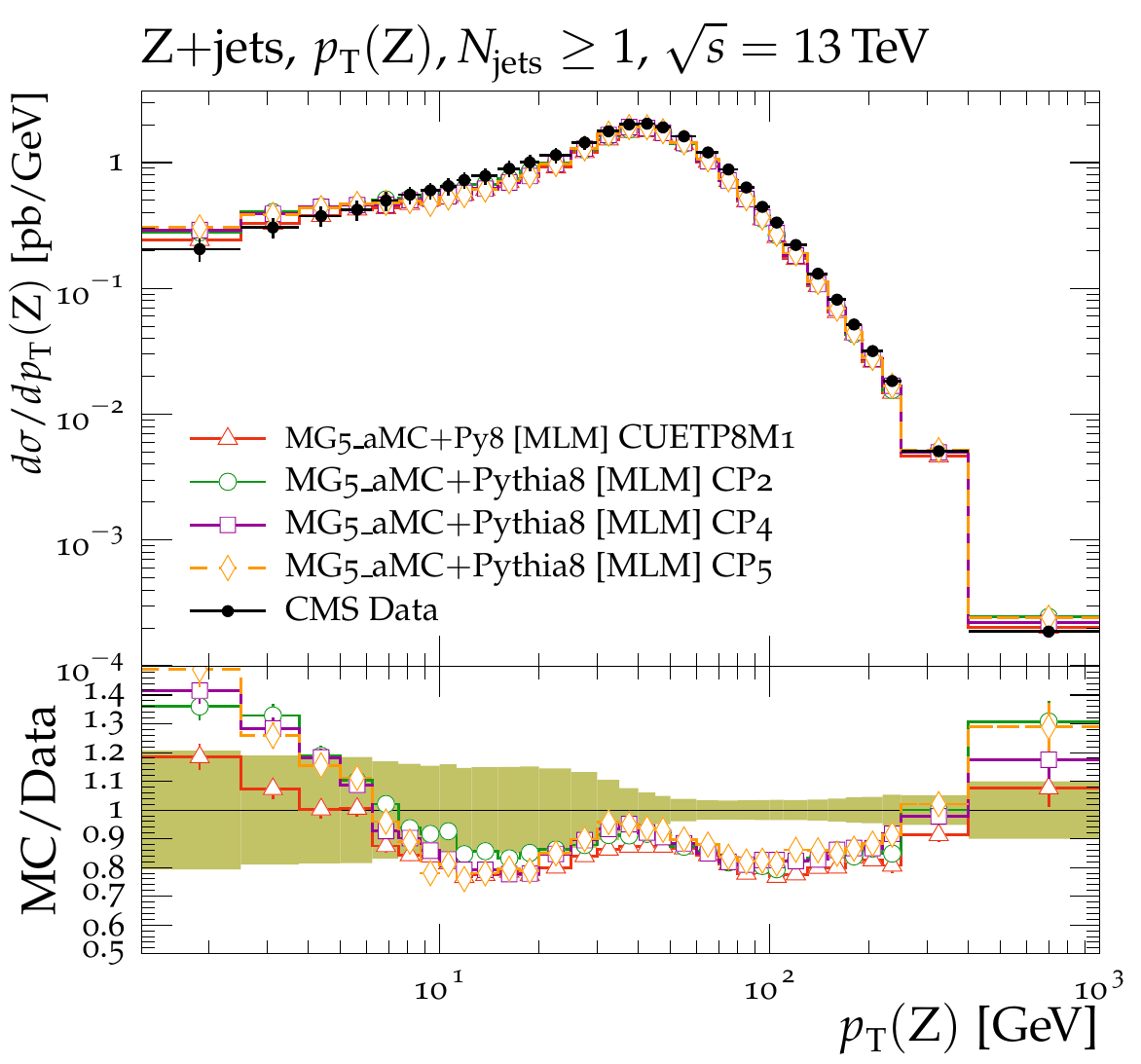}
\includegraphics[width=0.49\textwidth]{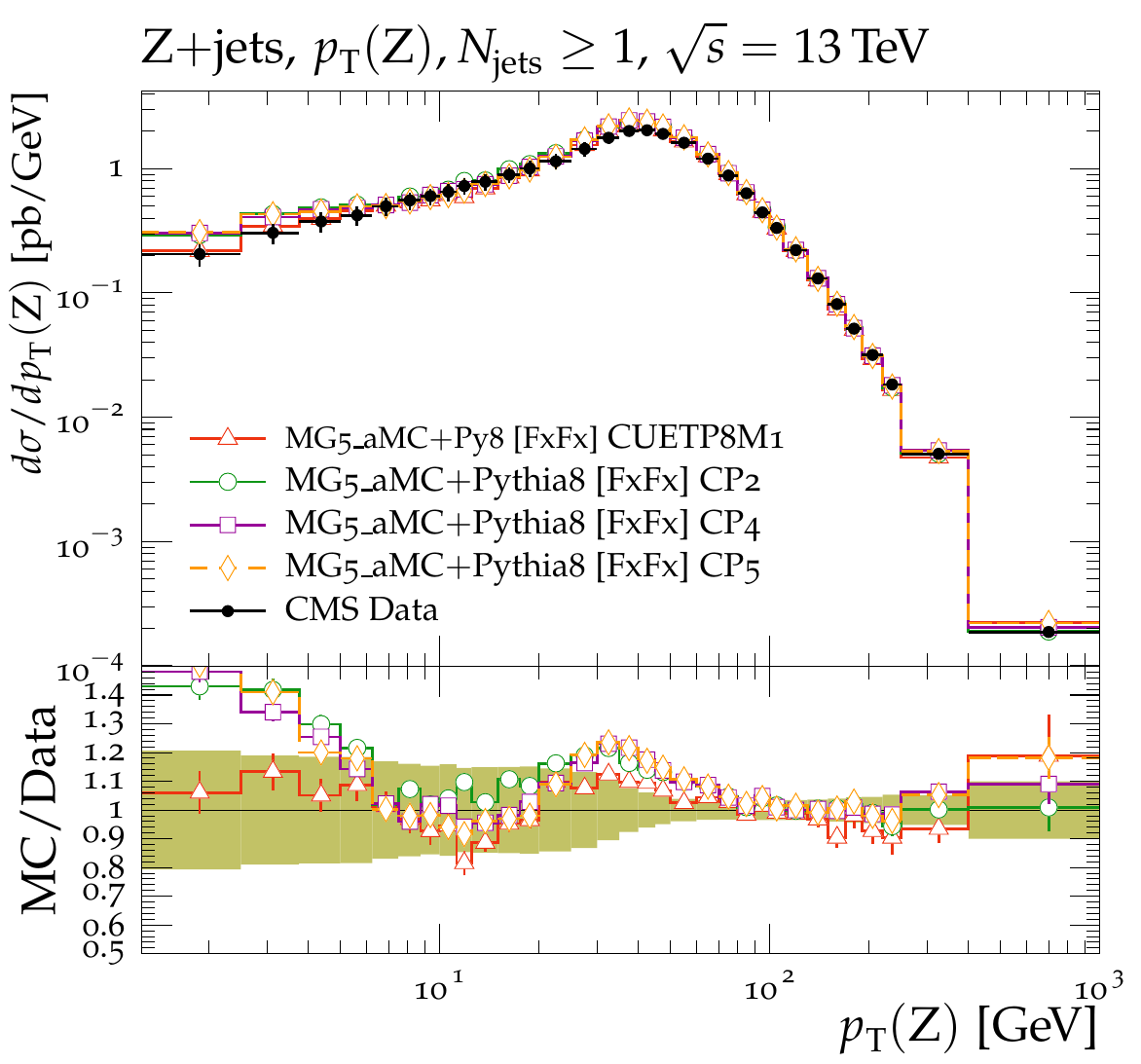}
\caption{Comparison with the measurement~\cite{Sirunyan:2018cpw} of the \pt(Z) predicted by \MG~+~\PYTHIAviii with \ktMLM merging (left) and \FxFx merging (right) for the different tunes. Tunes CP1 and CP3 are not shown in the plot but they present a similar behaviour as tunes CP2 and CP4, respectively. The ratios of simulations to the data (MC/Data) are also shown, where the shaded band indicates the total experimental uncertainty in data. Vertical lines drawn on the data points refer to the total uncertainty in the data. Vertical lines drawn on the MC points refer to the statistical uncertainty in the predictions. Horizontal bars indicate the associated bin width.}
\label{fig:zj_zpt}
\end{figure*}

Results of Ref.~\cite{Sirunyan:2018cpw} are also used to validate the description of the transverse momentum of the weak vector boson in  $\cPZ$ + $\ge 1$ jet events. The comparison is shown in Fig.~\ref{fig:zj_zpt}. The new tunes provide similar descriptions for this distribution. Predictions using \ktMLM achieve a poor agreement with the data, independently of the UE tune, with respect to \FxFx, which is able to describe the transverse momentum of the  \cPZ{} boson at $\pt >10\GeV$. The region below 10\GeV is poorly described for both \FxFx and \ktMLM and the new tunes, but is well-described by predictions using the CUETP8M1 tune.

To summarize the study of weak vector boson production, the CP2, CP4, and CP5 tunes provide similar descriptions of the UE observables with a reasonable agreement with the data. In general, the CP2 tune performs better in describing variables such as \pt$^{\text{bal}}$ and \pt(Z). For the jet multiplicity, the CP2 and CP4 tunes are equally good in describing the measurement, whereas CP5 tends to undershoot the PS dominated region with at least five jets with a significance of 3.5 standard deviations.

\section{Summary and conclusions}
\label{Summary}

{\tolerance=800 A new set of tunes for the underlying-event (UE) simulation in the \PYTHIAviii event generator is obtained by fitting various measurements sensitive to soft and semihard multipartonic interactions at different collision energies. To derive these tunes, the leading order (LO), next-to-leading order (NLO), or next-to-next-to-leading order (NNLO) versions of the NNPDF3.1 parton distribution function (PDF) set for the simulation of the underlying-event components are used. In these tunes, the values of the strong coupling, $\alpS(m_\cPZ)$, used for the simulation of hard scattering, initial- and final-state radiation, and multiple-parton interactions are chosen consistent with the order of the PDF used.
In the LO NNPDF3.1 set, $\alpS(m_\cPZ) = 0.130$, whereas for the NLO and NNLO NNPDF3.1 sets, $\alpS(m_\cPZ) = 0.118$. In general, the combination of contributions from multiple-parton interactions and parton-shower emissions is crucial to give a good description of variables measured in soft-collision events.
The infrared threshold is relatively independent of center-of-mass energy when using NLO or NNLO PDF sets.
Irrespective of the specific PDF used, predictions from the new tunes reproduce well the UE measurements at center-of-mass energies $\sqrt{s}=1.96$ and 7\TeV. A significant improvement in the description of UE measurements at 13\TeV is observed with respect to predictions from old tunes that were extracted using data at lower collision energies.
For the first time, predictions based on higher-order PDF sets are shown to give a reliable description of minimum-bias (MB) and UE measurements, with a similar level of agreement as predictions from tunes using LO PDF sets. \par}

Predictions of the new tunes agree well with the data for MB observables measured at pseudorapidities in the central ($\abs{\eta}< 2.4$) and forward ($3.2 <\abs{\eta}< 4.7$) regions.
The new CMS tunes simultaneously describe the number of charged particles produced in diffractive processes and MB collisions.
Neither the new CMS tunes nor the CUETP8M1 tune describe the very forward region ($-6.6<\eta<-5.2$) well.

{\tolerance=800 Measurements sensitive to double-parton scattering contributions are reproduced better by predictions using the LO PDF set in the UE simulation, without rapidity ordering of the initial-state shower. \par}

{\tolerance=800 The UE simulation provided by the new tunes can be interfaced to higher-order and multileg matrix element generators, such as \POWHEG and \MG, without degrading the good description of UE observables. Such predictions also reproduce well observables measured in multijet final states, Drell--Yan, and top quark production processes. The central values of the normalized \ttbar cross section in bins of the number of additional jets predicted by \POWHEG~+\PYTHIAviii overestimate the data when a high value of $\alpS^\mathrm{ISR}(m_\cPZ)\gtrsim0.130$ is used (CMS \PYTHIAviii CP1 and CP2 tunes). Even when $\alpS^\mathrm{ISR}(m_\cPZ)=0.118$ is used, the CP4 tune overestimates the data at high jet multiplicities. This is cured by the rapidity ordering of the initial-state shower (CP5 tune). Measurements of azimuthal dijet correlations are also better described when a value of $\alpS^\mathrm{ISR}(m_\cPZ)=0.118$ is used in predictions obtained with \POWHEG merged with \PYTHIAviii.  \par}

Comparisons with LEP event-shape observables and the distribution of the angle between two groomed subjets ($\Delta R_g$) in \ttbar events at the LHC show that in  ME-PS merged configurations CMW rescaling is disfavored. It is also found that $\Delta R_g$ is better described by tunes with $\alpS^\mathrm{FSR}(m_\cPZ)$ higher than $\sim$0.120 while LEP event-shape observables and UE event observables in \ttbar events prefer a central value $\sim$0.120~\cite{Sirunyan:2018avv}.

All of the new CMS tunes are supplied with their eigentunes, which can also be used to determine the uncertainties associated with the theoretical predictions.
We show that predictions using the new tunes based on PDFs determined at LO, NLO, and NNLO agree reasonably well with the measurements, and that the new tunes can also be applied to LO and NLO calculations merged with parton showers, multiple-parton interactions, and hadronization.

\begin{acknowledgments}
\hyphenation{Bundes-ministerium Forschungs-gemeinschaft Forschungs-zentren Rachada-pisek} We congratulate our colleagues in the CERN accelerator departments for the excellent performance of the LHC and thank the technical and administrative staffs at CERN and at other CMS institutes for their contributions to the success of the CMS effort. In addition, we gratefully acknowledge the computing centres and personnel of the Worldwide LHC Computing Grid for delivering so effectively the computing infrastructure essential to our analyses. Finally, we acknowledge the enduring support for the construction and operation of the LHC and the CMS detector provided by the following funding agencies: the Austrian Federal Ministry of Education, Science and Research and the Austrian Science Fund; the Belgian Fonds de la Recherche Scientifique, and Fonds voor Wetenschappelijk Onderzoek; the Brazilian Funding Agencies (CNPq, CAPES, FAPERJ, FAPERGS, and FAPESP); the Bulgarian Ministry of Education and Science; CERN; the Chinese Academy of Sciences, Ministry of Science and Technology, and National Natural Science Foundation of China; the Colombian Funding Agency (COLCIENCIAS); the Croatian Ministry of Science, Education and Sport, and the Croatian Science Foundation; the Research Promotion Foundation, Cyprus; the Secretariat for Higher Education, Science, Technology and Innovation, Ecuador; the Ministry of Education and Research, Estonian Research Council via IUT23-4, IUT23-6 and PRG445 and European Regional Development Fund, Estonia; the Academy of Finland, Finnish Ministry of Education and Culture, and Helsinki Institute of Physics; the Institut National de Physique Nucl\'eaire et de Physique des Particules~/~CNRS, and Commissariat \`a l'\'Energie Atomique et aux \'Energies Alternatives~/~CEA, France; the Bundesministerium f\"ur Bildung und Forschung, Deutsche Forschungsgemeinschaft, and Helmholtz-Gemeinschaft Deutscher Forschungszentren, Germany; the General Secretariat for Research and Technology, Greece; the National Research, Development and Innovation Fund, Hungary; the Department of Atomic Energy and the Department of Science and Technology, India; the Institute for Studies in Theoretical Physics and Mathematics, Iran; the Science Foundation, Ireland; the Istituto Nazionale di Fisica Nucleare, Italy; the Ministry of Science, ICT and Future Planning, and National Research Foundation (NRF), Republic of Korea; the Ministry of Education and Science of the Republic of Latvia; the Lithuanian Academy of Sciences; the Ministry of Education, and University of Malaya (Malaysia); the Ministry of Science of Montenegro; the Mexican Funding Agencies (BUAP, CINVESTAV, CONACYT, LNS, SEP, and UASLP-FAI); the Ministry of Business, Innovation and Employment, New Zealand; the Pakistan Atomic Energy Commission; the Ministry of Science and Higher Education and the National Science Centre, Poland; the Funda\c{c}\~ao para a Ci\^encia e a Tecnologia, Portugal; JINR, Dubna; the Ministry of Education and Science of the Russian Federation, the Federal Agency of Atomic Energy of the Russian Federation, Russian Academy of Sciences, the Russian Foundation for Basic Research, and the National Research Center ``Kurchatov Institute"; the Ministry of Education, Science and Technological Development of Serbia; the Secretar\'{\i}a de Estado de Investigaci\'on, Desarrollo e Innovaci\'on, Programa Consolider-Ingenio 2010, Plan Estatal de Investigaci\'on Cient\'{\i}fica y T\'ecnica y de Innovaci\'on 2013--2016, Plan de Ciencia, Tecnolog\'{i}a e Innovaci\'on 2013--2017 del Principado de Asturias, and Fondo Europeo de Desarrollo Regional, Spain; the Ministry of Science, Technology and Research, Sri Lanka; the Swiss Funding Agencies (ETH Board, ETH Zurich, PSI, SNF, UniZH, Canton Zurich, and SER); the Ministry of Science and Technology, Taipei; the Thailand Center of Excellence in Physics, the Institute for the Promotion of Teaching Science and Technology of Thailand, Special Task Force for Activating Research and the National Science and Technology Development Agency of Thailand; the Scientific and Technical Research Council of Turkey, and Turkish Atomic Energy Authority; the National Academy of Sciences of Ukraine, and State Fund for Fundamental Researches, Ukraine; the Science and Technology Facilities Council, UK; the US Department of Energy, and the US National Science Foundation.

Individuals have received support from the Marie-Curie programme and the European Research Council and Horizon 2020 Grant, contract Nos.\ 675440 and 765710 (European Union); the Leventis Foundation; the A.P.\ Sloan Foundation; the Alexander von Humboldt Foundation; the Belgian Federal Science Policy Office; the Fonds pour la Formation \`a la Recherche dans l'Industrie et dans l'Agriculture (FRIA-Belgium); the Agentschap voor Innovatie door Wetenschap en Technologie (IWT-Belgium); the F.R.S.-FNRS and FWO (Belgium) under the ``Excellence of Science -- EOS" -- be.h project n.\ 30820817; the Beijing Municipal Science \& Technology Commission, No. Z181100004218003; the Ministry of Education, Youth and Sports (MEYS) of the Czech Republic; the Lend\"ulet (``Momentum") Programme and the J\'anos Bolyai Research Scholarship of the Hungarian Academy of Sciences, the New National Excellence Program \'UNKP, the NKFIA research grants 123842, 123959, 124845, 124850, 125105, 128713, 128786, and 129058 (Hungary); the Council of Scientific and Industrial Research, India; the HOMING PLUS programme of the Foundation for Polish Science, cofinanced from European Union, Regional Development Fund, the Mobility Plus programme of the Ministry of Science and Higher Education, the National Science Center (Poland), contracts Harmonia 2014/14/M/ST2/00428, Opus 2014/13/B/ST2/02543, 2014/15/B/ST2/03998, and 2015/19/B/ST2/02861, Sonata-bis 2012/07/E/ST2/01406; the National Priorities Research Program by Qatar National Research Fund; the Programa de Excelencia Mar\'{i}a de Maeztu, and the Programa Severo Ochoa del Principado de Asturias; the Thalis and Aristeia programmes cofinanced by EU-ESF, and the Greek NSRF; the Rachadapisek Sompot Fund for Postdoctoral Fellowship, Chulalongkorn University, and the Chulalongkorn Academic into Its 2nd Century Project Advancement Project (Thailand); the Welch Foundation, contract C-1845; and the Weston Havens Foundation (USA).
\end{acknowledgments}
\bibliography{auto_generated}

\clearpage
\appendix

\section{Tables of tune uncertainties}
\label{TablesOfTuneUncertainties}
This section provides the values of the parameters corresponding to the uncertainties when the new CMS \PYTHIAviii tunes are used. The tune uncertainty is obtained by extracting the eigentunes, which are defined by a change in the $\chi^2$ of the fit that equals the absolute $\chi^2$ value obtained in the tune. The eigentunes refer to the variations of the tunes along each of the maximally independent directions in the parameter space, obtained by using the covariance matrix in the region of the best tune. The number of directions defined in the parameter space equals the number of free parameters $n$ used in the fit and results into 2$n$ parameter variations, \ie, eigentunes. These variations represent a good set of systematic uncertainties in the given tune.

The estimations of the tune uncertainties, which have 2$n$ parameter variations, \ie, 10 for the new CMS \PYTHIAviii tunes, are very time consuming in analyses, since for each variation separate samples must be produced.
Therefore, a lower number of variations is preferred. Hence, two variations, one ``up" and one ``down", are defined. For the definition of the two variations, predictions using the parameters of the eigentunes are implemented for the UE observables at $\sqrt{s}=13\TeV$ and their differences with respect to the central predictions are added in quadrature. This procedure is applied in each bin and tune uncertainties are estimated without any correlation across the different bins. Positive differences between central predictions and tune variations define the upper edge of the bin-by-bin uncertainty, while negative differences define the lower edge of the bin-by-bin uncertainty. By following the same approach used for the extraction of the central values of the new CMS \PYTHIAviii tunes, the upper edge is fitted to obtain the ``up variation", while the ``down variation" is obtained by fitting the lower edge. The parameters of the up- and down-variations are listed in Table~\ref{tableUpDown_lo} for the tunes using LO PDF sets, and in Table~\ref{tableUpDown} for the tunes using (N)NLO PDF sets. We checked that for a wide range of MB and UE observables at $\sqrt{s}=13\TeV$ predictions from up- and down-variations, obtained by including the full set of eigenvalues, reproduce well the upper and the lower edge of the predictions. Hence, tune uncertainties estimated by evaluating predictions of up- and down-variations represent a reliable way of estimating the systematic uncertainties in the tunes.
The correlation matrix for the fit of the CP5 tune is displayed in Table~\ref{tableCorr}.
It is retrieved by evaluating the correlation of the parameter variations obtained in the eigentunes.

\begin{table*}[htb]
\centering
\topcaption{Parameters of the ``up-" and ``down-"variation eigentunes for the \PYTHIAviii CP1, and CP2 tunes.}
\label{tableUpDown_lo}
\begin{tabular}{lcccc}
\PYTHIAviii Parameter  & CP1 & CP1 & CP2 & CP2 \\
					  & Up	& Down & Up	& Down \\ \hline
\texttt{MultipartonInteractions:pT0Ref} [\GeVns{}] & 2.30    & 2.40    & 2.34   & 2.33\\
\texttt{MultipartonInteractions:ecmPow}       & 0.15  & 0.15  & 0.14 & 0.14\\
\texttt{MultipartonInteractions:coreFraction} & 0.51  & 0.39 & 0.51 & 0.23\\
\texttt{MultipartonInteractions:coreRadius}   & 0.58  & 0.60 & 0.41  & 0.34\\
\texttt{ColourReconnection:range} 			 & 8.31  & 8.50    & 1.46  & 2.56\\
\end{tabular}
\end{table*}

\begin{table*}[htbp]
\centering
\topcaption{Parameters of the ``up-" and ``down-"variation eigentunes for the \PYTHIAviii CP3, CP4, and CP5 tunes.}
\label{tableUpDown}
\cmsTable{
\begin{tabular}{lcccccc}
\PYTHIAviii Parameter  & CP3 & CP3 & CP4 & CP4 & CP5  & CP5\\
					  & Up	& Down & Up	& Down & Up	& Down  \\ \hline
\texttt{MultipartonInteractions:pT0Ref} [\GeVns{}] & 1.48 & 1.54 & 1.48 & 1.54 &  1.41 & 1.46 \\
\texttt{MultipartonInteractions:ecmPow}       & 0.02 & 0.02 & 0.02 & 0.02 & 0.03 & 0.03 \\
\texttt{MultipartonInteractions:coreFraction} & 0.35 & 0.25 & 0.36 & 0.33 & 0.43 & 0.73 \\
\texttt{MultipartonInteractions:coreRadius}   & 0.49 & 0.35 & 0.58 & 0.57 & 0.67 & 0.69 \\
\texttt{ColourReconnection:range} & 8.15 & 3.96 & 7.93 & 6.88 & 4.88 & 4.69 \\
\end{tabular}
}
\end{table*}

\begin{table*}[htb]
\centering
\topcaption{The correlation matrix, retrieved when extracting the CP5 tune.
This is obtained by evaluating the correlation values of the parameter variations obtained in the  eigentunes.}
\label{tableCorr}
\cmsTable{
\begin{tabular}{lcccccc}
                      & \texttt{pT0Ref} & \texttt{ecmPow} & \texttt{coreFraction} & \texttt{coreRadius} & \texttt{range} \\ \hline
\texttt{pT0Ref}       & 1.00   &  -0.21 & -0.19        & -0.19      & 0.15  \\
\texttt{ecmPow}       & -0.21  & 1.00   & 0.30         & 0.69       & -0.21 \\
\texttt{coreFraction} & -0.19  & 0.30   & 1.00         & 0.32       & -0.64 \\
\texttt{coreRadius}   & -0.19  & 0.69   & 0.32         & 1.00       & -0.52 \\
\texttt{range}        &  0.15  & -0.21  & -0.64        & -0.52      & 1.00 \\
\end{tabular}
}
\end{table*}

Variations of the values of the ISR and FSR are also studied, in order to
check the consistency of the selected $\alpS^\mathrm{ISR}(m_\cPZ)$ and $\alpS^\mathrm{FSR}(m_\cPZ)$ values selected for the tunes and to estimate the allowed range of $\alpS^\mathrm{ISR}(m_\cPZ)$ and $\alpS^\mathrm{FSR}(m_\cPZ)$ values in the PS using the CP5 tune.
 Starting from tune CP5, the value of $\alpS^\mathrm{ISR}(m_\cPZ)$ is fitted to UE observables measured by CMS at $\sqrt{s}=13\TeV$. The same procedure is repeated when $\alpS^\mathrm{FSR}(m_\cPZ)$ is fitted. The parameters obtained from the fits are shown in Table~\ref{tableUpDownPS}, along with the up and down variation.

\begin{table*}[htbp]
\centering
\topcaption{``Up" and ``Down" ISR and FSR variations for CP5 when $\alpS^\mathrm{ISR}(m_\cPZ)$ or $\alpS^\mathrm{FSR}(m_\cPZ)$ is treated as a free parameter.}
\label{tableUpDownPS}
\begin{tabular}{l c c c c}
\PYTHIAviii Parameter & Central & Up & Down & $\chi^2$/dof \\ \hline
$\alpS^\mathrm{ISR}(m_\cPZ)$ value & 0.121 & 0.128 & 0.114 & 0.75 \\
$\alpS^\mathrm{FSR}(m_\cPZ)$ value & 0.119 & 0.122 & 0.115 & 0.78\\
\end{tabular}
\end{table*}

\begin{figure*}[ht!]
\centering
\includegraphics[width=0.49\textwidth]{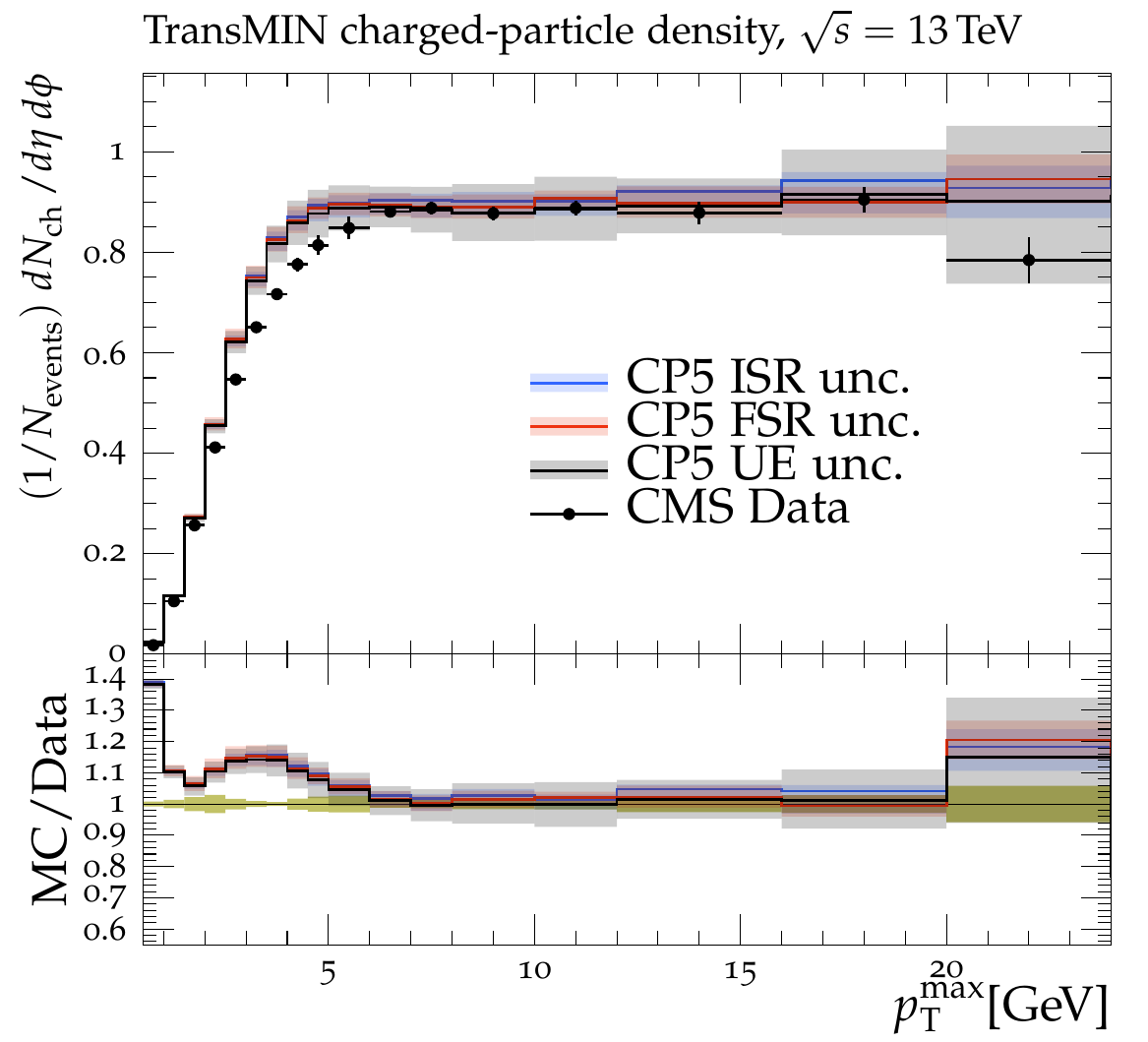}
\includegraphics[width=0.49\textwidth]{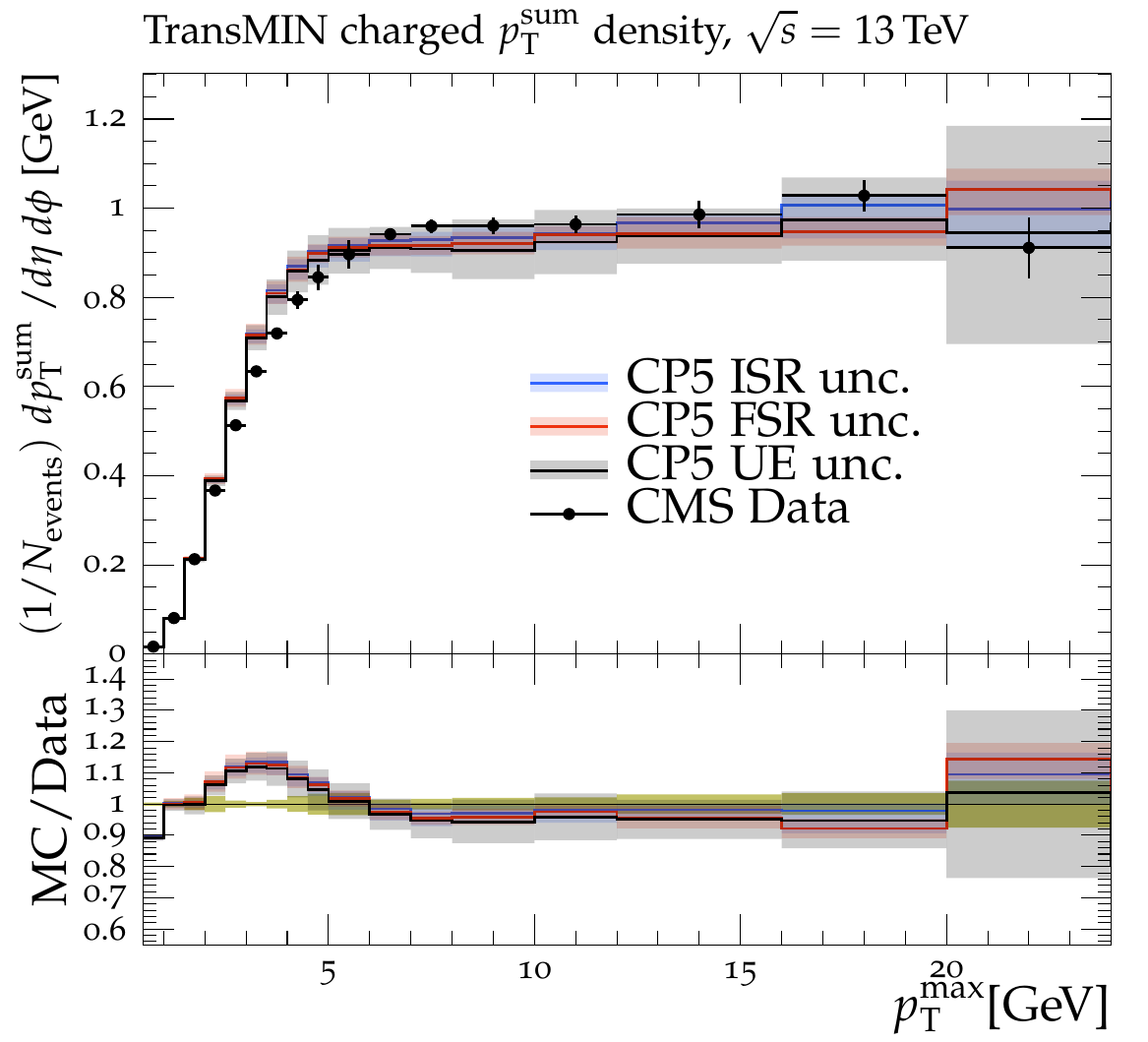}
\caption{The variations allowed by the CP5 tune when $\alpS^\mathrm{ISR}(m_\cPZ)$ (blue band) and $\alpS^\mathrm{FSR}(m_\cPZ)$ (red band) are left free in the fit for charged-particle (left) and charged \ptsum\ (right) density in the \tmin\ region at $\sqrt{s}=13\TeV$. Vertical lines drawn on the data points refer to the total uncertainty in the data. The grey band represents the total UE uncertainty for the tune CP5. Horizontal bars indicate the associated bin width.}
\label{fig:isrfsr}
\end{figure*}

Figure~\ref{fig:isrfsr} shows the predictions of the CP5 tune, with the corresponding variation bands relative to the UE parameters, and the $\alpS^\mathrm{ISR}(m_\cPZ)$ and $\alpS^\mathrm{FSR}(m_\cPZ)$ values for the charged-particle and \ptsum\ densities at $\sqrt{s}=13\TeV$ in the \tmin\ region.

\cleardoublepage \section{The CMS Collaboration \label{app:collab}}\begin{sloppypar}\hyphenpenalty=5000\widowpenalty=500\clubpenalty=5000\vskip\cmsinstskip
\textbf{Yerevan Physics Institute, Yerevan, Armenia}\\*[0pt]
A.M.~Sirunyan, A.~Tumasyan
\vskip\cmsinstskip
\textbf{Institut f\"{u}r Hochenergiephysik, Wien, Austria}\\*[0pt]
W.~Adam, F.~Ambrogi, E.~Asilar, T.~Bergauer, J.~Brandstetter, M.~Dragicevic, J.~Er\"{o}, A.~Escalante~Del~Valle, M.~Flechl, R.~Fr\"{u}hwirth\cmsAuthorMark{1}, V.M.~Ghete, J.~Hrubec, M.~Jeitler\cmsAuthorMark{1}, N.~Krammer, I.~Kr\"{a}tschmer, D.~Liko, T.~Madlener, I.~Mikulec, N.~Rad, H.~Rohringer, J.~Schieck\cmsAuthorMark{1}, R.~Sch\"{o}fbeck, M.~Spanring, D.~Spitzbart, W.~Waltenberger, J.~Wittmann, C.-E.~Wulz\cmsAuthorMark{1}, M.~Zarucki
% ...
\vskip\cmsinstskip
% ...
1:  Also at Vienna University of Technology, Vienna, Austria\\
% ...
\end{sloppypar}
\end{document}